\documentclass[twocolumn]{aastex7}

\usepackage{amsmath}
\usepackage{graphicx}
\usepackage{hyperref}
\usepackage{multirow}
\usepackage{etoolbox} 
\usepackage{lipsum} 

\usepackage{makecell}
\usepackage{subcaption}

\shorttitle{The QN and O\, $II$ lines in  SLSNe-I}
\shortauthors{R. Ouyed}

\begin{document}

\title{Late-Onset Energy Injection in Type Ic SNe and W-Shaped O\,{\small II}  Absorption in SLSNe-I}

\author{Rachid Ouyed}
\email{rouyed@ucalgary.ca}
\affiliation{Department of Physics and Astronomy, University of Calgary, 2500 University Drive NW, Canada}

\correspondingauthor{Rachid Ouyed}
\email{rouyed@ucalgary.ca}

\begin{abstract}
We show that delayed (weeks to months) energy injection into already expanding Type~Ic supernova (SN) ejecta can reproduce both the luminosity and spectral evolution of hydrogen-poor superluminous supernovae (SLSNe-I). Late-time reheating of the ejecta establishes the radiation temperature and density required for the emergence of the characteristic W-shaped O\,\textsc{ii} absorption complex near peak luminosity and naturally explains its disappearance as the ejecta expands and cools, without invoking departures from the nebular approximation or additional excitation mechanisms.
In our model, the neutron star (NS) formed in the core-collapse explosion undergoes a phase transition in its core to deconfined quark matter at time $t_{\rm QN}$. The transition is accompanied by rapid QCD-driven magnetic-field amplification, leading to the formation of a hybrid star (HS) that acts as a QCD-magnetar. This Quark-Nova (QN) event shortens the NS spin-down timescale and converts the remaining rotational energy into a renewed episode of energy injection, effectively resetting the central engine and producing two distinct powering epochs separated by a delay governed by the microphysics of the hadron-to-quark transition.
The model successfully reproduces the photometric and spectroscopic evolution of representative SLSNe-I, including iPTF13ajg, SN~2010gx, PTF09cnd, and PTF09atu. A key prediction is a systematic offset between spectroscopic expansion ages and photometric ages when the true explosion epoch is missed because the pre-QN emission lies below survey detection limits. We also discuss how QCD-magnetars can be distinguished observationally from standard magnetars and explore the implications for astrophysics and QCD. In particular, double-peaked SLSNe-I may serve as probes of the hadron-to-quark matter transition, enabling constraints on fundamental quark-matter parameters such as the deconfinement density and the hadron--quark surface tension.
\end{abstract}


\section{Introduction}
\label{sec:introduction}

SLSNe are among the most luminous stellar explosions in the Universe with luminosities $\sim 10$--$100$ times greater than those of normal SNe (\citealt{ofek_2007,quimby_2007,smith_2008,barbary_2009,young_2010,quimby_2011,galyam_2012,gomez_2020,moriya_2024}).  
 They are extremely rare events with  a volumetric rate of about 1 in every few
thousand SNe (\citealt{quimby_2013,frohmaier_2021}) with many
at redshifts exceeding 0.5 (e.g., \citealt{galyam_2019,nicholl_2021,moriya_2024} for a review).
 
A variety of physical models have been proposed to explain these extreme luminosities, including strong interaction with circumstellar material (CSM; \citealt{chevalier_2011,moriya_2013}), SN explosions with important radioactive material (\citealt{heger_2002,nomoto_2007,langer_2012}), jet-driven explosions (\citealt{soker_2017}),
fall-back accretion onto a black hole (\citealt{dexter_2013}), and sustained energy injection from the spin-down of a rapidly rotating magnetar \citep{ostriker_1971,kasen_2010,woosley_2010,inserra_2013}.
Although the different models proposed can reproduce the data (e.g. \citealt{moriya_2018} and references therein), the magnetar model \citep{duncan_1992} has emerged as the candidate capable of reproducing many of the global photometric properties of SLSNe \citep{nicholl_2021,chen_2023}. 

SLSNe can be separated into two main subclasses: the H-rich Type II SLSN group (SLSNe-II) and the H-poor  Type I (SLSNe-I)  (e.g., \citealt{branch_2017}).
 In this work we focus on SLSNe-I (\citealt{galyam_2009,barbary_2009,pastorello_2010,chomiuk_2011,quimby_2011,galyam_2012,inserra_2019,gomez_2024,aamer_2025}). 
 
 SLSNe-I are associated mainly with dwarf galaxies having low metallicity and high  star formation
rate (\citealt{neill_2011,stoll_2011,chen_2013,lunnan_2014,perley_2016,hatsukade_2018,schulze_2018,angus_2019}).
Observational and theoretical studies suggest these represent an  extension of the normal Type Ic population, with inferred ejecta masses  exceeding the few solar masses typical of ordinary Type Ic and broad-lined Ic SNe \citep[e.g.,][]{taddia_2015,nicholl_2016,blanchard_2020,gomez_2024}. In this interpretation, a massive stripped-envelope progenitor undergoes core collapse and is  re-energized by a central engine. The engine boosts the ejecta's peak luminosity by several magnitudes, broadens the light curve through an increased effective diffusion timescale, and can produce a relatively flat velocity evolution due to hydrodynamic interaction between the injected energy and the ejecta \citep{nicholl_2016}.  

Beyond their luminosity, SLSNe-I are characterized by distinctive early-time spectroscopic properties, most notably the presence of strong O\,\textsc{ii} absorption features in the $\sim(3600$–$5000)\text{\AA}$ range. Near peak brightness their spectra show extremely blue continua and broad, blended absorption features that distinguish them from ordinary stripped-envelope SNe (\citealt{dessart_2012,howell_2013,mazzali_2016}). In particular, pre-maximum spectra display a prominent broad, double-troughed (“W-shaped”) absorption complex around $\sim(4300$–$4700)\text{\AA}$ attributed to O\,\textsc{ii} transitions \citep{quimby_2011,leloudas_2012,inserra_2013,nicholl_2013,mazzali_2016,liu_2017,quimby_2018,galyam_2019,kumar_2020}. At later times, as the ejecta enters the nebular phase, their spectra evolve toward those of Type Ic and Type Ic-BL supernovae (e.g., \citealt{nicholl_2019} and references therein).

The W-shaped O\,\textsc{ii}  feature seems to be present only in a subset of events (Type~W SLSNe-I), while it is absent in others, such as SN~2015bn-like events \citep{quimby_2011,mazzali_2016,liu_2017,konyves-toth_2021,gutierez_2022}. Studies suggest differences in radiation temperature  and possibly ejecta geometry between these subclasses (\citealt{konyves-toth_2022}). Type~W events are generally hotter, with radiation temperatures $T_{\rm R} > 12{,}000$~K, whereas cooler events lack the characteristic O\,\textsc{ii} complex.  The suggested  bimodality is based on a small sample of SLSNe analyzed. 

\citet{mazzali_2016} argued that reproducing this feature requires a sustained ionization source beyond simple thermal excitation. 
\citet{saito_2024} on the other hand find that the observed features require only moderate departures from the nebular approximation
with the radiation temperature the primary controlling parameter. I.e., the O\,\textsc{ii} lines appear within a relatively narrow range of $T_{\rm R} \sim 14{,}000$-$16{,}000$~K near the photosphere.

While the origin of the W-shaped feature remains to be fully understood and explained, hints that 
 their presence or absence may be primarily a consequence of the thermal history of the ejecta is interesting.  We argue that this is 
 particularly possible if the SN ejecta is re-heated at an ``old age". Our motivation is to 
 consider energy-injection scenarios in which the timing of power deposition, relative to the expansion of the ejecta, plays a central role. 
 
 In contrast to standard central-engine models that assume prompt and continuous energy injection from birth, we explore a delayed-energy framework in which significant power is deposited only after the ejecta has substantially expanded, a few weeks
to a few months since the core-collapse SN explosion.  Specifically, here we focus on the Quark-Nova (QN) model where the
NS core experiences a conversion to a quark matter phase (made of up and down quarks) capable of sustaining $\sim 10^{18}$ G
magnetic fields (see \citealt{ouyed_2025a} and references therein).  The resulting hybrid star (HS), containing a quark-matter core described by Quantum Chromodynamics (QCD), can develop surface magnetic fields of $\gtrsim10^{15}$ G. Throughout this work we refer to such objects as QCD-magnetars, highlighting the QCD origin of their magnetic fields.

 For reviews of the microphysics and macrophysics of the QN, we refer the reader to \citet{ouyed_2022a} and \citet{ouyed_2022b}, respectively.
The delay time $t_{\rm QN}$, between the birth of the NS (i.e., the initial SN explosion) and its conversion to an HS,  is the key free parameter of the QN model. 
It is governed by quantum nucleation processes in the NS core once critical densities for deconfinement to (up and down) quark mattter, are reached. 

Our model invokes NSs on the heavy end of the mass spectrum allowed by the equation of state (i.e., $M_{\rm NS}\sim 2M_{\odot}$), capable of reaching a required critical core densities for deconfinement and quantum quark nucleation (e.g., \citealt{staff_2006}). These NSs are born with millisecond spin periods, providing the necessary reservoir of rotational energy to power the event. Finally, a magnetic field $B_{\rm NS}$ sets the spin-down timescale, which regulates the rate at which this rotational energy is injected into the ejecta prior to conversion to a HS.

 The transition to a HS and SN ejecta's rebrightening proceeds through the following stages:

\begin{itemize}

\item {\bf The QN ejecta:}  The conversion releases $\sim 100\,\mathrm{MeV}$ 
($\sim 10^{-4}\,\mathrm{erg}$) per baryon as hadronic matter transitions to deconfined quark matter \citep{glendenning_1997,weber_2005}. 
For a core of radius $R_{\rm c}$ containing 
$\sim 10^{57}\,(R_{\rm c}/R_{\rm NS})^{3}$ baryons, 
the averaged total liberated energy  therefore  exceeds
$10^{50}$-$10^{51}$ erg for typical core size of a few kilometres; $R_{\rm NS}$ is the NS radius. 

A fraction  is converted into thermal pressure that ejects the outermost layers of the star.
For $R_{\rm c}\sim 2$ km for example, typically 
$M_{\rm QN} \sim 10^{-2}\,M_{\odot}$ of NS material is ejected at characteristic velocities 
$v_{\rm QN} \sim 0.1c$ \citep{keranen_2005}; i.e. $E_{\rm QN}\sim 10^{50}$ erg in kinetic energy. This material constitutes the neutron-rich QN ejecta which
was shown to be a viable site for r-process nucleo-synthesis (\citealt{jaikumar_2007}).

\item {\bf QCD-magnetar}   At densities characteristic of massive NS cores, certain quark-matter phases can sustain internal magnetic fields up to 
$\sim 10^{18}\,\mathrm{G}$ \citep{iwazaki_2003,iwazaki_2005,ebert_2005,dvornikov_2016, efrain_2021}. The underlying reason is that quark matter is characterized by energy scales of order 
of hundreds of MeV, substantially exceeding the few MeV scales of hadronic matter, thereby allowing the support of much stronger internal magnetic fields (see Appendix \ref{sec:SLSNe-QCD}).  
Assuming approximate magnetic-flux conservation across the phase transition, leading to a dipole-like configuration, implies surface dipole fields exceeding
$10^{15}\,\mathrm{G}$ for the newly formed HS.  I.e., extreme surface magnetic fields arise naturally as a consequence of the quark-matter phase transition itself, rather than through
 the standard classical magnetic filed amplification during the 
proto-NS phase  such as convective dynamos (\citealt{thompson_1993,raynaud_2020,masada_2022}), and/or 
the magnetorotational instability (MRI; \citealt{akiyama_2003,obergaulinger_2009,reboul-salze_2021}).
 In this sense, our model is uniquely characterized by two distinct epochs of SN ejecta powering: a pre-$t_{\rm QN}$ (NS spin-down powered) phase and a post-$t_{\rm QN}$ (HS spin-down powered) phase. 

\item {\bf LFBOT production (thermalization of NS spin-down power in the QN ejecta):}   At $t_{\rm QN}$, the HS retains the residual rotational energy of the NS, but now spins down under a substantially stronger magnetic field.  The sudden field amplification drastically shortens the spin-down timescale, leading to the rapid release of the remaining rotational energy. 
If the QN ejecta efficiently absorbs and reprocesses this energy before interacting with the more massive SN ejecta, the resulting emission can manifest as a luminous fast blue optical transient (LFBOT; e.g., \citealt{drout_2014,pursiainen_2018,metzger_2022}).  The LFBOT emission is either embedded within the still optically thick SN ejecta (\citealt{ouyed_2025a}) or becomes directly observable once the ejecta has become optically thin, if the NS conversion occurs at a later time (\citealt{ouyed_2025b}).

The relatively small QN ejecta mass of $M_{\rm QN}\sim 10^{-2}M_{\odot}$,  implies short diffusion timescales, of a few days, and high peak luminosities ($\ge 10^{45}$ erg s$^{-1}$).
The reprocessing (effectively softening) of the hard radiation produced by the HS spin-down through interaction with the QN ejecta is a distinctive feature of our model. In this picture, the QN ejecta shields  the overlying material from the highly ionizing hard radiation that would otherwise arise from direct spin-down powering.

\item {\bf SLSN-I formation (thermalization of LFBOT power in the SN ejecta):}  The radiation emerging from the  LFBOT subsequently propagates into the more massive SN ejecta, characterized by mass $M_{\rm ej}$ and velocity $v_{\rm ej}$, with a diffusion timescale $t_{\rm ej,d}$. 
If $t_{\rm QN} \lesssim 2t_{\rm ej,d}$, the injected luminosity is efficiently trapped and thermalized \citep{ouyed_2025b}.  
The event is then observed as a SLSN-I, whose photometric and spectroscopic properties are governed primarily by the timing and magnitude of this delayed energy deposition \citep{ouyed_2025a}. 

\end{itemize}

Two dimensionless ratios govern the observable outcomes: 
(i) the ratio of the NS spin-down timescale to the delay time, $t_{\rm NS,SpD}/t_{\rm QN}$, and 
(ii) the ratio of the SN ejecta diffusion timescale to the delay time, $t_{\rm ej,d}/t_{\rm QN}$. 
The first determines the fraction of the NS rotational energy expended prior to conversion and the amount remaining to power the post-QN phase. 
The second controls the efficiency with which the injected energy is trapped and reprocessed by the expanding SN ejecta. 
Together, these ratios regulate the partition of the NS spin-down energy between the pre- and post-conversion epochs, thereby governing the relative peak luminosities, durations, and color evolution of the resulting light curve components.

At the time of conversion, $t_{\rm QN}$, the fraction of the initial (NS)  rotational energy remaining is
$(1 + t_{\rm QN}/t_{\rm NS,SpD})^{-2/(n-1)}$, 
where $n$ is the braking index (here, $n=3$ for a constant magnetic field) and $t_{\rm NS,SpD}$  the NS spin-down timescale (\citealt{manchester_1977,shapiro_1983,lyne_1993}). 
The corresponding spin period at conversion, inherited by the HS, is
\begin{equation}
P_{\rm HS} =
P_{\rm NS}
\left(1 + \frac{t_{\rm QN}}{t_{\rm NS,SpD}}\right)^{1/2}\ .
\label{eq:PHS-from-PNS}
\end{equation}

The condition $t_{\rm QN}/t_{\rm NS,SpD} < 1$ is necessary but not sufficient for a prominent post-QN signature; we additionally require $t_{\rm QN}/t_{\rm ej,d} < 2$ to ensure efficient trapping of the injected LFBOT luminosity by the SN ejecta. 

Focusing on the ratio $t_{\rm QN}/t_{\rm NS,SpD}$, we identify three limiting regimes that illustrate the diversity of outcomes in the model (discussed jointly with $t_{\rm QN}/t_{\rm ej,d}$ in \S~\ref{sec:model}):

\begin{itemize}

\item $t_{\rm QN} < t_{\rm NS,SpD}$:  
A large fraction of the rotational energy remains at conversion. 
The post-QN phase dominates the energetics, producing a luminous secondary peak and elevated radiation temperatures. As we show in this work, 
these conditions favor the emergence of W-shaped O\,\textsc{ii} absorption features.  The pre-$t_{\rm QN}$ phase would manifest itself 
 as lower-luminosity bumps in the light curve prior to the main event. Observations would likely detect only the peaks of the
 pre-$t_{\rm QN}$ light curve while the full underlying shape and emission remain hidden.

\item $t_{\rm QN} \sim t_{\rm NS,SpD}$:  
A good  fraction of the initial rotational energy has already been expended prior to conversion. 
Energy injection proceeds in two comparable stages. 
Radiation temperatures are lower,  thus suppressing high-ionization O\,\textsc{ii} features. 
Such events may retain a more classical Type~Ic spectroscopic appearance during both the pre- and post-$t_{\rm QN}$ phases while exhibiting a relatively symmetric double-peaked light curve. We suggest SN2019stc (\citealt{gomez_2021})  as a possible candidates.

\item $t_{\rm QN} > t_{\rm NS,SpD}$:  
The NS has largely spun down before conversion, leaving only a small residual rotational reservoir. 
The post-QN contribution is therefore weak, and no pronounced secondary peak is expected. 
In this limit, one recovers the classical magnetar scenario (involving a high-$B$ NS) with prompt energization of the
Type Ic ejecta. The energy injection in the post-$t_{\rm QN}$ phase is via the conversion of the
QN ejecta's kinetic energy into heating the SN ejecta resulting in late bumps in contrast to the $t_{\rm QN} < t_{\rm NS,SpD}$ case
above  (see \S \ref{sec:bumps} for further discussion).

\end{itemize}

In this work, we investigate the photometry and spectroscopy of Type~Ic SN ejecta experiencing a QN event occurring weeks to months after the explosion.
We explore conditions conducive to producing the characteristic W-shaped O\,\textsc{ii} absorption complex. 

The paper is organized as follows. In \S~\ref{sec:model}, we present the model and its governing equations in detail. We compute the luminosity contributions from $^{56}$Ni decay, NS spin-down power,  the LFBOT starting at $t_{\rm QN}$, together with their subsequent processing by the expanding SN ejecta. Pseudo-bolometric SLSN-I light curves are then calculated for a range of SN ejecta parameters by systematically varying the ratios $t_{\rm QN}/t_{\rm NS,SpD}$ and $t_{\rm QN}/t_{\rm ej,d}$, and are compared with observed SLSNe-I data.

In \S~\ref{sec:spectroscopy-ANALYTICAL}, we analytically estimate the physical conditions required for the formation of the W-shaped O\,\textsc{ii} absorption features predicted by the model. A detailed spectral analysis is presented in \S~\ref{sec:spectroscopy-TARDIS} using \texttt{TARDIS}. There, we simultaneously fit both the photometric light curves and spectral evolution of four well-observed SLSNe-I: iPTF13ajg, SN~2010gx, PTF09cnd, and PTF09atu spanning both the narrow- and broad-lined subclasses of SLSNe-I. 

We discuss the implications, limitations, and testable predictions of the model in \S~\ref{sec:discussion}. 
In that section we also refer the reader to Appendix~\ref{sec:SLSNe-QCD}, which is dedicated to demonstrating how SLSNe-I data can be used to probe the physics of the hadron-to-quark transition and  constrain quantities such as the transition density in NS cores and the hadronic-quark-matter surface tension.
 We conclude in \S~\ref{sec:conclusion}.
  
 %
\begin{table*}[t!]
\centering
\caption{Fiducial values of model parameters.}
\label{table:parameters}
\begin{tabular}{|c|c|c|c||c|c||c|c|c||c|}
\hline
\multicolumn{4}{|c||}{Supernova$^1$} & 
\multicolumn{2}{c||}{Neutron Star$^2$} & 
\multicolumn{3}{c||}{QN Ejecta$^3$} & 
Hybrid Star$^4$ \\
\hline
$v_{\rm ej}$ (cm s$^{-1}$) & $\kappa_{\rm ej}$ (cm$^2$ g$^{-1}$) & $M_{\rm Ni}(M_{\rm ej})$ & $n_{\rm ej}$ & 
$P_{\rm NS}$ (ms) & $B_{\rm NS}$ (G) & 
$M_{\rm QN}$ ($M_\odot$) & $v_{\rm QN}$ (c) & $\kappa_{\rm QN}$ (cm$^2$ g$^{-1}$) & $B_{\rm HS}$ (G) \\
\hline
$1.2\times 10^9$ & 0.1 & $10^{-2}$ & 8 & 5  & $10^{12.5}$ & $10^{-2}$ & 0.1 & 3.0 & $10^{15}$ \\
\hline
\end{tabular}
$^1$ $t_{\rm ej, d}\simeq 32.8~{\rm d}\times (M_{\rm ej}/10M_{\odot})^{1/2}$;  $^2$ $t_{\rm NS, SpD}\sim 33$ yr;  $^3$ $t_{\rm QN, d}\simeq 3.6$ d;  $^4$ $P_{\rm HS}= P_{\rm NS}\times (1+t_{\rm QN}/t_{\rm NS, SpD})^{1/2}\simeq 5$ ms.
\end{table*}
%

\section{Quark-Nova Framework for SLSNe-I}
\label{sec:model}

Hereafter, dimensionless quantities are defined as $Q_x \equiv Q/10^x$, with all quantities in cgs units unless stated otherwise. Calculations are performed in the NS rest frame, corresponding to the ambient medium. Subscripts ``ej'' and ``w'' refer to SN ejecta and the w-shaped O\,II absorption-line quantities, respectively, while subscripts for the QN event (e.g., $t_{\rm QN}$) are distinguished from those describing HS properties (e.g., $t_{\rm HS,SpD}$ and $P_{\rm HS}$).

We consider a Type~Ic progenitor with ejecta of mass $M_{\rm ej}$  expanding at $v_{\rm ej}$. The diffusion timescale is
$t_{\rm ej,d} = \sqrt{2\kappa_{\rm ej} M_{\rm ej}/\beta c v_{\rm ej}} \simeq 32.8~{\rm d} \times (\kappa_{\rm ej,-1} M_{\rm ej, 34.3}/v_{\rm ej,9.1})^{1/2}$ 
for the model's fiducial values of $M_{\rm ej}= 10M_{\odot}$ (i.e. $10^{34.3}$ g in our equations)\footnote{The corresponding cgs values appearing in the analytic expressions are indicated in parentheses as powers of ten when necessary.}  and $v_{\rm ej} = 12,000~{\rm km~s^{-1}}$ ($\sim 10^{9.1}$ cm s$^{-1}$); $\beta = 4\pi^3/9$ accounts for geometric effects \citep{arnett_1982}. We adopt an optical opacity $\kappa_{\rm ej} = 0.1~{\rm cm^2~g^{-1}}$, representative of stripped-envelope SNe (e.g., \citealt{wheeler_2015}). 
 The model's fiducial values are listed in Table \ref{table:parameters}.  

\subsection{The Hybrid Star (HS)}

We consider a NS with mass $M_{\rm NS}=2M_{\odot}$, radius $R_{\rm NS}=12$ km, birth period $P_{\rm NS}=5~{\rm ms}$ ($10^{-2.3}~{\rm s}$)  and 
surface magnetic field $B_{\rm NS}=10^{12.5}~{\rm G}$. Here, we estimate spin-down timescales and corresponding 
luminosity of both the NS and the  HS. We assume an aligned, force-free magnetized wind for both of them  
with a moment of inertia $\simeq 2\times 10^{45}$ g cm$^2$ (\citealt{michel_1970,manchester_1977,shapiro_1983,lyne_1993,contopoulos_1999,lattimer_2005}).

  The NS  spin-down timescale is $t_{\rm NS,SpD} \simeq 33\, {\rm yr} \, B_{\rm NS,12.5}^{-2} P_{\rm NS,-2.3}^{2}$. Its rotational energy is $E_{\rm NS,rot} \simeq 1.7 \times 10^{51}~P_{\rm NS,-2.3}^{-2}~{\rm erg}$.  At $t_{\rm QN}$, the HS inherits a period $P_{\rm HS} = P_{\rm NS}\, (1+t_{\rm QN}/t_{\rm NS,SpD})^{1/2}$, and its surface magnetic field is amplified to $B_{\rm HS} = 10^{15}~{\rm G}$. 
   The characteristic HS spin-down timescale is then
\begin{equation}
\label{eq:tHSspd}
t_{\rm HS,SpD} \simeq 0.3~{\rm d} \, B_{\rm HS,15}^{-2} P_{\rm HS,-2}^{2} \ .
\end{equation}
The HS period is expressed in units of 10~ms for clarity and to distinguish it from the progenitor NS period. 
For $B_{\rm NS}< \sim 10^{13.5}$ G, and for $t_{\rm QN}$ of weeks to a few months, 
 the spin period of the resulting HS at transition  will remain essentially the same as that of the progenitor NS period.
 
  The HS spin-down luminosity is
\begin{equation}
L_{\rm HS,SpD} \sim 1.7 \times 10^{46}~{\rm erg~s^{-1}}\, B_{\rm HS,15}^{2} P_{\rm HS,-2}^{-4} \left(1 + \frac{t}{t_{\rm HS,SpD}}\right)^{-2}
\end{equation}
This energy is injected behind the QN ejecta, powering the LFBOT phase \citep{ouyed_2025a,ouyed_2025b}.

\subsection{The LFBOT}

The photon diffusion timescale through the QN ejecta is 
\begin{equation}
t_{\rm QN,d} \simeq 3.6~{\rm d}\, \kappa_{\rm QN,0.5}^{1/2} M_{\rm QN,31.3}^{1/2} v_{\rm QN,9.5}^{-1/2},
\end{equation}
where $v_{\rm QN}=0.1c$ ($\simeq 10^{9.5}~{\rm cm~s^{-1}}$)
 and $M_{\rm QN}=10^{-2}M_{\odot}$ ($\simeq 10^{31.3}$ g). Given the heavy-element composition of the QN ejecta \citep{jaikumar_2007}, we take 
 its optical opacity to be $\kappa_{\rm QN} \sim 3~{\rm cm^2~g^{-1}}$ ($\simeq 10^{0.5}~{\rm cm^2~g^{-1}}$).

For $t_{\rm HS,SpD} < t_{\rm QN,d}$, the bolometric luminosity peaks near $t_{\rm QN,d}$ (i.e., at time $t_{\rm QN}+t_{\rm QN,d}$ since the
SN event) with $L_{\rm LFBOT,pk} \sim E_{\rm HS,SpD}/3 t_{\rm QN,d}$. I.e.,  
\begin{equation}
\label{eq:Llfbot}
L_{\rm LFBOT,pk} \sim 4.6 \times 10^{44}~{\rm erg~s^{-1}} \frac{v_{\rm QN,9.5}^{1/2}}{P_{\rm HS,-2}^{2} \kappa_{\rm QN,0.5}^{1/2} M_{\rm QN,31.3}^{3/4}}\ .
\end{equation}

The QN ejecta may eventually catch up with the homologously expanding SN ejecta when
$v_{\rm QN} t_{\rm coll} = v_{\rm ej,in}(t_{\rm QN}+t_{\rm coll})$,
where $v_{\rm ej,in}$ is the average velocity of the slower inner layers of the SN ejecta. Here $t_{\rm coll}$ denotes the time elapsed since $t_{\rm QN}$. 
To ensure that most of the LFBOT luminosity is reprocessed and emitted prior to this collision, we require $t_{\rm QN,d} < t_{\rm coll}$. Using the above relation, this condition yields
(for $v_{\rm QN}>> v_{\rm ej,in}$)
\begin{equation}
t_{\rm QN} > t_{\rm QN,min} = \left(\frac{v_{\rm QN}}{v_{\rm ej,in}}\right)t_{\rm QN,d}
\sim 10\,t_{\rm QN,d} \sim 36~{\rm days},
\end{equation}
for fiducial parameters and a reasonable estimate $v_{\rm QN}/v_{\rm ej,in}\sim10$. 
Minimum delay times of weeks therefore ensure that the QN ejecta, and the associated LFBOT emission, remain physically separated from the overlying SN ejecta.

Finally, the QN ejecta will decelerate after sweeping up an ambient mass comparable to its own, with $t_{\rm QN,dec} \sim (3 M_{\rm QN}/4\pi \rho_{\rm amb} v_{\rm QN}^3)^{1/3}$. For sufficiently low ambient densities $\rho_{\rm amb}$, $t_{\rm QN,dec} > t_{\rm coll}$, and the collision proceeds as described. If the collision occurs during the QN diffusion (i.e., LFBOT) phase, both HS spin-down energy and QN kinetic energy contribute to reheating the SN ejecta. This scenario will be explored elsewhere (see \S~\ref{sec:discussion} for further discussion).

\subsection{SLSNe as Reprocessed LFBOTs}
\label{sec:reprocessed-LFBOT}

While optically thick, the SN ejecta acts as an effective calorimeter, reprocessing the LFBOT power. To ensure efficient energy deposition, we consider the regime
\begin{equation}
t_{\rm QN,min} = \frac{v_{\rm QN}}{v_{\rm ej,in}} t_{\rm QN, d} \lesssim t_{\rm QN} \lesssim 2 t_{\rm ej,d}\,,
\end{equation}
Throughout this paper we adopt $t_{\rm QN}=40$~days  as a fiducial value, ensuring that most of the LFBOT radiation is processed by the SN ejecta prior to the collision while satisfying $t_{\rm QN} \sim t_{\rm ej,d}$. In general, 
we find that best fits to SLSNe-I photometric and spectroscopic evolution imply delays of a few weeks to a few months between the SN and QN events. 

The SLSN peak luminosity is $L_{\rm SLSN, pk} = E_{\rm HS, SpD}/3 t_{\rm ej, d}$, or,
\begin{align}
\label{eq:L-SLSN-peak}
L_{\rm SLSN, pk} &= \frac{E_{\rm HS, SpD}}{3 t_{\rm QN}} \times \frac{t_{\rm QN}}{t_{\rm ej,d}} \nonumber\\
&\sim \frac{4.3\times 10^{43}~{\rm erg~s^{-1}}}{P_{\rm HS, -2}^2} \times \frac{t_{\rm QN}}{t_{\rm ej,d}}\ .
\end{align}

The corresponding radiation temperature at the photosphere is $T_{\rm R,ph} \simeq 1.4 T_{\rm eff}$, where
\begin{equation}
T_{\rm eff} = \left(\frac{L_{\rm SLSN,pk}}{4 \pi \sigma_{\rm SB} [v_{\rm ej} (t_{\rm QN}+t_{\rm ej,d})]^2} \right)^{1/4} \ ,
\end{equation}
with $\sigma_{\rm SB} $ the Stefan-Boltzmann constant. 
The factor 1.4 accounts for nonuniform radial temperature structure in the ejecta based on 
 $T_{\rm R}(r) = (1.4 - 0.4 (1 - R_{\rm ph}/r)) T_{\rm eff}$ \citep[e.g.,][]{mazzali_1993}, where $R_{\rm ph}$ is the photospheric radius.
This yields
\begin{align}
\label{eq:SLSN-TR}
T_{\rm R,ph} &\sim1.1\times 10^4~{\rm K}\left( \frac{1}{P_{\rm HS,-2}^2 v_{\rm ej,9.1}^{2} t_{\rm QN, 40}^2} \right)^{1/4}\times \\\nonumber
&\times \frac{1}{ (t_{\rm ej,d}/t_{\rm QN})^{1/4} (1+t_{\rm ej,d}/t_{\rm QN})^{1/2}}\ .
\end{align}
The $t_{\rm NS, SpD}/t_{\rm QN}$ enters via Eq. (\ref{eq:PHS-from-PNS}). Here, $t_{\rm QN}$ is in units of 40 days rather than in the cgs form $10^x$ s
 used for other quantities.\\

\begin{figure*}[t!]
\centering
\includegraphics[scale=0.4]{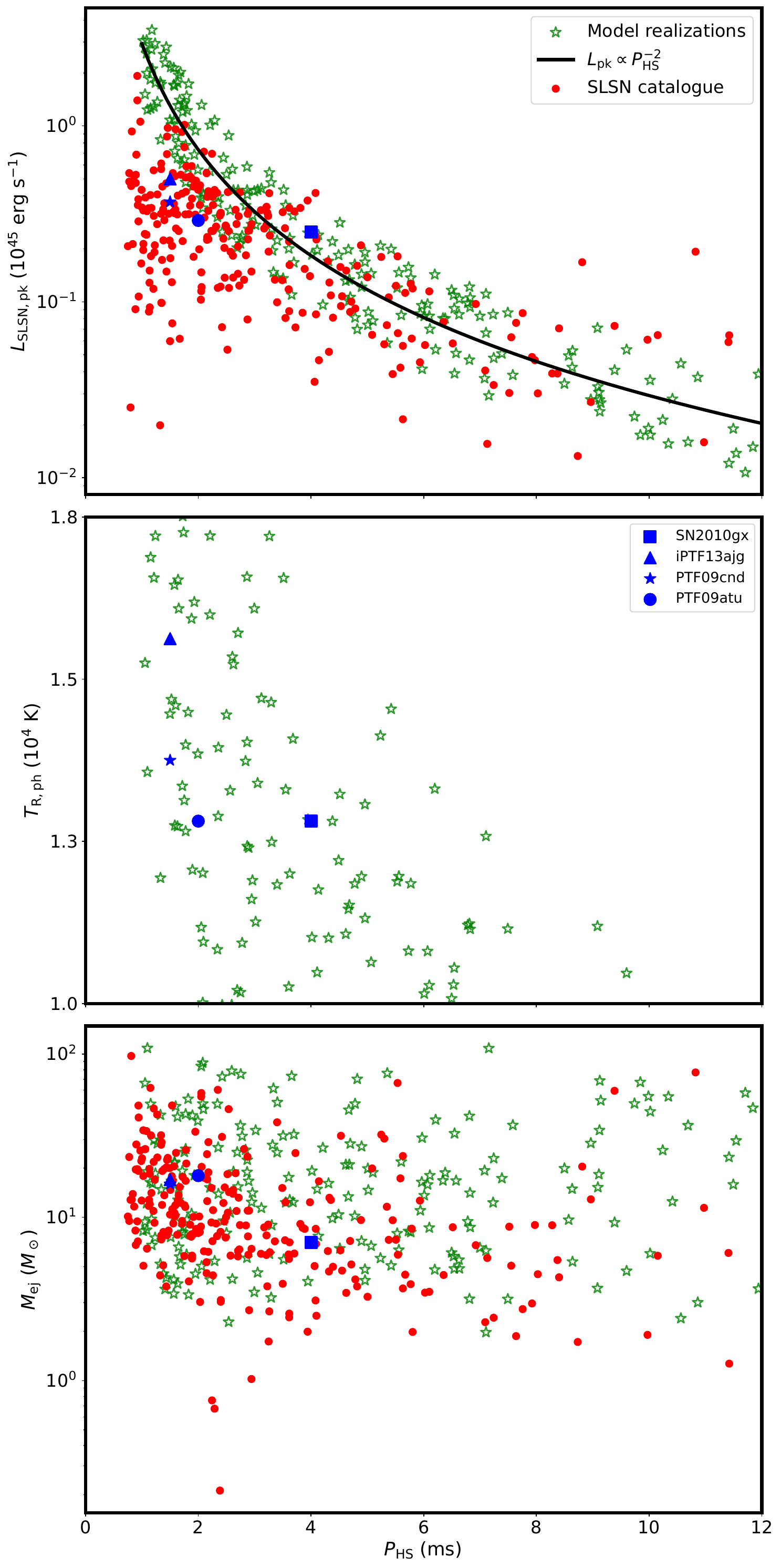}
\caption{
Peak luminosity (top panel), photospheric radiation temperature (middle panel), and ejecta mass (bottom panel) as functions of the HS spin period in our model. Green open stars show Monte Carlo realizations obtained by sampling $P_{\rm NS}$, $t_{\rm QN}/t_{\rm ej,d}$, and $v_{\rm ej}$ over the ranges adopted in the text (see \S \ref{sec:model}). Red points correspond to SLSNe-I from the observational catalogue of \citet{gomez_2024}, while blue symbols indicate representative best-fit parameters for four well-studied events (SN~2010gx, iPTF13ajg, PTF09cnd, and PTF09atu) within our model (see \S \ref{sec:spectroscopy-TARDIS-4SLSNe-I}). The solid black curve in the middle panel shows the analytic scaling $L_{\rm SLSN,pk} \propto P_{\rm HS}^{-2}$ for the case $t_{\rm ej,d}=t_{\rm QN}$ (see Eq.~\ref{eq:L-SLSN-peak}).
}
\label{fig:Mej-vsPms-TR-1}
\end{figure*}

Figure~\ref{fig:Mej-vsPms-TR-1} shows the results of Monte Carlo sampling,
where the parameters are independently drawn from log-uniform distributions within the ranges
$1 \le P_{\rm HS}~({\rm ms}) \le 12$,
$0.5 \le t_{\rm QN}/t_{\rm ej,d} \le 2$,
and $10{,}000 \le v_{\rm ej}~({\rm km~s^{-1}}) \le 15{,}000$.

The top and middle panels show the corresponding distributions of SLSN peak luminosities and photospheric radiation temperatures, respectively. The bottom panel shows the resulting $M_{\rm ej}$ versus $P_{\rm HS}$, computed from $t_{\rm ej,d} = \sqrt{2 \kappa_{\rm ej} M_{\rm ej} / \beta c v_{\rm ej}}$, with $0.05 \le \kappa_{\rm ej}~({\rm cm^2~g^{-1}}) \le 0.2$ also sampled from a log-uniform distribution.

The agreement with observed SLSNe data (\citealt{gomez_2024}) is encouraging. The scaling $L_{\rm SLSN,pk} \propto P_{\rm HS}^{-2}$ shown in the top panel assumes $t_{\rm QN} \sim t_{\rm ej,d}$ in Eq.~(\ref{eq:L-SLSN-peak}), corresponding to the regime of optimal reprocessing of the LFBOT and reheating of the SN ejecta, which leads to the maximum peak luminosity.

\begin{figure*}[t!]
\centering
\includegraphics[scale=0.4]{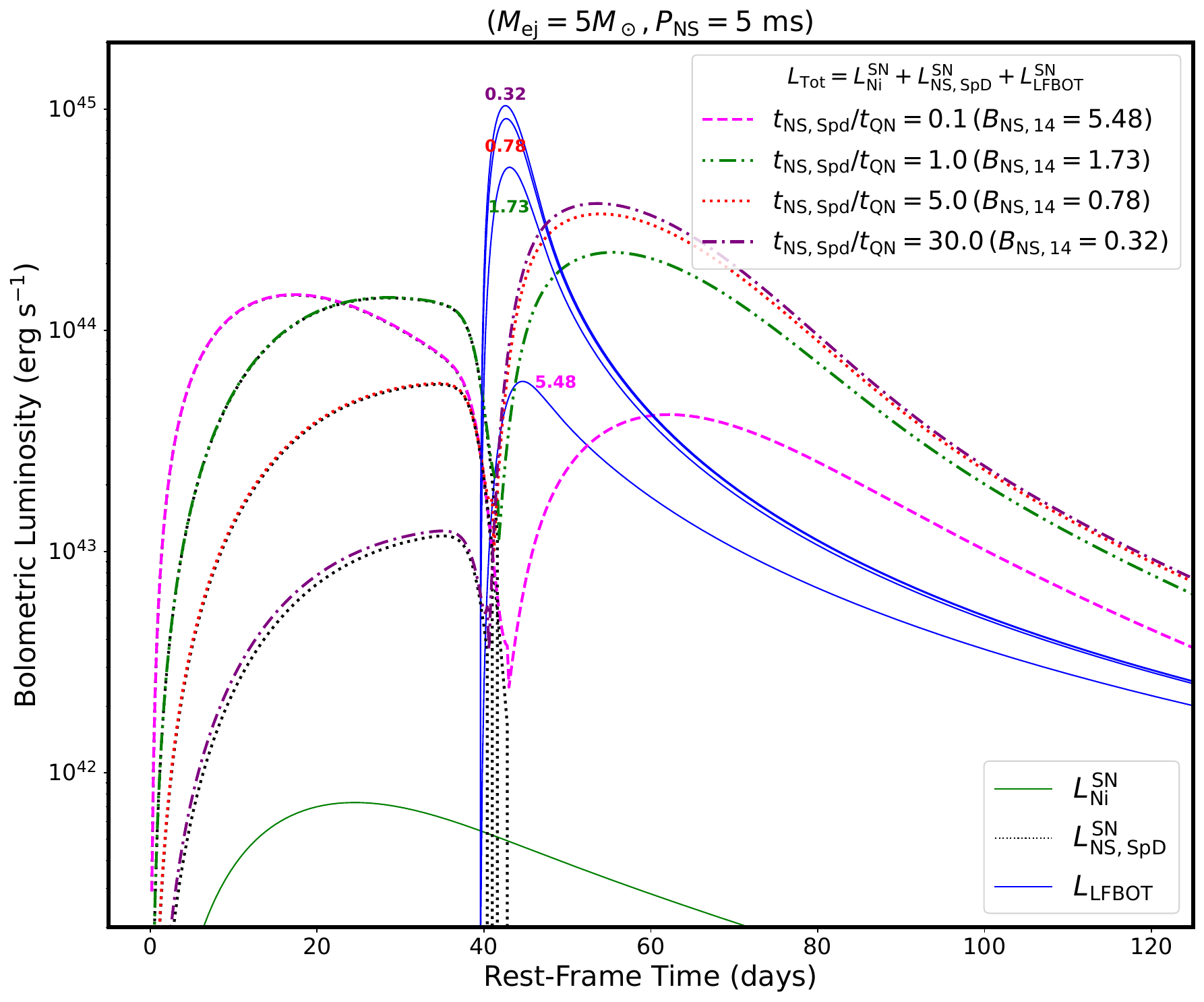}
\includegraphics[scale=0.4]{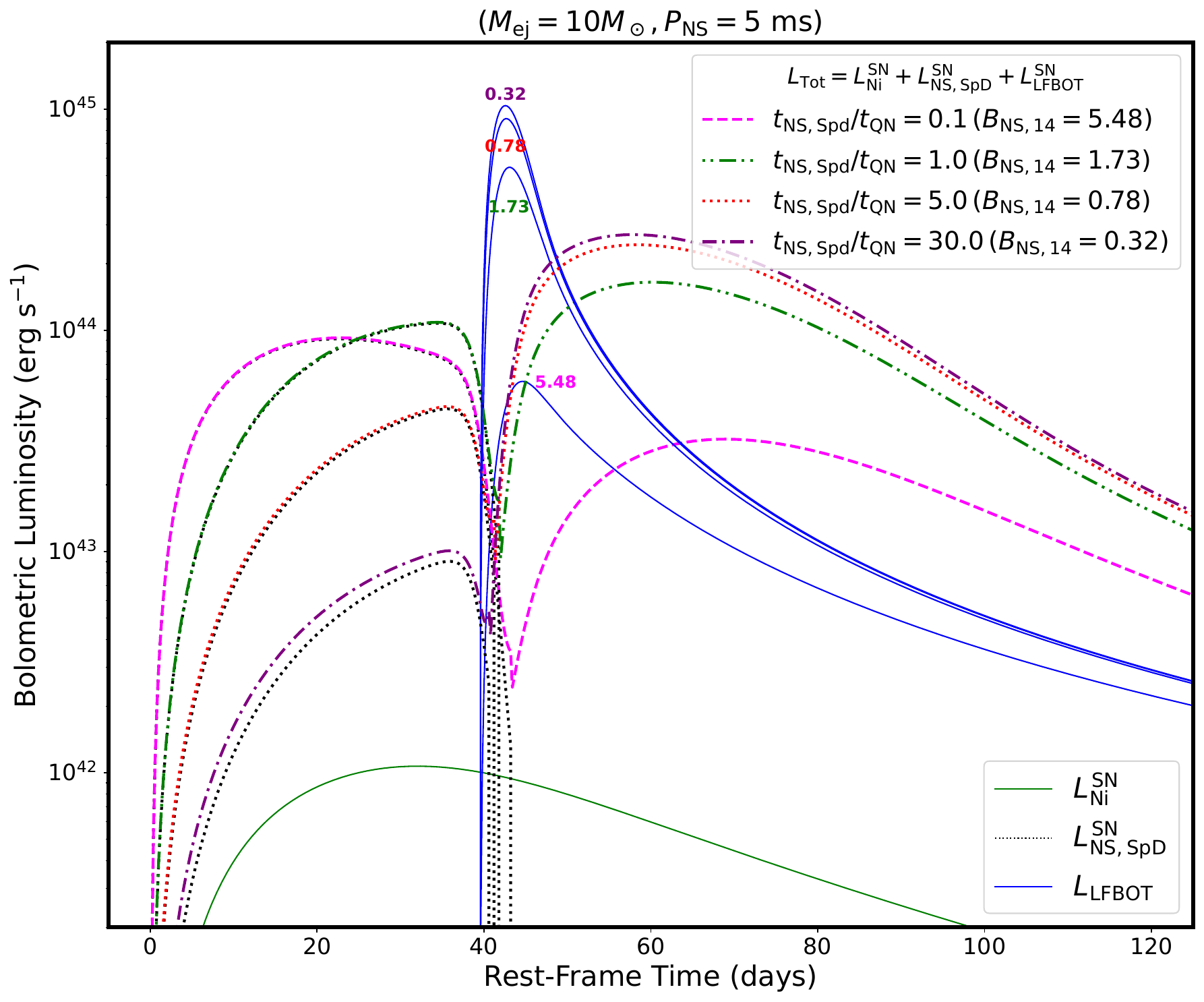}
\caption{
Bolometric light curves in our model for ejecta masses $M_{\rm ej}=5M_{\odot}$ (top) and $M_{\rm ej}=10M_{\odot}$ (bottom). Colored curves show the total luminosity $L_{\rm tot}$ for different ratios $t_{\rm NS,SpD}/t_{\rm QN}$, obtained by varying the NS magnetic field strength (Eq.~\ref{eq:Ltotal}). The HS forms at $t_{\rm QN}=40$ days after the Type Ic SN. We refer to the HS as a QCD-magnetar, with $B_{\rm HS}=10^{15}$ G and $P_{\rm HS}=P_{\rm NS}(1+t_{\rm QN}/t_{\rm NS,SpD})^{1/2}$. The thin green curve shows the purely $^{56}$Ni-powered SN component, while the blue curve indicates the intrinsic LFBOT luminosity prior to reprocessing by the ejecta. Larger $t_{\rm NS,SpD}/t_{\rm QN}$ implies more NS rotational energy available to reheat the ejecta at $t_{\rm QN}$. When $t_{\rm QN}\sim t_{\rm NS,SpD}$, the pre- and post-QN bumps have comparable brightness and width, while smaller ratios produce hotter post-QN ejecta (i.e., more luminous SLSNe) and favor the appearance of W-shaped O\,\textsc{ii} absorption features (see \S \ref{sec:spectroscopy-TARDIS}).
}
\label{fig:tSpD-over-tQN}
\end{figure*}
 \begin{figure*}[t!]
\centering
\includegraphics[scale=0.8]{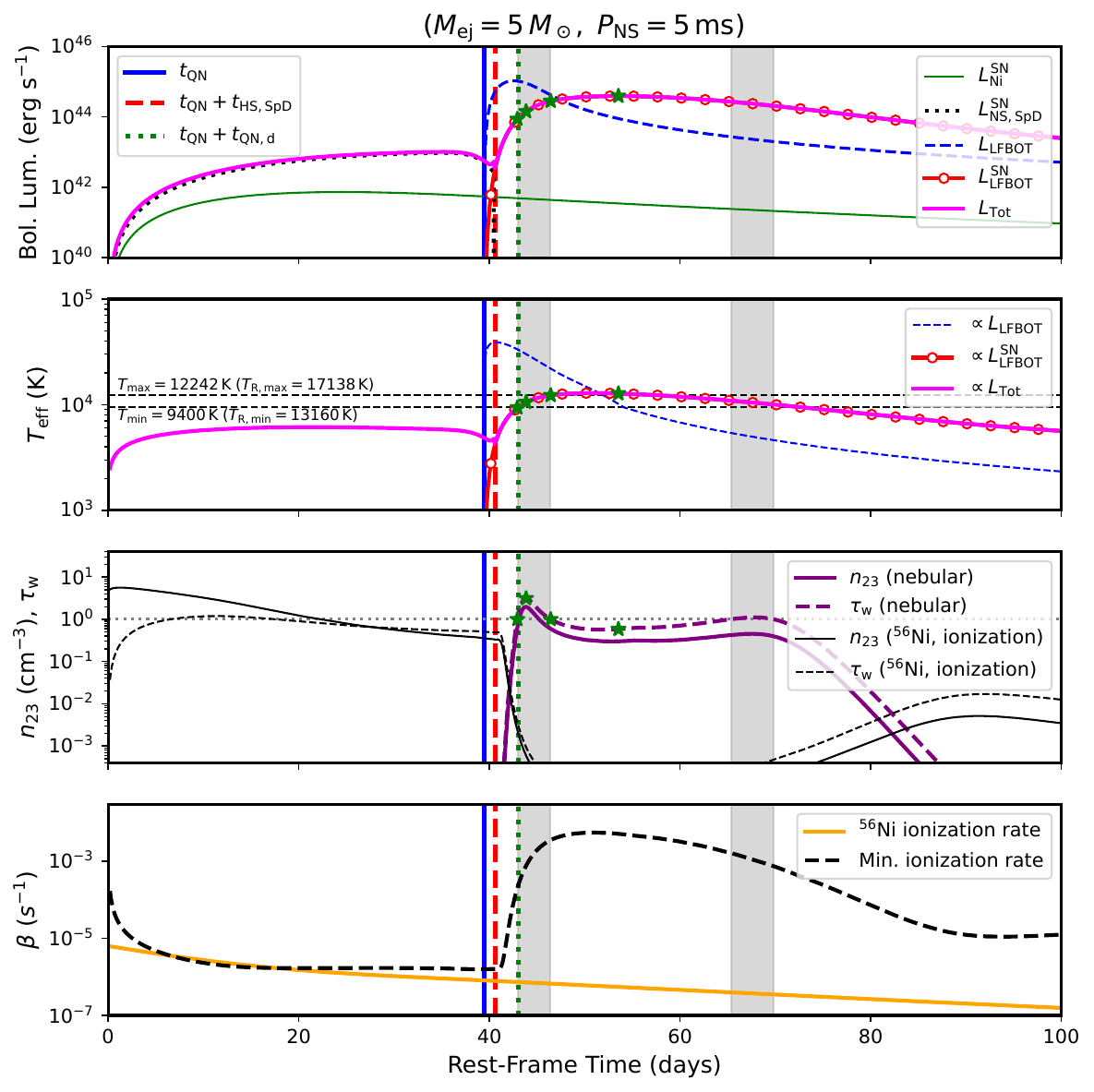}
\caption{
Time evolution of SN properties for an ejecta mass $M_{\rm ej}=5M_{\odot}$. All other parameters are fixed to the fiducial values listed in Table~\ref{table:parameters}.
{\bf Top panel:} Bolometric luminosity from individual power sources and their sum. Contributions from $^{56}$Ni radioactive decay (thin green), neutron-star (NS) spin-down power (black dotted), and reprocessed LFBOT emission (red open symbols) are shown separately. The intrinsic LFBOT luminosity is shown by the dashed blue curve, and the total bolometric luminosity by the solid magenta curve.
{\bf Second panel:} Effective temperature $T_{\rm eff}$ computed from the total bolometric luminosity. The temperature corresponding to the intrinsic (non-reprocessed) LFBOT emission is shown for comparison.
{\bf Third panel:} Sobolev optical depth $\tau_{\rm w}$ of the O\,\textsc{ii} $\lambda4649$\AA\ absorption line (dashed purple) and the number density of the 23 eV O\,\textsc{ii} level (solid purple). Thin curves show the contribution from $^{56}$Ni-powered ionization.
{\bf Bottom panel:} Ionization rate from $^{56}$Ni decay (solid) compared with the minimum rate required to maintain $\tau_{\rm w}=1$ (dashed). Green stars mark the epochs when $\tau_{\rm w}=1$ during the rise and decline, the time of maximum $\tau_{\rm w}$, and the epoch of peak total luminosity $L_{\rm Tot,pk}$. The shaded region in all panels indicates the interval where $\tau_{\rm w}>1$, during which the $\lambda4649$\AA\ absorption line (and the associated W-shaped feature) is expected to be observable. Vertical lines mark $t_{\rm QN}$, the HS spin-down timescale, and the QN ejecta diffusion timescale.
}
\label{fig:LC-5Mej}
\end{figure*}
\begin{figure*}[t!]
\centering
\includegraphics[scale=0.8]{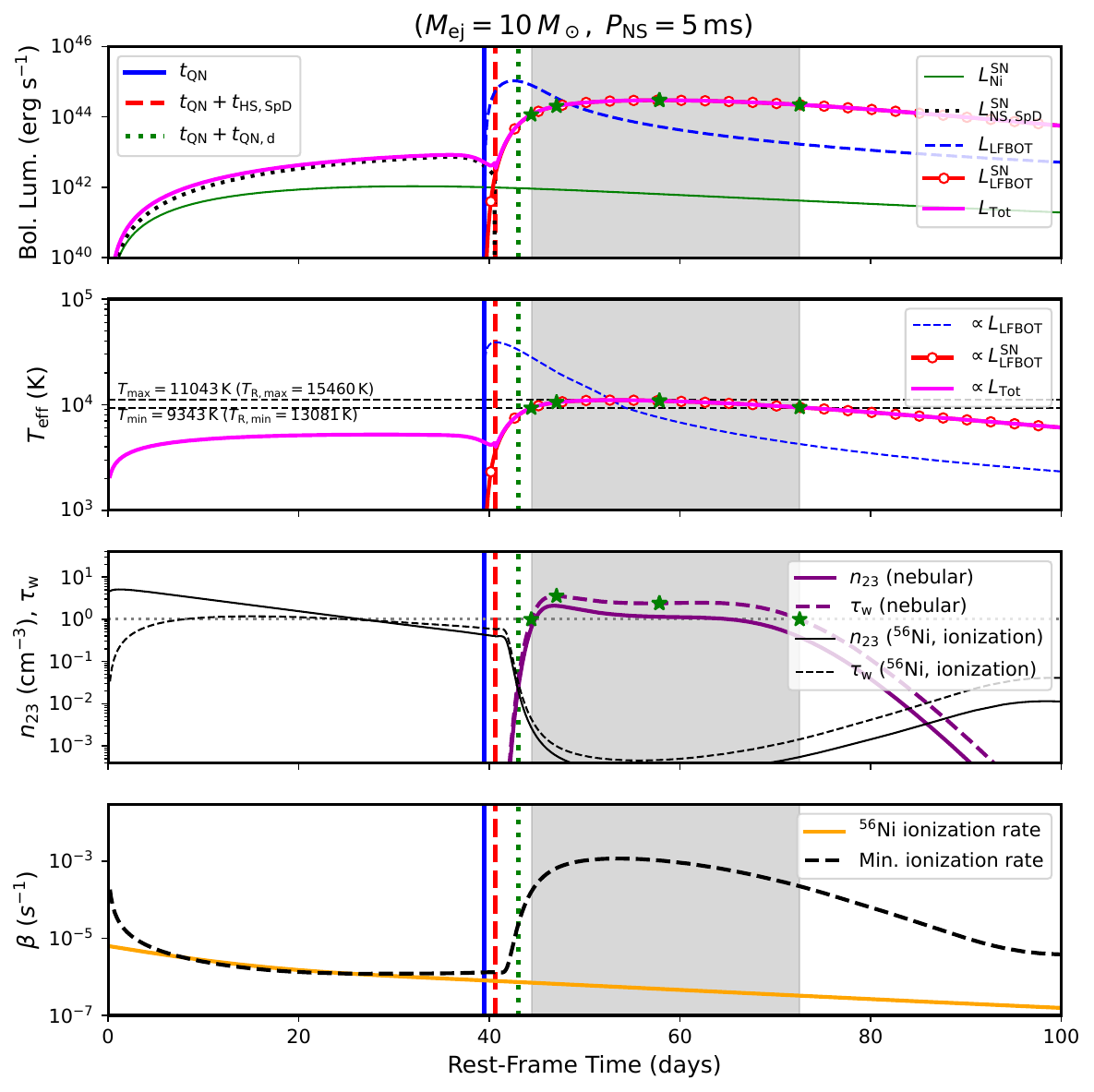}
\caption{Same as Figure~\ref{fig:LC-5Mej}, but for $M_{\rm ej}=10M_{\odot}$.}
\label{fig:LC-10Mej}
\end{figure*}

\subsection{Multi-Component Light Curve Modelling}
\label{sec:multi-L}

The bolometric luminosity in our model includes four power sources and two main phases (we do not consider CSM interaction):

\begin{enumerate}
    \item \textbf{Radioactive decay:} For $t \ge 0$, decay of $^{56}$Ni and $^{56}$Co in the SN ejecta produces an output luminosity $L_{\rm Ni}^{\rm SN}$.
    
    \item \textbf{NS spin-down:} For $0 \le t \le t_{\rm QN}$, the NS injects rotational energy into the SN ejecta, emerging as $L_{\rm NS, SpD}^{\rm SN}$.
    
    \item \textbf{LFBOT emission:} For $t \ge t_{\rm QN}$, residual NS rotational energy powers the QN ejecta, producing $L_{\rm LFBOT} \equiv L_{\rm HS, SpD}^{\rm QN}$.  
    
    \item \textbf{Reprocessing of LFBOT:} $L_{\rm LFBOT}$ is absorbed and reprocessed by the overlying SN ejecta, producing an output luminosity $L_{\rm LFBOT}^{\rm SN}$.
\end{enumerate}

The $^{56}$Ni powered LC is calculated using Eq. (9) in \citet{chatzopoulos_2012} while for spin-down powered lightcurves we use Eq.~(13)  in that paper. 
 In both ejecta, the gamma-ray optical depth is chosen as to ensure that most gamma rays and positrons are trapped.  
  For the SN processing of the LFBOT we used Eq.~(3) in \citep{chatzopoulos_2012}  with the LFBOT
  luminosity as input (see \citealt{ouyed_2025a} for details). 
Results are insensitive to the progenitor's initial radius $R_0$ as long as $R_0 < 10^3\,R_{\odot}$.  

The total output luminosity is
\begin{equation}
\label{eq:Ltotal}
L_{\rm Tot} = L_{\rm Ni}^{\rm SN} + L_{\rm NS, SpD}^{\rm SN} + L_{\rm LFBOT}^{\rm SN} \ .
\end{equation}
For $t \ge t_{\rm QN}$, 
\begin{equation}
L_{\rm SLSN} = L_{\rm Tot}  (t\ge t_{\rm QN}) = L_{\rm Ni}^{\rm SN} + L_{\rm LFBOT}^{\rm SN}\ .
\end{equation}

For the fixed $M_{\rm Ni}= 0.01M_{\rm ej}$ (see Table~\ref{table:parameters}), 
we have $L_{\rm Ni}^{\rm SN}  << L_{\rm LFBOT}^{\rm SN}$ so that  $L_{\rm SLSN}\simeq L_{\rm LFBOT}^{\rm SN}$.

To illustrate the effect of the ratio $t_{\rm NS, SpD}/t_{\rm QN}$, 
we vary the NS magnetic field $B_{\rm NS}$ while fixing all other parameters to fiducial values with $t_{\rm QN} = 40$~days.  
We analyze two  SN ejecta masses, which sets the diffusion timescale $t_{\rm ej, d}\simeq 32.8\ {\rm d}\times (M_{\rm ej}/10 M_{\odot})^{1/2}$. 

 In our implementation, the onset of the additional power at $t_{\rm QN}$ is modeled using a smooth temporal transition rather than a sharp step, allowing the energy to be gradually deposited into the ejecta. This treatment captures the progressive thermal response of the expanding material while avoiding artificial discontinuities in the numerical evolution. It is also
 necessary during the spectral analysis as discussed in  \S~\ref{sec:spectroscopy-ANALYTICAL}.

Figure~\ref{fig:tSpD-over-tQN} shows pseudo bolometric light curves for $M_{\rm ej} = 5\,M_\odot$ (top) and $10\,M_\odot$ (bottom).  
The thin green curve shows purely $^{56}$Ni-powered luminosity, the blue curve shows intrinsic LFBOT prior to SN reprocessing, and non-solid colored curves show $L_{\rm Tot}$ (i.e. Eq. (\ref{eq:Ltotal})) for different $t_{\rm NS, SpD}/t_{\rm QN}$ ratios, varied via the NS magnetic field. 

When $t_{\rm NS, SpD} \gg t_{\rm QN}$, most of the NS's rotational energy is retained at the time of conversion, producing a luminous post-QN peak (dwarfing the pre-$t_{\rm QN}$ emission) and, as shown in \S \ref{sec:spectroscopy-ANALYTICAL},  high radiation temperatures that favor the formation of W-shaped O\,\textsc{ii} absorption features.  Conversely, when $t_{\rm NS, SpD} \ll t_{\rm QN}$, most rotational energy has already been expended prior to the QN, producing a single-peaked light curve dominated by the pre-QN emission; i.e., the classical magnetar-powered SLSNe.  

In the regime $t_{\rm NS, SpD} \sim t_{\rm QN}$, the energy injections before and after the QN are comparable, resulting in double-peaked light curves with similar luminosities; the lower radiation temperatures in this regime suppress O\,\textsc{ii} features in the pre-$t_{\rm QN}$ and post-$t_{\rm QN}$ preserving its Type Ic
signature all along (this is reminiscent of SN2019stc; \citealt{gomez_2021}).
These trends help account for the observed diversity of SLSNe-I light curves in our model.

Figures~\ref{fig:LC-5Mej} and \ref{fig:LC-10Mej} present the pseudo bolometric light curves for the fiducial value of the NS magnetic field, $B_{\rm NS} = 10^{12.5}$~G, for two ejecta masses ($M_{\rm ej} = 5M_{\odot}$ and $10M_\odot$, top panels). Individual contributions are: $^{56}$Ni decay (thin green), reprocessed NS spin-down (dotted), LFBOT emission from the QN ejecta (blue dashed), reprocessed LFBOT by the SN ejecta (red solid with markers), and the total luminosity (magenta).  Displaying the light curves in the top panels sets the stage for the subsequent comparison of other quantities (such as the photospheric  and radiation temperature as well as the Sobolev optical depth of the O\,\textsc{ii} lines; see \S \ref{sec:spectroscopy-ANALYTICAL}) which are shown in the lower two panels. This allows a direct assessment of how ejecta mass and timing affect the photometric  and spectral evolution.

The second panels display effective temperatures per powering source $L_{\rm x}$ with the resulting emitted luminosity being $L_{\rm x}^{\rm SN}$. I.e.,
\begin{equation}
T_{\rm eff, x} = \left( \frac{L_{\rm x}^{\rm SN}}{4 \pi \sigma_{\rm SB} R_{\rm ph}^2} \right)^{1/4},
\end{equation}
with the photospheric radius $R_{\rm ph}$ defined via $\int_{R_{\rm ph}}^\infty \kappa_{\rm ej} \rho_{\rm ej}(r)\, dr = 2/3$.

We assume a power-law density profile  given as (\citealt{chevalier_1977}) 
\begin{equation}
\label{eq:rhoenv}
\rho_{\rm ej}(r,t)=
\begin{cases}
\rho_{\rm Plat.}(t), & r < R_{\rm Plat.},\\
\rho_{\rm Plat.}(t) \left( \frac{R_{\rm Plat.}}{r} \right)^{-n_{\rm ej}}, & r > R_{\rm Plat.},
\end{cases}
\end{equation}
with $R_{\rm Plat.}(t) = v_{\rm t} t$ and $\rho_{\rm Plat.}(t) = A t^{-3}$. Here, 
$A = \frac{5 n_{\rm ej}-25}{2 \pi n_{\rm ej}}\, E_{\rm ej} v_{\rm t}^{-5}$ and 
$v_{\rm t} = \left(\frac{10 n_{\rm ej}-50}{3 n_{\rm ej}-9} \frac{E_{\rm ej}}{M_{\rm ej}}\right)^{1/2}$ with 
$E_{\rm ej}$  the ejecta's kinetic energy and $n_{\rm ej}$ the outer density slope; we adopt $n_{\rm ej} = 8$ as the fiducial value.

For the same LFBOT energy input and fixed $t_{\rm QN}$, $5\,M_\odot$ ejecta is more luminous (Eq.~\ref{eq:L-SLSN-peak}) and hotter (Eq.~\ref{eq:SLSN-TR}) than the $10\,M_\odot$ ejecta.  
Shaded regions indicate epochs with visible W-shaped O\,\textsc{ii} lines (see below), with horizontal dotted lines marking the corresponding photospheric temperature range ($T_{\rm min} \le T_{\rm eff} \le T_{\rm max}$).  
Local radiation temperature, $T_{\rm R}=1.4T_{\rm eff}$, is shown in brackets.

\section{Analytical Treatment of the O\,\textsc{ii} Absorption Lines}
\label{sec:spectroscopy-ANALYTICAL}

We now investigate analytically the conditions required for the formation of the W-shaped O\,\textsc{ii} absorption features in the re-heated Type Ic ejecta at $t_{\rm QN}$. 
We follow the formalism of \citet[and references therein; see also \citealt{mazzali_2016}]{saito_2024} to compute the relevant O\,\textsc{ii} level populations.
 The Sobolev optical depth evaluated at time $t$ since explosion is (\citealt{sobolev_1957})
\begin{equation}
\tau_{\rm w} = \frac{\pi e^2}{m_e c} f_{\rm w} \lambda_{\rm w} n_{\rm 23} t \ ,
\end{equation}
where $e$ and $m_e$ are the electron charge and mass, respectively. 
We focus on the prominent O\,\textsc{ii} transition at $\lambda_{\rm w}=4649\,\text{\AA}$ with oscillator strength $f_{\rm w}=0.34$, taken as representative of the blended multiplet that produces the observed W-shaped feature. 
Here, $n_{\rm 23}$ is the number density of the 23 eV excited state of O\,\textsc{ii} (3s $^4$P), with $f_{23}=0.043$ and $\lambda_{23}=539\,\text{\AA}$ (Wiese 1996). 
We denote the O\,\textsc{ii} and O\,\textsc{iii} number densities by $n_{\rm OII}$ and $n_{\rm OIII}$, respectively.

The population of the 23 eV excitation level is computed using the standard nebular approximation, while the ionization balance is determined with the modified nebular approximation \citep{lucy_1991,mazzali_1993,lucy_1999}. The local electron temperature is taken as $T_{\rm e} = 0.9\,T_{\rm R}$, and a correction factor $\eta \sim 0.4$ accounts for optically thick effects and recombination fractions (e.g., \citealt{mazzali_2016}). Geometrical dilution is included through $W(r)=\frac{1}{2}(1-\sqrt{1-(R_{\rm ph}/r)^2})$ \citep{mihalas_1978,lucy_1991,lucy_1999}.

We also include non-thermal excitation and ionization from $^{56}$Ni decay.  Combining equations (6) and (8) in \citet{saito_2024}, the number density of the 23 eV excited state is
\begin{equation}
n_{\rm 23} = \frac{\beta_{\rm Ni}  n_{\rm OII}\alpha_{\rm ratio}}{R_{\rm PI} + n_{\rm e} (q_{\rm 23\rightarrow OII}+ q_{\rm 23\rightarrow OIII}) +\beta_{\rm esc}A_{\rm 23\rightarrow gs} } ,
\label{eq:n23}
\end{equation}
where:
(i) $\beta_{\rm Ni}$ is the ionization rate due to $^{56}$Ni decay (their Eqs.~7 and 12); 
(ii) $q_{\rm 23\rightarrow OII}$ and $q_{\rm 23\rightarrow OIII}$ are electron-collision transition rates (their Eq.~10); 
(iii) $R_{\rm PI}$ is the photoionization rate (correcting for a missing $4\pi$ in their Eq.~9); 
(iv) $A_{\rm 23\rightarrow gs}$ is the Einstein coefficient for decay to the O\,\textsc{ii} ground state (gs); 
(v) $\beta_{\rm esc}$ is the escape probability (their Eq.~11); and 
(vi) $\alpha_{\rm ratio} = \alpha_{\rm OIII\rightarrow OII,23}/\alpha_{\rm OIII\rightarrow OII}$, the ratio of recombination coefficients from O\,\textsc{iii} to the 23 eV level and to the ground state of O\,\textsc{ii} (from \citealt{nahar_1999}). 

We apply this formalism to the full temporal evolution of the SN ejecta in our model, from $t=0$, and including the pre- and post-$t_{\rm QN}$ phases.  
We assume no departures from the nebular approximation during LFBOT reprocessing, but we include non-thermal excitation and ionization from $^{56}$Ni decay throughout.

We compute the total population of the 23 eV level for the models shown in the top panels of Figures~\ref{fig:LC-5Mej} and \ref{fig:LC-10Mej}. 
The corresponding results are presented in the bottom two panels:

(1) The third panel shows the evolution of $n_{\rm 23}$ and the Sobolev optical depth $\tau_{\rm w}$. 
The $\lambda_{\rm w}=4649\,\text{\AA}$ absorption line we assume is observable when $\tau_{\rm w}>1$, indicated by shaded regions. 
Horizontal dashed lines mark the corresponding radiation temperatures, $T_{\rm R}=1.4T_{\rm ph}$ (values shown in brackets), at the photosphere during the $\tau_{\rm w}>1$ phase. 
   
For $t_{\rm QN}=40$ days (for fiducial parameter values), the density at the photosphere is 
$\sim10^{-14}\,\mathrm{g\,cm^{-3}}$. Combined with reheating to temperatures in the range 
$13{,}000\,\mathrm{K} \le T_{\rm R} \le 17{,}000\,\mathrm{K}$, within the $\tau_{\rm w}>1$ region, 
our results are consistent with those of \citet{saito_2024}, who found a similar $T_{\rm R}$ range 
when considering conditions at a fixed time since explosion. This agreement is therefore a 
self-consistency result, as the physical conditions at the photosphere in our model are comparable 
to those considered in their study.

In agreement with their Figure~8, we confirm that for $T_{\rm R}\lesssim13{,}000\,\mathrm{K}$, 
$\tau_{\rm w}$ decreases due to Boltzmann suppression of the 23\,eV level, while for 
$T_{\rm R}\gtrsim17{,}000\,\mathrm{K}$, O\,\textsc{ii} becomes thermally ionized to O\,\textsc{iii}.
   
For $M_{\rm Ni}/M_{\rm ej}=10^{-2}$, non-thermal ionization alone yields $\tau_{\rm w}<1$, indicating that radioactive input is insufficient to produce observable O\,\textsc{ii} absorption, specifically the $\lambda_{\rm w}=4649\,\text{\AA}$ transition.

(2) The bottom panel shows $^{56}$Ni ionization rate. 
The minimum ionization rate $\beta_{\rm Ni,min}$ required for $\tau_{\rm w}=1$ is obtained by inserting $n_{\rm 23}=1/(\alpha_{\rm w} t)$ into Eq.~(\ref{eq:n23}), where 
$\alpha_{\rm w}=\frac{\pi e^2}{m_e c} f_{\rm w}\lambda_{\rm w}$. 
This critical value is shown as a dashed line in the bottom panels of Figures~\ref{fig:LC-5Mej} and \ref{fig:LC-10Mej}. 
For $M_{\rm Ni}/M_{\rm ej}=10^{-2}$, the ionization rate remains sub-critical, implying that O\,\textsc{ii} absorption in our models is primarily driven by SN reheating due to reprocessed LFBOT power at $t>t_{\rm QN}$. This is also in agreement with \citet{saito_2024} and the results shown in their Figure 11 when using conditions of the Type Ic
representative  of our ejecta at $t_{\rm QN}$.

The onset of LFBOT power during the transition phase at around $t_{\rm QN}$ induces several coupled effects. Reheating raises $T_{\rm R}$, enhancing excitation rates and modifying collisional coefficients. Partial re-ionization alters the O\,\textsc{ii}/O\,\textsc{iii} balance, temporarily increasing $n_{\rm OII}$ within the relevant temperature window, while the photoionization rate initially remains modest, reducing the denominator in Eq.~(\ref{eq:n23}). Together, these effects increase $n_{\rm 23}$ over a few days. As the ejecta subsequently expands and cools, photoionization strengthens and $n_{\rm 23}$ declines. 
 In our implementation, this onset is modeled using a smooth temporal transition rather than a sharp step, allowing the additional power to be gradually deposited into the ejecta. This approach captures the progressive thermal response of the expanding material while avoiding artificial discontinuities in the numerical evolution. 
 A fully self-consistent treatment would require hydrodynamic coupling of the energy deposition associated with HS formation and spin-down injection, which lies beyond the scope of this work.

For fixed fiducial parameters (i.e., fixed LFBOT energy input and $t_{\rm QN}$), increasing the ejecta mass lowers the radiation temperature. 
For $M_{\rm ej}>10M_{\odot}$, the system enters the $\tau_{\rm w}<1$ regime and the $\lambda_{\rm w}=4649\,\text{\AA}$ O\,\textsc{ii} absorption disappears
(but not necessarily the other O\,\textsc{ii}  lines in the W-shaped range; see \S \ref{sec:spectroscopy-TARDIS}). 
Conversely, for $M_{\rm ej}<5M_{\odot}$, the temperature is $T_{\rm R}\gtrsim17{,}000$ K during most of the peak luminosity phase (lasting $\sim 20$ days), 
which thermally ionizes O\,\textsc{ii} to O\,\textsc{iii} and suppressing the 23 eV transition until the ejecta cool below this threshold (see Figure~\ref{fig:LC-5Mej}). 
The $5\,M_{\odot}$ model enters the $\tau_{\rm w}>1$ regime twice during the light-curve evolution. During the second phase, however, $\tau_{\rm w}\sim1$, and the $\lambda_{\rm w}=4649\,\text{\AA}$ line is therefore significantly weaker. These constraints may be relaxed by varying other parameters in the analytical approach given here, but robust conclusions ultimately require detailed spectral modelling, presented next.

\begin{deluxetable*}{lcccccc}[t!]
\centering
\tablecaption{Physical conditions at the photosphere at selected epochs measured since explosion during the evolution of the re-heated SN ejecta, for $M_{\rm ej}=5 M_\odot$ and $10 M_\odot$ models. Four epochs are marked by stars in Figures~\ref{fig:LC-5Mej} and \ref{fig:LC-10Mej}. Two additional epochs, corresponding to $L_{\rm Tot, pk}/5$ and $L_{\rm Tot, pk}/10$ during the post-peak decline, are listed here only.\label{table:the-Sobolev-times}}
\tablehead{
\colhead{Epochs} & \colhead{$\tau_{\rm w}=1$ (rising)} & \colhead{$\tau_{\rm w, max}$} & 
\colhead{$\tau_{\rm w}=1$ (declining)} & \colhead{$L_{\rm Tot, pk}$} & 
\colhead{$L_{\rm Tot, pk}/5$} & \colhead{$L_{\rm Tot, pk}/10$}
}
\startdata
\multicolumn{7}{c}{\textbf{$M_{\rm ej} = 5\,M_\odot$}} \\
\hline
Time [days] & 43.0 & 43.8 & 46.6 & 53.6 & 83.0 & 92.7 \\
$\tau_{\rm w}$ & 1.0 & 3.1 & 1.0 & 0.6 & $<<1$ & $<<1$ \\          
$L$ [$10^{44}$ erg s$^{-1}$] & 0.87 & 1.4 & 2.8 & 3.8 & 0.76 & 0.38 \\
$\rho_{\rm ph}$ [$10^{-14}$ g cm$^{-3}$] & 2.6 & 2.5 & 2.3 & 2.0 & 0.56 & 0.40 \\
$v_{\rm ph}$ [$10^3$ km s$^{-1}$] & 9.8 & 9.8 & 9.7 & 9.4 & 7.9 & 7.3 \\
$R_{\rm ph}$ [$10^{15}$ cm] & 3.7 & 3.7 & 3.9 & 4.3 & 5.7 & 5.8 \\
\hline
\multicolumn{7}{c}{\textbf{$M_{\rm ej} = 10\,M_\odot$}} \\
\hline
Time [days] & 44.6 & 47.6 & 71.2 & 57.8 & 98.9 & 111.8 \\
$\tau_{\rm w}$ & 1.0 & 3.5 & 1.0 & 2.6 & 0.0 & 0.0 \\
$L$ [$10^{44}$ erg s$^{-1}$] & 1.2 & 2.0 & 2.2 & 2.8 & 0.56 & 0.28 \\
$\rho_{\rm ph}$ [$10^{-14}$ g cm$^{-3}$] & 2.8 & 2.5 & 1.5 & 1.9 & 0.66 & 0.46 \\
$v_{\rm ph}$ [$10^3$ km s$^{-1}$] & 10.5 & 10.4 & 9.5 & 9.9 & 8.6 & 8.1 \\
$R_{\rm ph}$ [$10^{15}$ cm] & 4.0 & 4.2 & 5.8 & 5.0 & 7.4 & 7.9 \\
\enddata
\end{deluxetable*}
\vspace{2em} 
%
\begin{table*}[t!]
\caption{ \texttt{TARDIS}, Parameter Settings for the $5M_{\odot}$ and $10M_{\odot}$ runs and 
for the Sobolev points shown in Figures \ref{fig:LC-5Mej} and \ref{fig:LC-10Mej}  (see Table \ref{table:the-Sobolev-times} for relevant physical quantities). 
 We use a  homogeneous chemical composition.}
\begin{minipage}{\textwidth}
\begin{tabular}{|c|c|c|c|c|c|c|c|c|c|c|c|}\hline
\multicolumn{12}{|c|}{Composition (mass fraction)}  \\\hline
He & C  & O & Ne & Mg & Si & S & Ca & Ti & Fe & Co & Ni  \\\hline
 0.1& 0.4& 0.475& 0.02& $5\times 10^{-4}$ & $2\times 10^{-3}$ & $5\times 10^{-4}$ & $5\times 10^{-5}$& $2\times 10^{-5}$& $5\times 10^{-4}$& $3\times 10^{-4}$& $ 10^{-5}$\\\hline
\end{tabular}
\hfill
\centering
 \begin{tabular}{|c|c|c|c|}\hline
\multicolumn{4}{|c|}{SN ejecta parameters at the 6 epochs listed in Table \ref{table:the-Sobolev-times}}  \\\hline
Luminosity requested ($L_{\rm Tot}$) &  \multicolumn{3}{|c|}{Row 3 in Table \ref{table:the-Sobolev-times}} \\\hline
Density at the edge of the photosphere  ($\rho_{\rm ph}$) &  \multicolumn{3}{|c|}{Row 4 in Table \ref{table:the-Sobolev-times}}  \\\hline
Velocity at the edge of the photosphere ($v_{\rm in}=v_{\rm ph}$)  &  \multicolumn{3}{|c|}{Row 5 in Table \ref{table:the-Sobolev-times}}  \\\hline
Velocity at the edge of the ejecta ($v_{\rm out}$) & \multicolumn{3}{|c|}{$2\times 10^4$ km s$^{-1}$} \\\hline 
\multicolumn{4}{|c|}{Plasma parameters}\\\hline
Ionization mode & \multicolumn{3}{|c|}{nebular}\\\hline
Excitation mode & \multicolumn{3}{|c|}{dilute-lte} \\\hline
\end{tabular}\\
We set 10 velocity shells for all simulations with $\rho_{\rm shell} = \rho_{\rm ph}\times (v_{\rm ph}/v_{\rm shell})^8$.
 \end{minipage}
 \label{table:TARDIS-uniform-composition}
\end{table*}

\begin{figure*}[t!]
\centering
\includegraphics[scale=0.45]{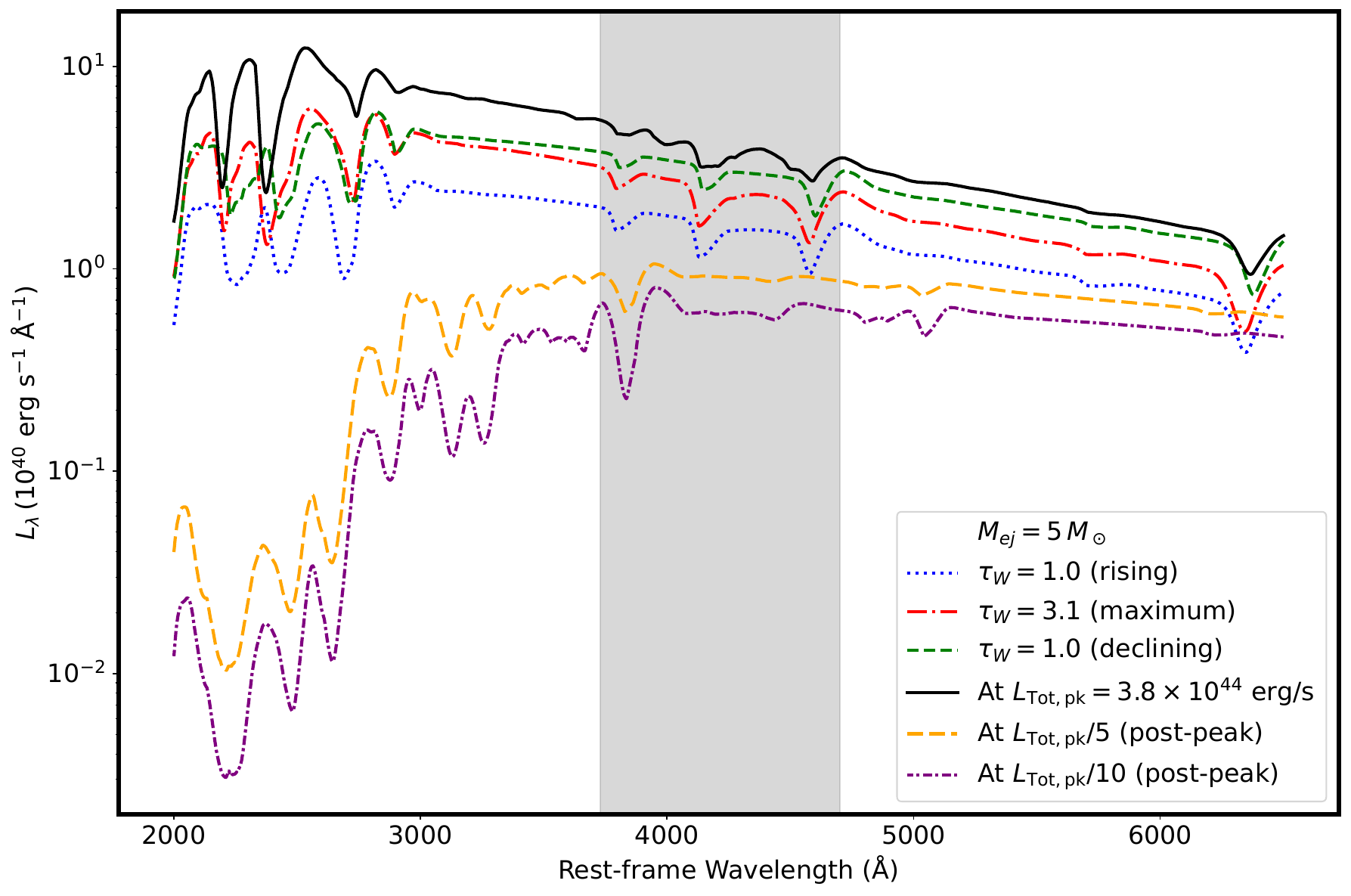}
\includegraphics[scale=0.45]{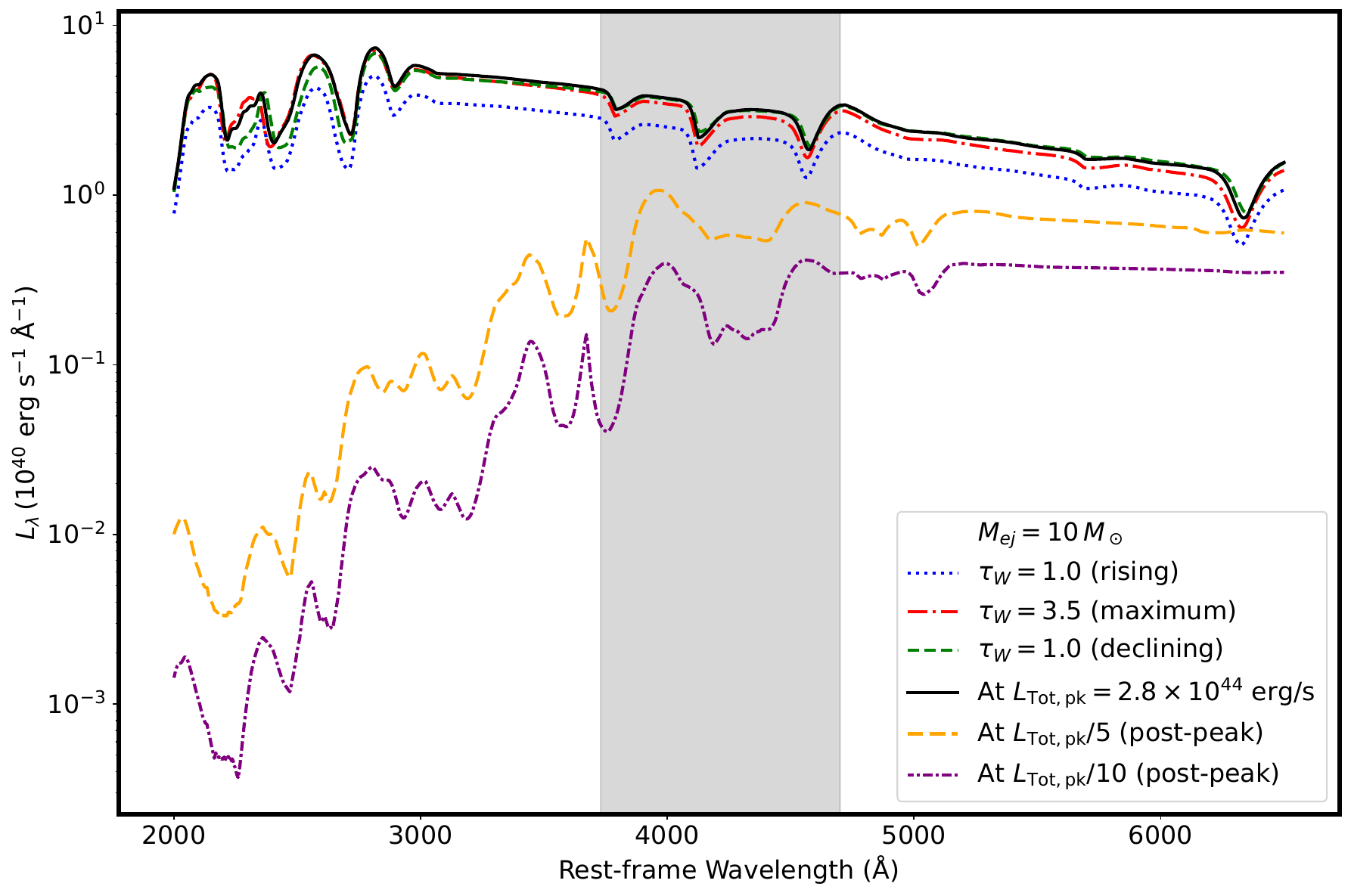}
\caption{Synthetic spectra at the epochs marked by green stars in Figures~\ref{fig:LC-5Mej} and \ref{fig:LC-10Mej}, with the corresponding photospheric conditions listed in Table~\ref{table:the-Sobolev-times}, for the $M_{\rm ej}=5M_{\odot}$ and $10M_{\odot}$ SN models. The characteristic W-shaped O~\textsc{ii} absorption features are prominent at early times when the ejecta temperature exceeds $\sim12{,}000$ K and fade as the ejecta expands and cools, eventually revealing the underlying Type Ic spectrum.
}
\label{fig:computed-spectra}
\end{figure*}

\begin{figure*}[t!]
\centering
\includegraphics[scale=0.6]{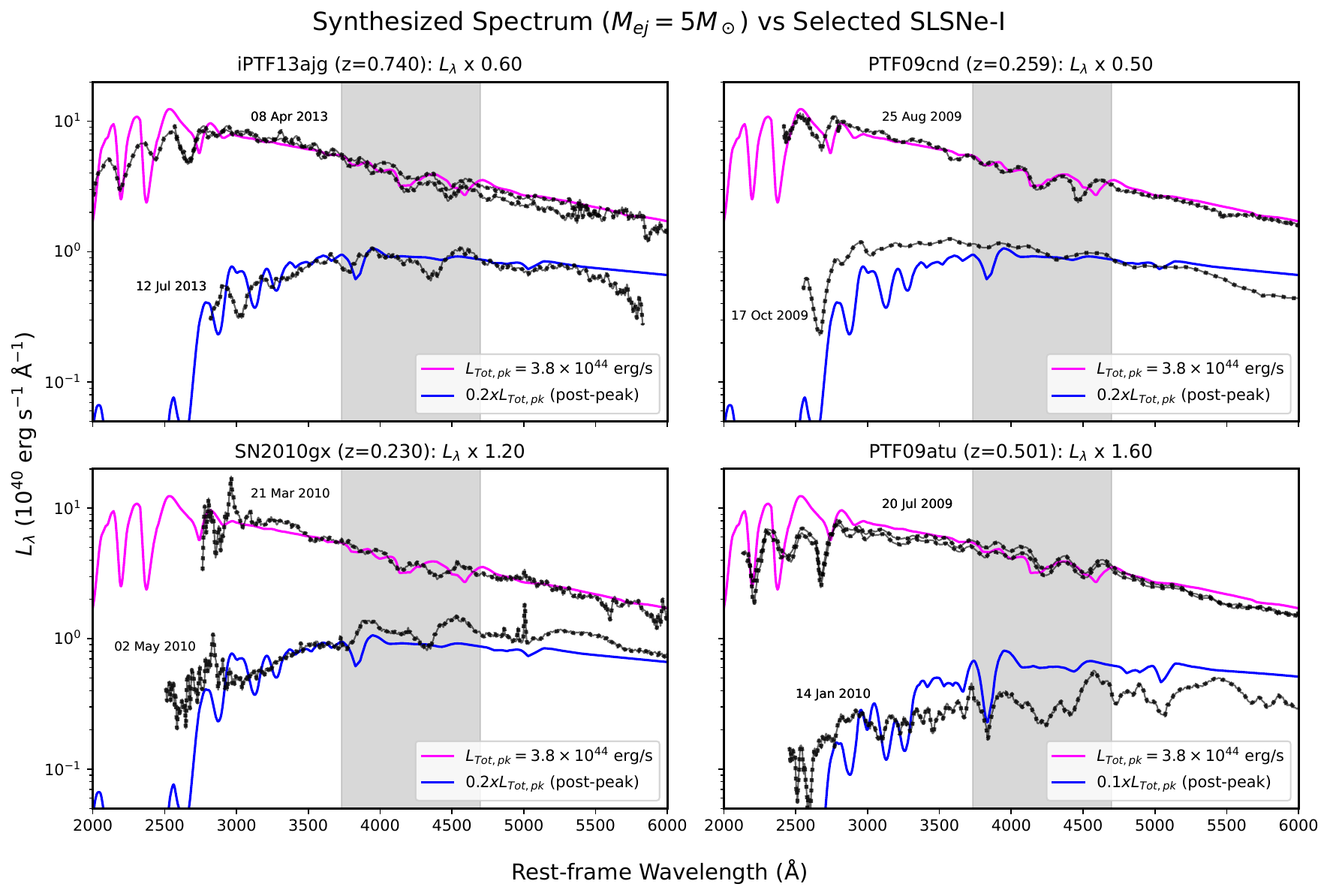}
\caption{Comparison between synthetic spectra (magenta and blue solid lines) for the $M_{\rm ej}=5M_{\odot}$ model and four selected SLSNe-I. The two model spectra correspond to the $L_{\rm Tot, pk}$ and $L_{\rm Tot,pk}/5$ epochs listed in Table~\ref{table:the-Sobolev-times} ($L_{\rm Tot, pk}/10$ for PTF09atu). The observed spectra (labeled by date of observation) are shown near peak luminosity and at late times (several months post-peak) for each event. The model spectra reproduce the overall spectral morphology, including the characteristic W-shaped O~\textsc{ii} feature and the broad continuum shape at early times. See \S~\ref{sec:spectroscopy-TARDIS-4SLSNe-I} and Figure~\ref{fig:compare-Tardis-Spectrum-to-SNLSNe} for tailored fits in which photometry and spectra are modeled simultaneously.
}
\label{fig:compare-5Msun-to-SNLSNe}
\end{figure*}

\begin{figure*}[t!]
\centering
\includegraphics[scale=0.6]{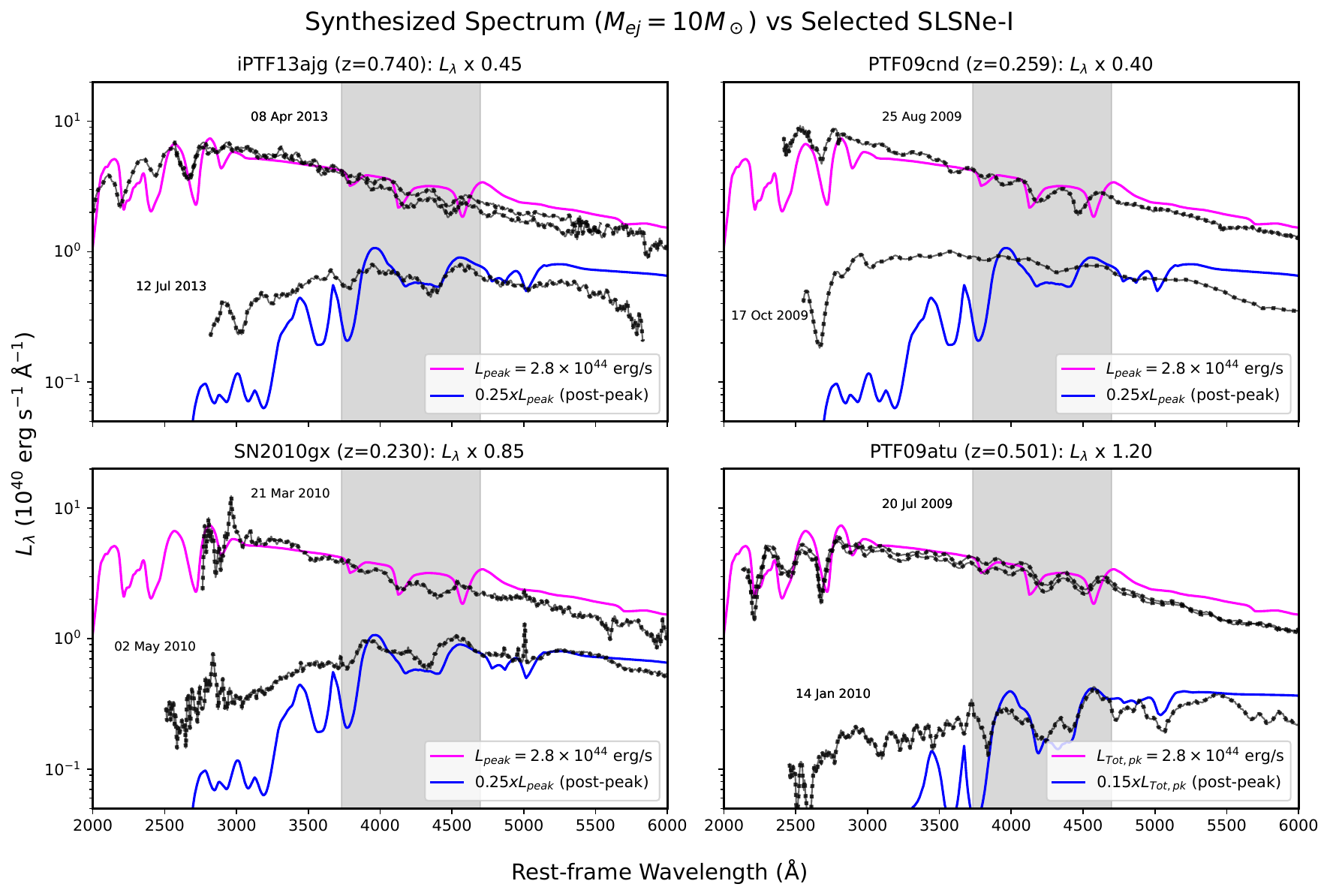}
\caption{Same as Figure~\ref{fig:compare-5Msun-to-SNLSNe}, but for $M_{\rm ej}=10M_{\odot}$.}
\label{fig:compare-10Msun-to-SNLSNe}
\end{figure*}

\section{Spectrum Modelling of the O\,\textsc{ii} Absorption Features}
\label{sec:spectroscopy-TARDIS}

We compute synthetic spectra for the models shown in Figures~\ref{fig:LC-5Mej} and \ref{fig:LC-10Mej} at six representative epochs during the post-$t_{\rm QN}$ phase.
Four of these epochs are highlighted by green asterisks in the four panels and correspond to:
(i) the first time $\tau_{\rm w}=1$ is reached during the rising phase;
(ii) the epoch of maximum Sobolev optical depth, $\tau_{\rm w,max}$;
(iii) the time when $\tau_{\rm w}=1$ during the declining phase; and
(iv) the epoch of peak total luminosity, $L_{\rm Tot,pk}$.

For each epoch, the photospheric quantities required for spectral synthesis (i.e. the luminosity, density, velocity, and time since explosion)  are listed in Table~\ref{table:the-Sobolev-times}.
Two additional snapshots are selected at $L_{\rm Tot,pk}/5$ and $L_{\rm Tot,pk}/10$ on the post-peak decline, enabling direct comparison with late-time spectra of the four selected SLSNe-I.

We employ the one-dimensional, time-independent Monte Carlo radiative-transfer code  \texttt{TARDIS}, \citep{kerzendorf_2014, vogl_2019}.
 \texttt{TARDIS}, assumes a sharp inner boundary representing an opaque, blackbody-emitting photosphere, where all energy deposition is thermalized.
The line-forming region is modelled as a series of optically thin shells above this inner boundary.
Photon packets are emitted from the photosphere according to a blackbody distribution and propagated through the ejecta, undergoing electron scattering and bound–bound interactions.
Escaping packets are collected to construct the emergent synthetic spectrum.
Although the code is time-independent, a sequence of snapshots with evolving input parameters provides a self-consistent temporal evolution.

User-defined inputs include the density structure, composition, velocity boundaries (the photosphere at $v_{\rm in}$ and the ejecta edge at $v_{\rm out}$), time since explosion, and bolometric luminosity.
Taking the explosion time ($t_{\rm expl}$) as a reference and defining $v_{\rm in} = v_{\rm ph}$ and $\rho_{\rm in} = \rho_{\rm ph}$ at the photosphere, the density structure is prescribed, for each epoch,  as
\begin{equation}
\rho(v, t_{\rm expl}) = \rho_{\rm ph} \left(\frac{v_{\rm ph}}{v}\right)^{n_{\rm ej}},
\end{equation}
consistent with our light-curve modeling.

For the six epochs listed in Table~\ref{table:the-Sobolev-times}, ionization is treated using the \texttt{nebular} mode, while excitation is handled using \texttt{dilute-lte}.
Radiative rates are computed from the Monte Carlo–estimated radiation field, and line interactions are treated with the \texttt{macroatom} formalism, accounting for scattering and fluorescence.
The Monte Carlo configuration is fixed at 15 iterations, $10^5$ packets per iteration, $10^6$ packets in the final iteration, and 5 virtual packets.
We adopt the full Kurucz GFALL atomic dataset \citep{kurucz_1995}, and the final spectrum is generated using the formal integral method \citep{vogl_2019}, minimizing Monte Carlo noise over the wavelength range $2000\text{\AA} \le \lambda \le 7000\text{\AA}$.

We initially assume a homogeneous ejecta composition representative of Type Ic explosions, adopting abundances similar to those inferred by \citet{mazzali_2016} and \citet{saito_2024} to enable direct comparison with their spectral analysis of SLSNe-I including the four we study here.
A summary of the \texttt{TARDIS} setup we use  is provided in Table~\ref{table:TARDIS-uniform-composition}.

The resulting synthetic spectra are shown in Figure~\ref{fig:computed-spectra} for the $5M_{\odot}$ (top panel) and $10M_{\odot}$ (bottom panel) ejecta models.
The shaded region marks the wavelength range of the blended O\,\textsc{ii} absorption lines including the W-shaped feature.
In both models, the feature is prominent at early epochs, during the re-heating phase, and progressively weakens after peak as the ejecta expands and cools, with the spectra evolving toward a classical Type Ic appearance.
The $10M_{\odot}$ model exhibits a less pronounced temporal evolution of the W-shaped feature around peak luminosity, reflecting the narrower range of radiation temperatures compared to the $5M_{\odot}$ case. 

Figure~\ref{fig:compare-5Msun-to-SNLSNe} compares the $5M_{\odot}$ synthetic spectra with observed spectra of our SLSNe-I sample. 
 They are  iPTF13ajg (\citealt{vreeswijk_2014}), SN~2010gx (\citealt{pastorello_2010}), PTF09cnd (\citealt{quimby_2011}), and PTF09atu (\citealt{quimby_2011}). 
 These are well observed and span both the narrow- and broad-lined subclasses of SLSNe-I with existing  high-quality photometric and spectroscopic coverage at both early and late times. Data we obtained from the Weizmann Interactive Supernova Data Repository (\citealt{yaron_2012}). 

The synthetic spectra shown correspond to $L_{\rm Tot, pk}$ and to a post-peak phase at $L_{\rm Tot,pk}/5$.
We do not attempt detailed object-by-object fits (see also \S \ref{sec:spectroscopy-TARDIS-4SLSNe-I}); the goal is to assess whether the global morphology and strength of the W-shaped feature are naturally reproduced. For best comparison, observed fluxes required only minor, time-independent scaling of the observed fluxes for each object with scaling factors  indicated in the  sub-panel titles.

Remarkably, the $5M_{\odot}$ model reproduces the general appearance of the W-feature at peak across the four SLSNe-I, suggesting these events converge to a relatively narrow range of physical conditions at maximum light. This may be because  the peak luminosity and radiation temperature occupy a restricted parameter space (cf. Eqs.~(\ref{eq:SLSN-TR}) and (\ref{eq:L-SLSN-peak})).

Reproducing observed post-maximum spectra, typically obtained weeks to months after peak, requires selecting model snapshots at $L_{\rm Tot, pk}/5$ (and $L_{\rm Tot, pk}/10$ for PTF09atu).
In our simulations, such luminosities occur slightly earlier than the observational epochs, indicating that the model ejecta cool somewhat more rapidly than inferred from the data.
Moreover, in the fiducial runs, the ejecta remain hotter for longer than in some observed events, explaining the persistence of the W-shaped feature relative to observations.
The $10M_{\odot}$ models provide less satisfactory agreement at late times, suggesting that the conditions represented by the $5M_{\odot}$ runs (despite the relatively low mass compared to what is observed in SLSNe-I) are more characteristic of Type-W SLSNe.
This conclusion will be reinforced next when performing object-specific fits, by varying additional parameters, to both light curves and spectra of the SLSNe-I sample.

\subsection{Detailed analysis of the selected SLSNe}
\label{sec:spectroscopy-TARDIS-4SLSNe-I}

Here, we simultaneously fit both the photometric light curves and spectral evolution of SLSNe-I sample. 
The best-fit parameters for each object are listed in Table~\ref{table:SLSNe-fits}.  
These parameters were obtained through iterative forward modeling aimed at simultaneously reproducing the observed light-curve morphology (rise time, peak luminosity, and decline rate) and the spectral evolution.  We do not perform a formal $\chi^2$ minimization because the model is computationally intensive, involves correlated parameters, and is intended to capture the underlying physical regime rather than provide a purely statistical fit.  
Instead, we adopt a physically guided fitting procedure that prioritizes consistency between the derived ejecta properties and the spectral constraints.

The time shifts (see last column in Table \ref{table:SLSNe-fits}) were applied to align the model’s $t_{\rm QN}$ with the observed onset of the second re-brightening. 
In the observational data, the time axis is defined relative to the first measurement such that the light curve starts at $t=0$. 
For PTF09cnd and PTF09atu, the earliest points before the QN-powered rise were not included (reasons given below), so the shifts reflect the relative timing of the re-brightening rather than absolute calendar time. Adjusting $t_{\rm QN}$ to remove the shifts is not feasible because we fit all four SLSNe-I with the same $t_{\rm QN}$. The shifts therefore primarily reflect differences in observational cadence and early-time coverage.

\begin{figure*}[t!]
\centering
\includegraphics[scale=0.8]{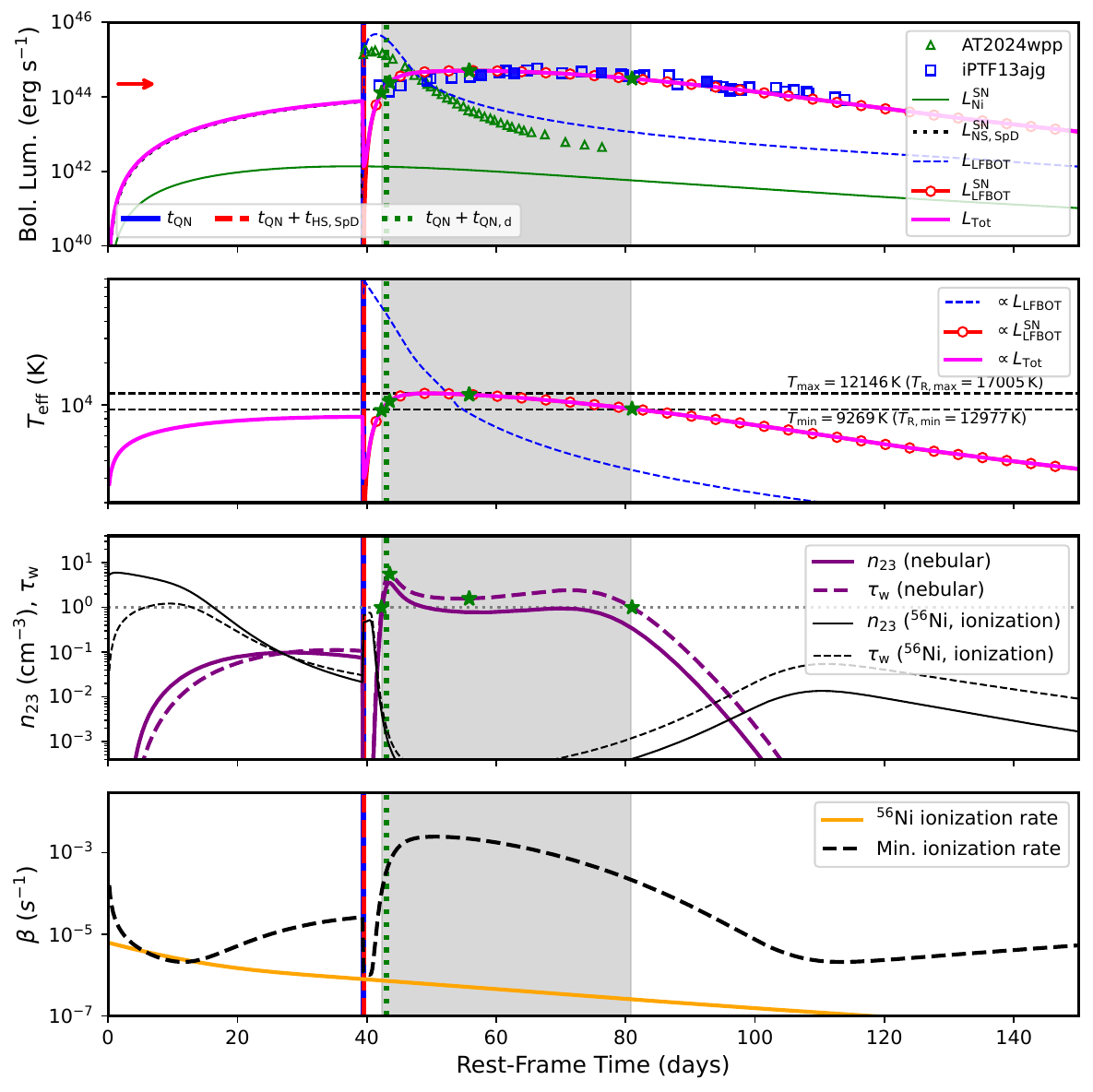}
\caption{Time evolution of the Type Ic SN ejecta properties in four panels for the best-fit model of iPTF13ajg (open blue squares). The panels are defined as in Figure~\ref{fig:LC-5Mej}. For reference, the LFBOT AT2024wpp is shown as open green triangles. Solid blue squares mark the epochs analyzed spectroscopically by \citet{saito_2024}. In our model, synthetic spectra at these epochs are computed with \texttt{TARDIS} using the corresponding photospheric properties listed in Table~\ref{table:saito-points}, derived from the best-fit light-curve model. The red right-pointing arrow indicates the approximate PTF detection limit at the redshift of the SLSN.
}
\label{fig:LC-iPTF13ajg}
\end{figure*}

\begin{figure*}[t!]
\centering
\includegraphics[scale=0.8]{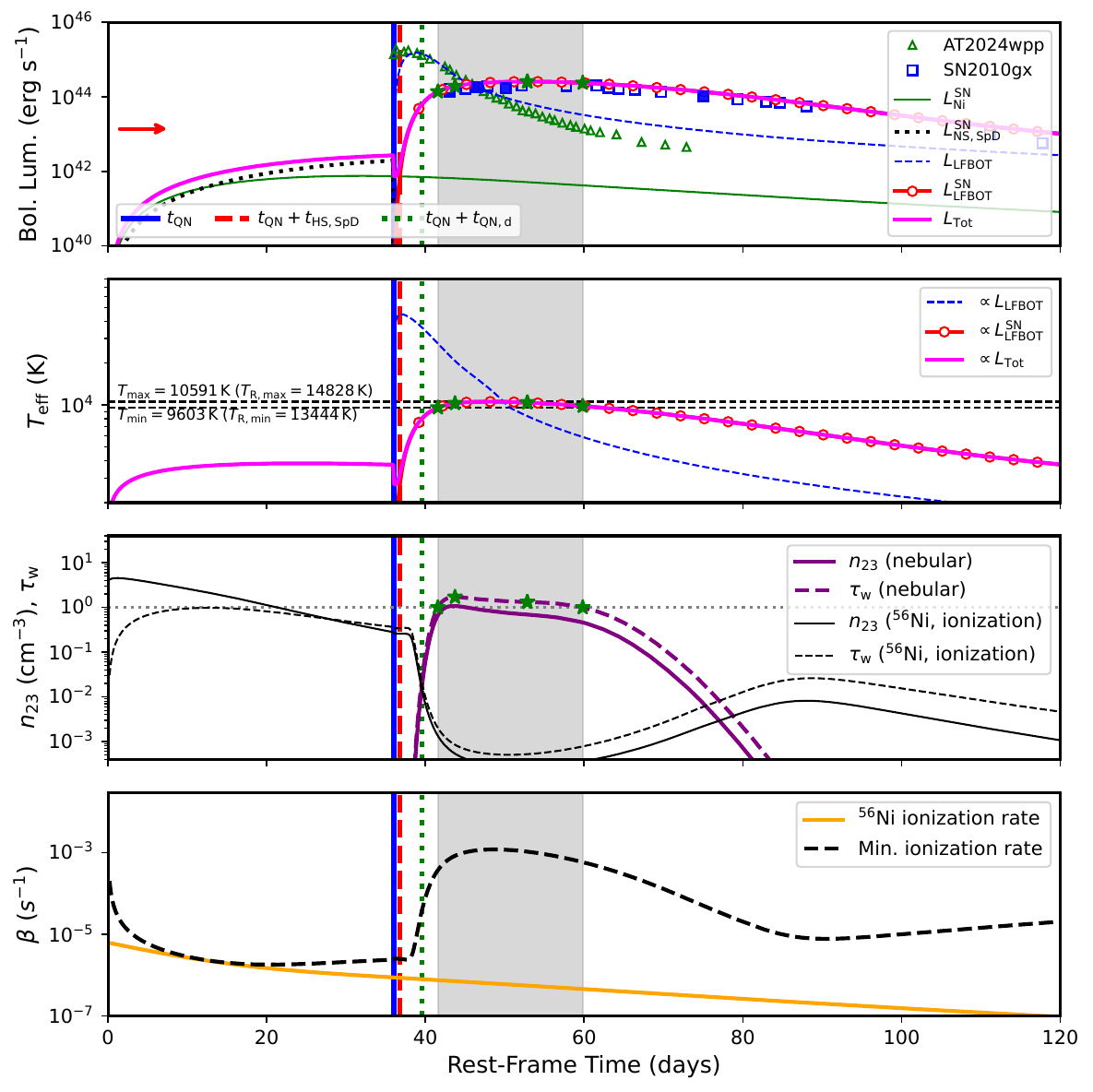}
\caption{Same as Figure~\ref{fig:LC-iPTF13ajg}, but for SN2010gx.
}
\label{fig:LC-SN2010gx}
\end{figure*}

\begin{figure*}[t!]
\centering
\includegraphics[scale=0.8]{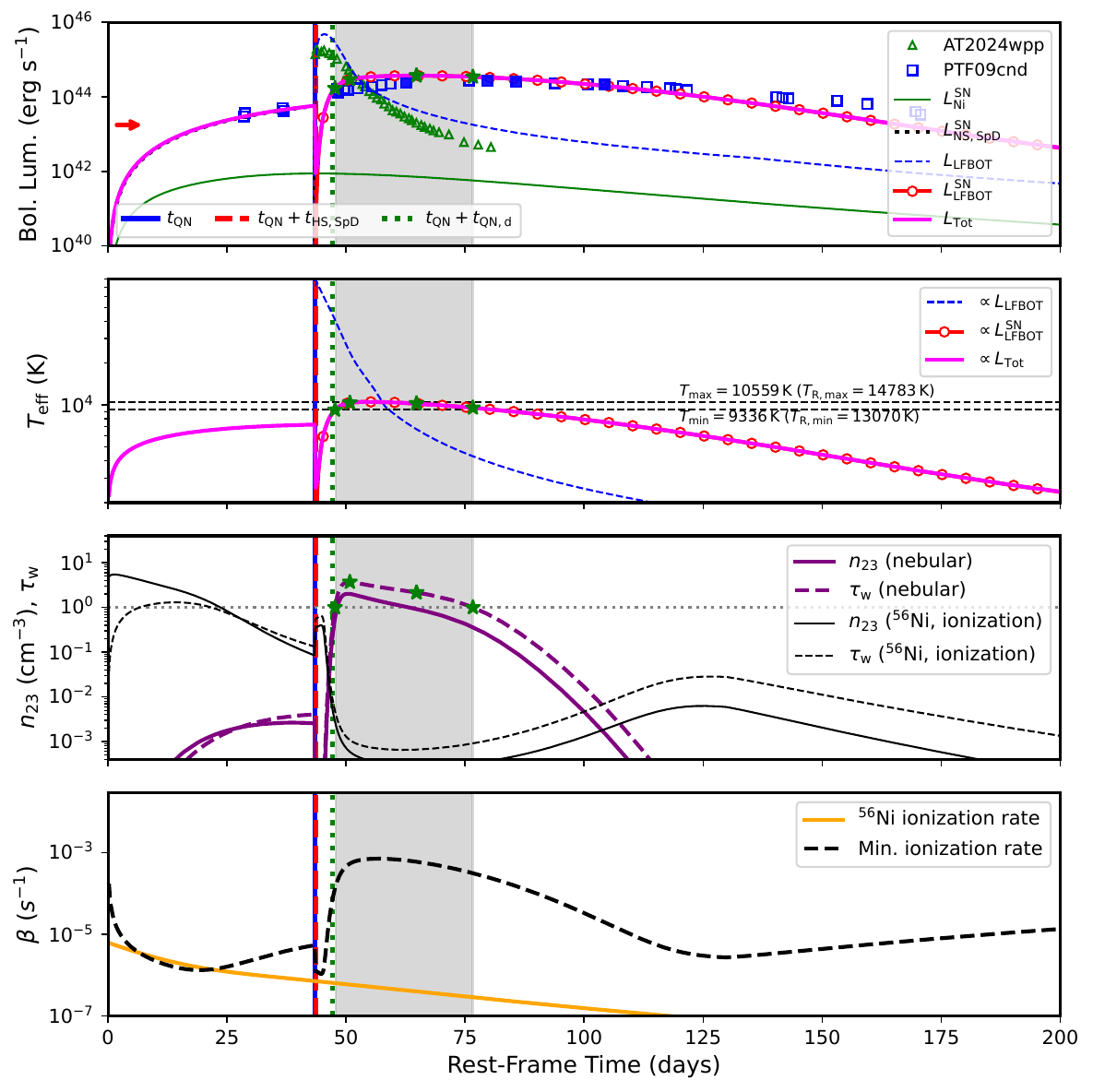}
\caption{Same as Figure~\ref{fig:LC-iPTF13ajg}, but for PTF09cnd.
}
\label{fig:LC-PTF09cnd}
\end{figure*}

\begin{figure*}[t!]
\centering
\includegraphics[scale=0.8]{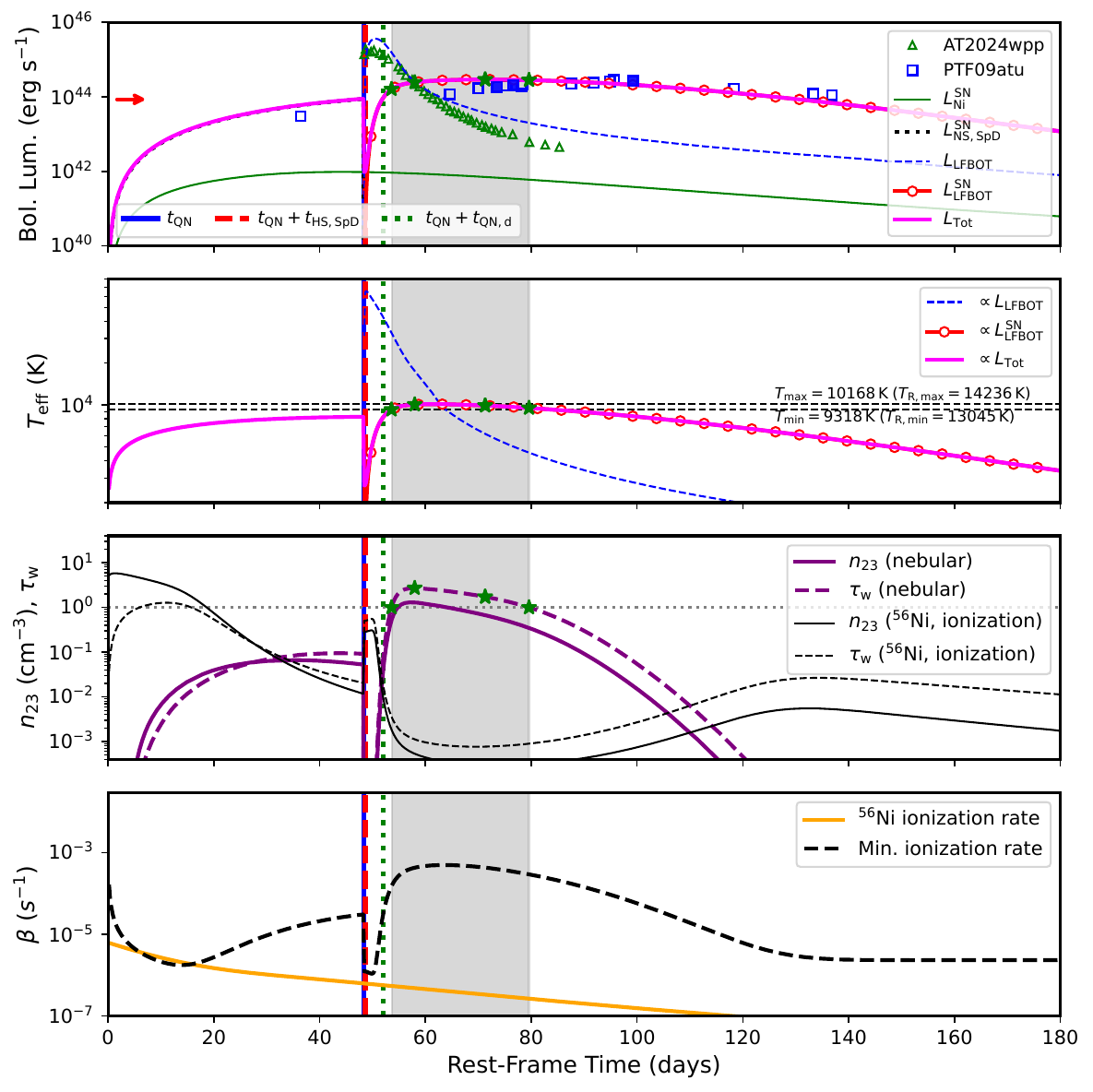}
\caption{Same as Figure~\ref{fig:LC-iPTF13ajg}, but for PTF09atu.
}
\label{fig:LC-PTF09atu}
\end{figure*}

The resulting light-curve fits are shown in the top panels of Figures~\ref{fig:LC-iPTF13ajg}--\ref{fig:LC-PTF09atu}.  
Solid squares indicate the epochs selected by \citet[see their Figure 4]{saito_2024} for spectral modeling, based on the observed spectra and corresponding light-curve phases.

For each of these epochs, we extract the physical conditions predicted by our light curve model namely, the total bolometric luminosity, photospheric velocity and density; listed in Table~\ref{table:saito-points}. These values serve as inputs to  \texttt{TARDIS}, for computing synthetic spectra.  

The ejecta composition is assumed to be radially stratified, roughly following the
general abundance structures inferred from spectral modelling of
stripped-envelope SNe  (e.g., \citealt{shivvers_2019} and references therein), in which the outer layers are dominated by C and O while intermediate-mass
and iron-group elements are progressively more abundant at lower velocities.

\begin{figure*}[t!]
\centering
\includegraphics[scale=0.6]{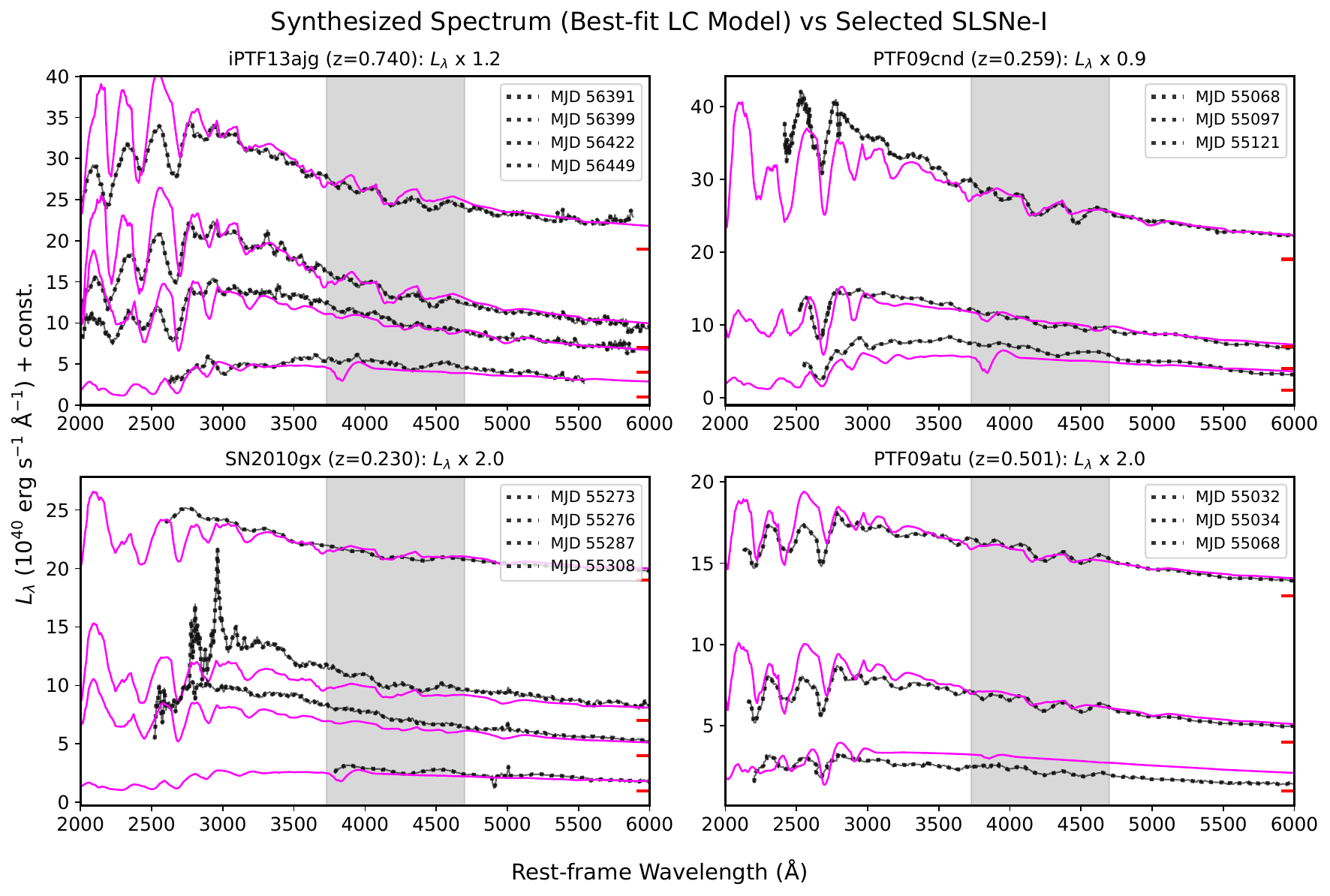}
\caption{Comparison between our model's synthetic spectra (magenta solid lines) from the best-fit light-curve models of the selected SLSNe-I (see Figures~\ref{fig:LC-iPTF13ajg}–\ref{fig:LC-PTF09atu}) and the observed spectra. The MJDs shown correspond to the epochs analyzed by \citet{saito_2024} (their Figure~4 and Table~A1), increasing from top to bottom in each panel. The observed spectra were uniformly scaled to obtain the best match with the model, with the applied scaling factors indicated in each panel. Vertical offsets between spectra are indicated by the red horizontal markers on the right-hand $y$-axis. Despite adopting the same stratified composition for all events (Table~\ref{table:TARDIS-stratified-composition}), the model reproduces the W-shaped O\,\textsc{ii} features at early times and their disappearance as the ejecta cools. More accurate fits can be obtained by modeling each event individually, particularly for SN2010gx, but such detailed tailoring is beyond the scope of this work.
}
\label{fig:compare-Tardis-Spectrum-to-SNLSNe}
\end{figure*}

A summary of the \texttt{TARDIS} setup used is shown in Table \ref{table:TARDIS-stratified-composition}.
The resulting spectra are shown as magenta curves in Figure~\ref{fig:compare-Tardis-Spectrum-to-SNLSNe}, alongside the observed spectra and corresponding MJDs (see Figure 4 and Table~A1 of \citealt{saito_2024}).

Overall, the model reproduces the main spectral features and their temporal evolution, requiring only minor, time-independent scaling of the observed fluxes for each object (the scaling factors are indicated in the sub-panel titles). Agreement is particularly encouraging near peak luminosity, where the W-shaped O\,\textsc{ii} absorption feature is naturally reproduced.  
At later epochs, the fits become less precise, although the model continues to capture the overall spectral morphology and the progressive weakening of the O\,\textsc{ii} feature.
 SN2010gx yielded the poorest fit and would benefit from a more detailed parameter exploration (see \S~\ref{sec:SN2010gx-fits}).

The physical conditions derived from the object-specific fits are close to those obtained in the representative $5\,M_{\odot}$ models discussed earlier.  
The main difference is that the present fits incorporate a chemically stratified ejecta with an important reduction in He (see Table \ref{table:TARDIS-stratified-composition}). In concert with improved light-curve matching, this allowed for a more realistic reproduction of the fading W-shaped feature and the gradual emergence of a Type Ic--like spectrum at late times.  
The central result is that the observed spectra overall can be reproduced, albeit with some limitations, without invoking departures from the nebular approximation or additional excitation mechanisms during the duration of the light-curve evolution.

As shown in Figures~\ref{fig:LC-iPTF13ajg}-\ref{fig:LC-PTF09atu}, the four objects in our sample exhibit differing levels of pre-maximum coverage.  
Unlike iPTF13ajg and SN2010gx, PTF09cnd  and PTF09atu display photometric detections several weeks prior to maximum light.  
If these early points are interpreted as part of the main SLSN rise, the inferred light curves would appear significantly broader than those of typical SLSNe-I.

Within our framework, however, these early detections are naturally interpreted as belonging to the pre-QN (pre-HS) phase, during which the luminosity 
is dominated by NS spin-down prior to the delayed energy injection at $t_{\rm QN}$.  
When the early-time measurements of PTF09cnd and PTF09atu are forced into the main SLSN component, simultaneous
fits to the observed light-curve morphology and spectral evolution became more challenging and  less convincing.  
Modeling these detections as a physically distinct pre-QN phase yields more natural and more consistent fits to both photometry and spectra.

Under this interpretation, the post-reheating light-curve widths of PTF09cnd and PTF09atu are comparable to those of iPTF13ajg and SN2010gx.  
Their spectral evolution likewise shows the characteristic W-shaped O\,\textsc{ii} absorption feature near peak and its subsequent fading, consistent with the broader SLSN-I population.  
This indicates that PTF09cnd and PTF09atu are not intrinsically broader or physically distinct events, but rather members of the same underlying class, with early-time sampling capturing an additional pre-reheating phase.

The red right-pointing arrows in the figures indicate approximate Palomar Transient Factory (PTF) detection limits at the corresponding redshifts (\citealt{law_2009,rau_2009}).  
For iPTF13ajg and SN2010gx, the modeled pre-QN emission lies below these limits, which may explain the absence of early detections.  
For fiducial parameters, the pre-SLSN phase is typically a few magnitudes fainter than the nominal survey sensitivity.  
Such measurements would therefore have low signal-to-noise ratios and would likely only be recoverable through forced photometry
 or post-discovery image subtraction, rather than blind, high-significance detections (see \S \ref{sec:pre-tQN-discussion}).

\begin{deluxetable*}{lccccccc}
\tablecaption{Best-fit model parameters for the selected SLSNe-I sample adopted in the light-curve modeling and \texttt{TARDIS} spectral calculations.\label{table:SLSNe-fits}}
\tablehead{
\colhead{SN} & 
\colhead{$M_{\rm ej}\,(M_\odot)$} & 
\colhead{$v_{\rm ej}\,(10^4~{\rm km~s^{-1}})$} & 
\colhead{$\kappa_{\rm ej}\,({\rm cm^2~g^{-1}})$} & 
\colhead{$n_{\rm ej}$} & 
\colhead{$P_{\rm NS}\,({\rm ms})$} & 
\colhead{$B_{\rm NS}\,(10^{12}~{\rm G})$} & 
\colhead{$t_{\rm shift}\,({\rm days})^\dagger$}
}
\startdata
iPTF13ajg & 17.0 & 1.2 & 0.10 & 8.0 & 1.5 & 3.2 & 4.0 \\
SN2010gx  & 7.0  & 1.3 & 0.15 & 8.0 & 4.0 & 3.2 & 7.0 \\
PTF09cnd  & 16.0 & 1.1 & 0.18 & 9.0 & 1.5 & 3.2 & -15.0 \\
PTF09atu  & 18.0 & 1.0 & 0.15 & 7.0 & 2.0 & 7.0 & -12.0 \\
\enddata
\tablenotetext{}{\footnotesize Other parameters are fixed to the fiducial values listed in Table~\ref{table:parameters}. $^{\dagger}$ Data shifted relative to $t_{\rm QN}$.}
\end{deluxetable*}

\section{Discussion}
\label{sec:discussion}

Here we discuss the principal physical implications of our model, its limitations, and its observationally testable predictions.
In no particular order:

  \subsection{Departure from the nebular approximation}
  \label{sec:age-difference}

Reported departures from the nebular approximation in spectral modeling of the O\,\textsc{ii} W-shaped feature in SLSNe-I may reflect assumptions about the temporal anchoring of the ejecta evolution. In models where delayed energy injection reheats already expanding ejecta, the radiative state near peak luminosity does not necessarily correspond to the true hydrodynamical age of the explosion. If spectral modeling implicitly ties the ejecta age to this reheated radiation field, the inferred thermodynamic conditions may not accurately represent the actual expansion history.
 What appears as a significant departure from the nebular approximation could therefore arise from a mismatch between the radiative state and the underlying dynamical age, rather than from intrinsically strong non-thermal effects.

\subsection{The Pre-QN (pre-HS) phase}
\label{sec:pre-tQN-discussion}

The apparent scarcity of detected pre-reheating phases may be explained by survey depth, cadence, and selection effects.  
For fiducial parameters, the luminosity during the pre-$t_{\rm QN}$ phase is $L \sim 10^{42}\,\mathrm{erg\,s^{-1}}$, typically $\sim 4$–5 magnitudes fainter than the superluminous peak.  
At redshifts characteristic of current SLSN samples, such emission generally lies near or below single-epoch detection limits of wide-field surveys.  
Consequently, the onset of the reheated, superluminous phase is often identified observationally as the explosion epoch, even though it corresponds physically to delayed energy deposition into already expanding ejecta.

This selection bias is further reinforced by the operational definition of SLSNe, which preferentially selects events based on extreme peak luminosity, blue continua, and slow photometric evolution.  
Events for which the delayed energy injection is weak, mistimed, or inefficiently coupled to the ejecta would fail to reach the canonical superluminous threshold and would likely be classified as ordinary stripped-envelope SNe.  
As a result, magnitude-limited surveys preferentially detect systems in which the delayed engine operates near optimal efficiency (i.e., when $t_{\rm QN}\sim t_{\rm ej, d}$ in our model).

\subsection{Type-W and 2015bn-like SLSNe-I}

It has been suggested that SLSNe-I can be broadly divided into two spectroscopic subclasses (see \S~\ref{sec:introduction}): 
(i) events exhibiting a prominent W-shaped O\,\textsc{ii} absorption complex (“Type-W”), and 
(ii) those typified by SN~2015bn, in which this feature is weak or absent. We note that this classification is based on a small sample.

Within the delayed-energy framework presented here, both subclasses arise naturally from different regions of parameter space rather than from distinct physical mechanisms (see Figure \ref{fig:Mej-vsPms-TR-1}). 
The key controlling parameters are the ejecta diffusion time $t_{\rm ej,d}$, the timing of delayed energy injection $t_{\rm QN}$, and the total spin-down energy available for reheating.
In this interpretation, the presence or absence of the W-shaped feature is primarily a thermometer rather than an indicator of fundamentally different engine physics. 
Both subclasses represent delayed reheating of expanding SN ejecta; they differ in ejecta mass, timing, and thermal coupling efficiency. 
This unified explanation avoids the need to invoke separate progenitor channels while naturally accounting for the observed diversity in light-curve width and spectral morphology.

\subsection{Radiative filtering of the NS spin-down energy}

\citet{saito_2024} examine the role of non-thermal excitation and ionization in shaping the O\,\textsc{ii} W-shaped feature in SLSNe-I. While they find that large departures from the nebular approximation are generally not required, their analysis shows that the presence (or absence) of the W-shaped feature constrains the non-thermal ionization rate near the photosphere, thereby placing limits on the amount of $\gamma$-ray energy that can be deposited in the line-forming region.
In classical magnetar models, spin-down energy is injected directly into the SN ejecta, where high-energy photons and secondary electrons may alter ionization and excitation balances. Detailed radiative-transfer calculations are therefore needed to determine whether sufficient UV/optical photons can be produced without over-ionizing the O\,\textsc{ii} line-forming region or violating the ionization constraints inferred from the spectra.

In the QN framework, this potential tension is alleviated by the geometric and dynamical separation between the HS spin-down engine and the bulk SN ejecta. The HS spin-down energy is first absorbed by the optically thick QN ejecta, producing an LFBOT that acts as a calorimeter. In this configuration, the high-energy radiation from HS spin-down is efficiently thermalized and reprocessed into softer UV/optical photons before interacting with the SN ejecta. As a result, the line-forming region is exposed primarily to reprocessed radiation rather than to direct $\gamma$-ray deposition, naturally limiting the non-thermal ionization rate while maintaining the temperatures required for O\,\textsc{ii} excitation.

\subsection{Imprints of proto-NS magnetization on SLSNe-I light curves}

A central requirement of the delayed HS scenario is the formation of a massive ($M_{\rm NS} \sim 2\,M_{\odot}$), rapidly rotating (millisecond) NS. However, the subsequent light-curve morphology  also depends on the NS magnetic-field strength acquired at birth.

In our model, the morphology of SLSNe-I light curves reflects the efficiency of early magnetic braking. To have both fast rotation and a strong magnetic field during the proto-NS phase, an in-situ magnetic field amplification must be considered. Mechanisms include: convective dynamos (\citealt{thompson_1993,raynaud_2020,masada_2022}), and
MRI (\citealt{akiyama_2003,obergaulinger_2009,reboul-salze_2021}).

Pre-$t_{\rm QN}$ rapid magnetic braking extracts a substantial fraction of the NS rotational energy, leading to strong early energy injection into the SN ejecta. I.e., 
strongly magnetized NSs are expected to produce prominent early peaks  with comparatively weaker late-time. pots-$t_{\rm QN}$, re-brightening, whereas weak-field cases may yield double-peaked events or dominant late-time energy release. The observed diversity in single-peaked, double-peaked, and asymmetric light curves may therefore encode information about the NS progenitor core’s rotation profile and the efficiency of the underlying classical magnetic field amplification during the proto-NS phase.

\subsection{Early bumps, late bumps and undulations in light curves}
\label{sec:bumps}

\subsubsection{The early bumps as the visible peak of the primary SN emission}

Early-time photometry of SLSNe-I frequently reveals a short-lived luminous precursor or ``bump'' prior to the main peak. In most existing interpretations this feature is attributed to secondary processes such as shock cooling of extended material or CSM interaction. In this work we adopt a different interpretation and identify the early bump with the luminosity peak of the underlying core-collapse explosion itself. In this picture the bump corresponds to the ordinary stripped-envelope SN emission, analogous to a Type Ic event, powered primarily by radioactive decay of $^{56}$Ni together with modest early spin-down heating from the newly formed neutron star. The observed diversity in bump luminosity and temperature can therefore arise naturally from variations in the NS initial spin period and magnetic field strength.

\subsubsection{Late bumps from the QN–SN ejecta collision}

A key feature of the model, not considered here, is the collision of the QN ejecta with the slower-moving SN ejecta. 
Although the kinetic energy of the QN ejecta ($\sim 10^{50}\,\mathrm{erg}$) is smaller than the total rotational energy available from the spin-down of the HS the resulting shock can locally reheat the SN ejecta, producing noticeable structure in the light curve.  
Depending on the ejecta mass and density profile, this shock can generate short-lived bumps or inflections late in the light-curve evolution, and may subtly modify the spectral evolution.    In this sense, the QN–SN collision acts as a secondary engine, injecting energy on shorter timescales and potentially explaining fine-structure features in SLSN light curves that are not captured by the smooth spin-down luminosity alone.

Figure~4 in \citealt{ouyed_2025a} shows the manifestation of the collision between the two ejecta as a late-time bump in our model (see discussion therein).
Thus, in addition to the magnetar-like energy source of our model, it naturally incorporates a CSM-like interaction, where the SN ejecta itself acts as the pre-explosion shell. 
In our model, this shell is hydrogen-free, and no narrow lines are to be expected. 

The combination of multiple energy sources in our model provides a natural explanation for both early- and late-time bumps, 
and can reproduce the variety of SLSNe-I light curves observed in the literature, 
including single-peaked, double-peaked, and more complex morphologies.

\subsubsection{Late-time undulations}
\label{sec:undulations}

Oscillatory features have been reported in the late-time evolution of several SLSNe-I light curves (e.g., ???).  The QN model
offers plausible sources for such modulations. For example, it may arise if the newly formed HS is born precessing. Strong internal magnetic fields associated with the QCD-magnetar can induce a quadrupolar deformation of the HS, leading to a finite ellipticity and free precession of the rotation axis (e.g., \citealt{haskell_2008} and references therein). 

The characteristic precession period is approximately
\[
P_{\rm HS,prec} \sim \frac{P_{\rm HS}}{e_{\rm HS}},
\]
with ellipticity $e_{\rm HS} << 10^{-6}$ expected for the HS. This can yield precession periods significantly longer than the HS spin period.

Following the phenomenological  prescription in \citet{ouyed_2025a}, the luminosity of a precessing HS can be written as
\begin{align}
L_{\rm HS,SpD}^{\rm prec}(t) &= L_{\rm HS,SpD}(t)\times \\\nonumber
&\times \left[1+\alpha_{\rm HS}\exp\left(-\frac{1}{2}
\left(\frac{t-P_{\rm HS,prec}}{\beta_{\rm HS}P_{\rm HS,prec}}\right)^2\right)\right],
\end{align}
where $\alpha_{\rm HS}$ determines the amplitude of the modulation and $\beta_{\rm HS}$ controls its temporal width. Within the delayed energy-injection framework, and when applied to SN\,2023aew, this prescription reproduces the undulating features present in the late-time light curve (see Fig.~3 of \citealt{ouyed_2025a}). The same mechanism could therefore account for similar structures observed in the light curves of SLSNe-I.

An alternative mechanism for generating late-time oscillations involves torques exerted on the HS by fallback material from the QN ejecta. If the debris forms a transient Keplerian torus around the compact object (see \S\ref{sec:QN-and-HEA}), the interplay between radiation torques and viscous stresses within the disk can produce quasi-periodic variations in the accretion rate and energy output. The resulting timescales and energetics can be comparable to those required to explain the late-time undulations observed in some SLSNe light curves (see \citealt{ouyed_2014} for details).

\subsection{Beyond Type Ic Progenitors}

The key requirement of the delayed HS formation scenario is not the specific SN subtype but rather the birth of a sufficiently massive NS whose central density can reach the threshold for the hadron--quark phase transition. Although we have focused in this work on Type~Ic progenitors, the mechanism is therefore not intrinsically restricted to this subclass of stripped-envelope SNe. Our model naturally allows for complex multi-peaked light-curve morphologies to occur even in explosions whose peak luminosities are comparable to those of ordinary core-collapse SNe.

This could be the case for massive NSs born with moderate rotation. An example is SN~2023aew discussed in \citep{ouyed_2025a} and referenced when and where necessary in this work. In this framework, such events may represent manifestations of the same underlying engine operating under different environmental conditions, such as larger ejecta masses, lower rotational energies, or less efficient reprocessing of the injected energy. We have also suggested SN~2019stc (\citealt{gomez_2021}) as a plausible transitional candidate between these different regimes.

In this sense, the SLSNe-I considered in this work could represent the most luminous end of a wider class of explosions in which the formation of a HS (the QCD-magnetar) and the associated QN ejecta modify the late-time evolution of the SN.

\subsection{The QN and high-energy astrophysics}
\label{sec:QN-and-HEA}

\subsubsection{SGRs and AXPs}

Gamma-ray repeaters (SGRs) and anomalous X-ray pulsars (AXPs) have been interpreted as the HS (i.e. QCD-magnetar) progressively adjusting toward its new equilibrium configuration
(\citet{ouyed_2007a,ouyed_2007b}). 

The fallback debris from the QN ejecta mediates both the quiescent and bursting phases of the HS phase in its new found state. As shown in \citet{ouyed_2007a,ouyed_2007b}, the rotational state of the HS at birth (i.e., at $t_{\rm QN}$) determines the configuration of the QN fallback material. This material, with a typical mass of $\sim10^{-7}\,M_\odot$, is physically analogous to SN fallback, albeit at significantly lower mass accretion rates and under the strong influence of the HS magnetic field.

 Rapidly rotating HSs can redistribute the fallback debris into a Keplerian torus located at several HS radius. For longer initial spin periods, the material instead forms a magnetically confined, co-rotating shell.   The Keplerian  torus in our model is unlike a fall-back disk around a NS. The ring is rich in heavy elements, is very
close to the HS (a few stellar radii away) and is degenerate. Similar ring formation when a NS is born appears
implausible since the proto-NS is too large
 
 Using an angular momentum conservation argument, one finds that the upper value of the HS period capable of sustaining a Keplerian torus is 
 
 \begin{equation}
 P_{\rm HS, t} < \sim 4\ {\rm ms} \frac{B_{\rm HS, 15}^{3/2}}{(m_{\rm torus}/10^{-7}M_{\odot})^{3/4}}\ ,
 \end{equation}
 for $M_{\rm HS}=2M_{\odot}$ and $R_{\rm HS}=12$ km (see \citealt{ouyed_2007a,ouyed_2007b} for details).

The resulting high-energy phenomenology differs between these configurations: systems hosting Keplerian torii exhibit X-ray activity analogous to SGRs, while those with co-rotating shells display behavior reminiscent of AXPs.

We propose that less energetic SLSNe-I, associated with the slowest HSs in our model, would harbor SGRs as their compact remnants, whereas the most luminous SLSNe-I would be linked to AXPs. A statistical comparison between these phenomena, as well as an examination of their environmental properties, could provide an observational test of this scenario (see other testable predictions listed below).

\subsubsection{GRBs and FRBs}

Owing to the steep density gradients near the NS surface, a velocity stratification naturally arises when the outermost layers are ejected during the QN \citep{keranen_2005,ouyed_leahy_2009}. While the bulk of the ejecta is mildly relativistic, a small component with high-Lorentz-factor ($\Gamma_{\rm QN}> 10^2$) may accompany the outflow. This relativistic tail was connected to other classes of high-energy transients.
In particular, interaction between the relativistic part of the QN ejecta and the SN ejecta can lead to internal shocks capable of producing gamma-ray emission with spectral properties yielding a Band function (\citealt{ouyed_2020}; see especially Fig.~6 therein). 

 If, instead, the QN occurs in relative isolation, at sufficiently late times after the SN ejecta has dispersed, the freely expanding relativistic component may become susceptible to plasma instabilities. Under suitable conditions, such instabilities could generate coherent radio emission with properties reminiscent of Fast Radio Bursts \citep{ouyed_et_al_2025}. 
 
 \subsection{The QN and nuclear astrophysics}
\label{sec:QN-and-nuclear}
 
Previous studies have shown that the physical conditions in the QN ejecta are favorable for r-process nucleosynthesis, implying that QNe could contribute to the cosmic production of heavy elements in addition to core-collapse SNe and compact binary mergers \citep{jaikumar_2007,ouyed_2009}.  Furthermore, the neutron-rich QN ejecta cna act as a spallation source capable of modifying the composition of the SN ejecta late in its evolution if the QN ejecta catches up with the SN ejecta (see \citealt{ouyed_2011,ouyed_2015}).

\subsubsection{Observational discriminants between the QN-powered and classical magnetar scenarios}
\label{sec:QCD-magneatr-vs-standard-magnetar}

An important distinction between the delayed hybrid-star formation model proposed here and the standard magnetar-powered scenario lies in the nature of the outflow injected into the SN ejecta. In classical magnetar models, the rotational energy of the compact object is predominantly carried by a relativistic electron–positron pair wind and high-energy radiation. The interaction with the SN ejecta therefore primarily involves electromagnetic energy deposition and does not introduce significant additional baryonic material.

In contrast, in the QN framework the transition of the neutron star to a hybrid star is accompanied by the ejection of a small amount of neutron-rich material originating from the outermost layers of the neutron star. This QN ejecta carries both kinetic energy and baryonic matter. As it propagates through the previously expanding SN ejecta, it can reheat the ejecta and generate the late-time bump observed in some SLSNe light curves through the collision between the QN and SN ejecta. Importantly, the QN ejecta also acts as a baryonic buffer that can moderate the hard radiation produced by the spin-down of the newly formed hybrid star.

Because the QN ejecta is extremely neutron-rich, its interaction with the SN ejecta can lead to additional nucleosynthetic signatures. In particular, neutron spallation and related nuclear reactions may modify the chemical composition of the outer SN layers as the QN ejecta propagates through them. The subsequent cooling and mixing of the SN and QN material could therefore imprint distinctive abundance patterns or unusual spectral features at late times.

These processes provide potential observational diagnostics that could distinguish the QN-powered scenario from standard magnetar models. In the QN framework, the late-time bump would be associated not only with additional energy injection but also with the dynamical and chemical consequences of the neutron-rich QN ejecta. Detecting compositional anomalies, signatures of spallation products, or other spectroscopic evidence of baryon-loaded ejecta interacting with the SN envelope would therefore offer a direct test of the delayed hybrid-star formation scenario.

~\\

\begin{deluxetable*}{lcccccc}
\tablecaption{Best-fit model parameters at selected epochs for the SLSNe-I sample, adopted in \texttt{TARDIS} spectral modeling.\label{table:saito-points}}
\tablehead{
\colhead{SLSN-I} & 
\colhead{$t$ (days)} & 
\colhead{$L_{\rm Tot}\,(10^{44}~{\rm erg~s^{-1}})$} & 
\colhead{$\rho_{\rm ph}\,(10^{-14}~{\rm g~cm^{-3}})$} & 
\colhead{$v_{\rm ph}\,(10^4~{\rm km~s^{-1}})$} & 
\colhead{$R_{\rm ph}\,(10^{15}~{\rm cm})$} 
}
\startdata
PTF13ajg    & 57.7 & 5.0 & 2.2  & 1.05 & 5.2  \\
iPTF13ajg   & 62.8 & 4.8 & 1.9  & 1.03 & 5.6  \\
iPTF13ajg   & 76.3 & 3.6 & 1.5  & 1.00 & 6.5  \\
iPTF13ajg   & 92.5 & 2.0 & 1.1  & 0.95 & 7.6  \\
\hline
SN2010gx    & 46.6 & 2.3 & 1.6  & 1.11 & 4.5  \\
SN2010gx    & 50.1 & 2.5 & 1.4  & 1.10 & 4.8  \\
SN2010gx    & 57.7 & 2.5 & 1.2  & 1.06 & 5.3  \\
SN2010gx    & 75.0 & 1.4 & 0.83 & 1.00 & 6.5  \\
\hline
PTF09cnd    & 62.6 & 3.7 & 1.7  & 1.01 & 5.5  \\
PTF09cnd    & 79.6 & 3.4 & 1.1  & 0.98 & 6.7  \\
PTF09cnd    & 85.7 & 3.0 & 0.97 & 0.97 & 7.2  \\
PTF09cnd    & 104.3 & 2.0 & 0.73 & 0.94 & 8.4 \\
\hline
PTF09atu    & 73.5 & 2.9 & 1.3  & 0.88 & 5.6  \\
PTF09atu    & 76.6 & 2.9 & 1.2  & 0.87 & 5.8  \\
PTF09atu    & 99.2 & 2.1 & 0.83 & 0.83 & 7.1  \\
\enddata
\end{deluxetable*}
\vspace{2em} 
 %
\begin{table*}[t!]
\centering
\caption{\texttt{TARDIS} parameter settings corresponding to the best light-curve fits 
at the epochs listed in Table~\ref{table:saito-points} and shown as solid blue squares 
in Figures~\ref{fig:LC-iPTF13ajg}--\ref{fig:LC-PTF09atu}. 
The derived values of $L_{\rm Tot}$, $\rho_{\rm ph}$, and $v_{\rm ph}$ are used as inputs to \texttt{TARDIS}. 
The best spectral fits require a stratified density structure and a significantly reduced helium mass fraction compared to the uniform-composition models listed in Table~\ref{table:TARDIS-uniform-composition}.}
\begin{minipage}{\textwidth}
\begin{tabular}{|c|c|c|c|c|c|c|c|c|c|c|c|c|}\hline
\multicolumn{13}{|c|}{Composition (mass fraction) per velocity shell}  \\\hline
Velocity shell$^\dagger$ & He & C  & O & Ne & Mg & Si & S & Ca & Ti & Fe & Co & Ni  \\\hline
  $v_{\rm in}= v_{\rm ph}$  &  0.01 & 0.08 & 0.35 & 0.05 & 0.03 & 0.12 & 0.08 & 0.02 & 5$\times 10^{-4}$ & 0.13 & 0.09 & 0.04 \\\hline
$v_1$ &  0.005 & 0.18 & 0.48 & 0.05 &0.04 & 0.13 &0.07 & 0.015 & 3$\times 10^{-4}$ &0.02 &0.007 &0.003 \\\hline
$v_2$ &  0.002& 0.32& 0.58& 0.04& 0.025& 0.025& 0.008& 0.003& $10^{-4}$& 0.003& 5$\times 10^{-4}$& $10^{-4}$ \\\hline
$v_3$ & 0.001& 0.50& 0.40& 0.035& 0.015& 0.005& 0.002& 5$\times 10^{-4}$& 5$\times 10^{-5}$& 5$\times 10^{-4}$&  $10^{-4}$& $10^{-5}$ \\\hline
\end{tabular}
\hfill
\centering
 \begin{tabular}{|c|c|c|c|}\hline
\multicolumn{4}{|c|}{SN ejecta parameters at the epochs listed in Table \ref{table:saito-points}}  \\\hline
Luminosity requested  ($L_{\rm Tot}$) &  \multicolumn{3}{|c|}{Column 3 in Table \ref{table:the-Sobolev-times}} \\\hline
Density at the edge of the photosphere  ($\rho_{\rm ph}$) &  \multicolumn{3}{|c|}{Column 4 in Table \ref{table:the-Sobolev-times}}  \\\hline
Velocity at the edge of the photosphere ($v_{\rm in}=v_{\rm ph}$)  &  \multicolumn{3}{|c|}{Column 5 in Table \ref{table:the-Sobolev-times}}  \\\hline
Velocity at the edge of the ejecta ($v_{\rm out}$)   & \multicolumn{3}{|c|}{$2\times 10^4$ km s$^{-1}$} \\\hline 
\multicolumn{4}{|c|}{Plasma parameters}\\\hline
Ionization mode & \multicolumn{3}{|c|}{nebular}\\\hline
Excitation mode & \multicolumn{3}{|c|}{dilute-lte} \\\hline
\end{tabular}\\
$^\dagger$ The first shell is at $v_{\rm ph}$ with a corresponding density $\rho_{\rm ph}$. Subsequent shells are
at velocity $v_i = v_{\rm ph} + i\times \Delta v$ with $\Delta v = (v_{\rm out}-v_{\rm in})/N_{\rm shells}$ and $i=0,1,.., (N_{\rm shells}-1)$. The density scales as 
$\rho_i = \rho_{\rm ph} (v_{\rm ph}/v_i)^{n_{\rm ej}}$. Here, $N_{\rm shells}=4$. 
 \end{minipage}
  \label{table:TARDIS-stratified-composition}
\end{table*}

We now discuss some model's limitations before listing testable predictions


\subsection{Metallicity, rotation, and the onset of the delayed HS engine}

Extreme SLSNe-I require rotational energies of order $\sim 10^{52},\mathrm{erg}$, corresponding to initial HS spin periods
$\lesssim 2$ ms.  The parent NS period may be even smaller. Forming such rapidly rotating NS is challenging, particularly for hydrogen-poor progenitors that have undergone substantial envelope stripping prior to collapse.

The observed preference of SLSNe-I for sub-solar metallicity hosts (see \S~\ref{sec:introduction}) provides an important environmental constraint. Lower metallicity reduces wind-driven angular momentum losses, allowing massive stars to retain more rapidly rotating cores at collapse. These progenitors are also expected to harbour more massive stellar cores, which in turn are more likely to produce massive NSs which are more likely to trigger quark nucleation in their centers (e.g., \citealt{staff_2006}). In this sense, metallicity regulates both the retention of angular momentum and the probability of crossing the quark deconfinement threshold, naturally linking the environmental preference of SLSNe-I to the operation of a delayed central engine.

We emphasize, however, that once both conditions are satisfied, the delay itself is controlled by the microphysics of quark nucleation in the NS core. Our fits to light curves and spectra of SLSNe-I and SLSNe-II  in general (e.g., \citealt{leahy_ouyed_2008,ouyed_leahy_2013,ouyed_2013,ouyed_2016}) suggest that this process requires several weeks to a few months to reach percolation and convert the NS core to quark matter. As discussed in Appendix \ref{sec:SLSNe-QCD}, the duration of this delay may therefore provide indirect constraints on key parameters of the physics underlying the hadronic-to-quark matter transition.

Systems that satisfy the rotation requirement but fall short of the mass threshold, even if they experience efficient dynamo or MRI amplification, may instead produce ordinary magnetar-powered Type Ic events without delayed re-brightening. Conversely, systems lacking sufficiently rapid rotation will resemble conventional stripped-envelope SNe.

A limitation of the delayed hybrid-star model is therefore its reliance on a restricted region of progenitor parameter space. The model does not eliminate the angular-momentum problem, but rather shifts part of the required tuning from magnetic-field strength to a microphysically determined density threshold. Improved observational constraints on host metallicity, ejecta mass, and remnant properties will be essential to assess how frequently such conditions are realized.

\subsection{Spectral fitting strategy and caveats}
\label{sec:SN2010gx-fits}

For simplicity, we adopted a single stratified chemical composition for all four SLSNe analyzed here. While this common abundance structure provides encouraging fits to the observed spectra, broadly consistent with the corresponding light curve properties, certain objects may benefit from tailored adjustments to their composition profiles. For instance, SN2010gx appears to require a modified stratification to achieve improved agreement with the three observed spectra we selected. A comprehensive study in which each event is modelled with its own optimized abundance profile will be addressed in future work. Nonetheless, the application of a unified composition demonstrates that the overall spectral characteristics can already be reproduced within a consistent framework.

\subsection{Reduced parameter survey}
\label{sec:limitations-parameter}

In this work we have adopted a reduced parameter survey in order to illustrate the main physical features of the model. In particular, the delay time $t_{\rm QN}$ was kept fixed. Exploring this additional degree of freedom together with variations in the QN ejecta's properties such as mass, velocity, and energy
 may therefore broaden the range of observable light-curve morphologies produced by the model.

We have also assumed a fixed HS surface magnetic field of $B_{\rm HS}=10^{15}$~G. In reality, the surface field may vary between events. In QCD-based models of magnetized quark matter, the internal magnetic field in the quark core may be set by the microphysics of the phase and could therefore be approximately universal. However, the observable surface field depends on the size of the quark core through magnetic flux conservation and under the assumption of a dipolar field geometry, variations in the core radius would naturally lead to different surface magnetic fields from source to source.

A more complete exploration of the parameter space is required to assess the robustness and range of applicability of the model.

\subsection{Other limitations}
\label{sec:other-limitations}

We did not include ionization effects from NS spin-down power in the pre-$t_{\rm QN}$ phase. 
The O\,\textsc{ii} population could be influenced by non-thermal excitation and ionization whenever 
$\gamma$-rays from the NS spin-down are present in the ejecta, potentially modifying the early-time spectral features. 
While NS spin-down primarily affects the pre-$t_{\rm QN}$ phase, $^{56}$Ni decay could also contribute in the post-$t_{\rm QN}$ phase. 
In our models, we assumed a low $^{56}$Ni content in the ejecta ($M_{\rm Ni}/M_{\rm ej} = 0.01$), 
which may be lower than typical for Type Ic SNe.  This remains to be investigated.

~\\
Some testable predictions include

\begin{itemize}

\item {\bf Observational signatures of the delayed energy input:} A key discriminator is that the effective expansion age inferred spectroscopically, $\equiv R_{\rm ph}/v_{\rm ph}$, exceeds the photometric age inferred from the light curve.  Late-time spectra provide an independent and largely model-insensitive constraint on the true explosion epoch. Once the effects of delayed energy injection have faded and the evolution is dominated again by radioactive decay, the inferred ejecta radius, velocity, and temperature should reflect the actual time since explosion. If late-time spectral dating implies an explosion epoch much earlier than that inferred by anchoring the timeline to the luminous peak, this should support delayed-energy injection.

\item {\bf Pre-$t_{\rm QN}$:} The epoch preceding the delayed energy injection at $t_{\rm QN}$ corresponds to a conventional Type Ic supernova powered by a combination of $^{56}$Ni decay and NS spin-down.   Spectra obtained during this pre-reheating phase (e.g. in the case for the earliest detections of PTF09cnd and PTF09atu) are therefore expected to resemble those of ordinary Type Ic SNe.  
In particular, the characteristic W-shaped O\,\textsc{ii} absorption complex should be absent.  
Instead, the spectra should be consistent with expanding, cooling SN ejecta powered by internal radioactive and NS rotational energy deposition.

Systematic early-time spectroscopic campaigns could therefore directly test the existence of the proposed pre-QN phase.

\item {\bf Expansion velocities:} Assuming the opacity and ejecta mass remain approximately constant between the two phases, the relative widths of the two peaks depend primarily on the change in ejecta velocity produced by the delayed engine activity. Denoting the characteristic velocities before and after the transition by $v_1$ and $v_2$, respectively, the ratio of the diffusion timescales should satisfy
\begin{equation}
\frac{ (t_{\rm ej, d})_2 }{  (t_{\rm ej, d})_1 }  \sim \left( \frac{ v_{\rm ej, 1} }{ v_{\rm ej, 2} }\right)^{1/2}\ ,
\end{equation}
and
\begin{align}
\frac{ L_{\rm pk, 2} }{  L_{\rm pk, 1} }  &\sim \frac{E_{\rm HS, SpD}/(t_{\rm ej, d})_2}{E_{\rm NS, SpD}/(t_{\rm ej, d})_1}\\\nonumber
& \sim  \left(\frac{P_{\rm NS}}{P_{\rm HS}}\right)^2 \left( \frac{ v_{\rm ej, 2} }{ v_{\rm ej, 1} }\right)^{1/2}\sim \frac{1}{2}\left( \frac{ v_{\rm ej, 2} }{ v_{\rm ej, 1} }\right)^{1/2}\ ,
\end{align}
where $P_{\rm HS}/P_{\rm NS}= (1+t_{\rm QN}/t_{\rm NS, SpD})^{1/2} \sim 2^{1/2}$ is used which is when pre-QN phase is bright enough to be detected
and yielding a double-peaked SLSN-I. This can be tested using photometric and spectroscopic observations, from which the peak luminosities and ejecta velocities during the two phases can be inferred.

\item {\bf Unusual lines at unusual imes:}  Emission lines of [Ca\,\textsc{ii}] during the nominally photospheric phase (i.e., on top of the photospheric spectral features)
has been reported in some SLSNe-I \citep{galyam_2009,inserra_2017,anderson_2018}.  Their origin remains to be explained. 
Such lines would suggest the ejecta has reached sufficiently low densities to permit forbidden transitions. 
Clumping, engine-driven instabilities, or shell fragmentation have been proposed as mechanisms to produce localized low-density regions \citep{chen_2016,nicholl_2019}. 

The delayed reheating model should allow for the co-existence of highly ionized absorption features and nebular-like emission at epochs that, under standard assumptions, correspond to relatively young ejecta ages.  More generally, the coexistence of apparently “old” spectroscopic features with “young” photometric ages reflects the decoupling of expansion time and luminous reheating. 
Systematic searches for unexpected early forbidden lines in SLSNe-I therefore provide a direct and testable probe of delayed-engine scenarios.

\item {\bf LFBOTs and SLSNe-I:} For $2t_{\rm ej,d} < t_{\rm QN} < t_{\rm NS,SpD}$, the LFBOT will be directly visible 
(e.g., AT2024wwp (\citealt{lebaron_2026}) shown in the top panels of Figures~\ref{fig:LC-iPTF13ajg}--\ref{fig:LC-PTF09atu}; see also \citealt{ouyed_2025b}). 
The total energy released in these events should resemble that of SLSNe because the only difference is that the LFBOT is reprocessed by the SN ejecta.

The connection is
\begin{equation}
L_{\rm SLSN, pk} \sim L_{\rm LFBOT, pk} \times \frac{t_{\rm QN,d}}{t_{\rm ej,d}}\ ,
\end{equation}
where the corresponding peak luminosities and diffusion or rise timescales ($t_{\rm QN,d}$ and $t_{\rm ej,d}$) can be estimated from LFBOT and SLSN light curves.

\item {\bf SLSNe-I vs SLSNe-II:}  In the QN model, both hydrogen-poor SLSNe-I and hydrogen-rich SLSNe-II can be powered by a delayed HS engine. 
Because SLSNe-II progenitors retain massive hydrogen envelopes, the energy injected at $t_{\rm QN}$ is distributed over a larger mass. 
As a result, the radiation temperature is lower, the photosphere expands more slowly, and the light curve is broader and smoother. This may  explain the absence of the high-ionization W-shaped O\,\textsc{ii} absorption complex in SLSNe-II spectra.  

In essence, the same engine operates in both classes, but the envelope modifies the observable properties such as rise, peak temperature and photospheric velocities. Subtle imprints of the engine (and may be even of the hidden LFBOT) may still be detectable in late-time spectra, but these should be far less pronounced than in SLSNe-I.

\item {\bf HSs at the high-mass end of the NS population}:  Our model invokes NSs on the high-mass end of the distribution permitted by the equation of state (i.e., $M_{\rm NS}\sim 2\,M_{\odot}$). Because the transition to a HS ejects only a negligible amount of mass ($M_{\rm QN} \ll M_{\rm NS}$), the resulting HS should retain nearly the same mass as its parent NS. Consequently, HSs (i.e., QCD-magnetars) are expected to occupy the upper end of the NS mass distribution rather than forming a distinct population of much lighter remnants. Future measurements of the masses and radii of compact remnants associated with SLSNe-I, if they become feasible, could test this prediction.
 
\item {\bf Magnetic field extremes and a characteristic $B_{\rm HS}$:}  If indeed QCD-driven magnetic-field amplification operates during the formation of the HS, the resulting dipole fields may systematically exceed those achievable through conventional convective dynamo and MRI mechanisms. If the quark matter phase established in the HS core is universal (see Appendix \ref{sec:SLSNe-QCD}), the amplification process may saturate at a characteristic field strength $B_{\rm HS, c}$ in the HS core. In that case, QCD-magnetars produced through this channel would be expected to cluster around similarly extreme field values. Observationally, a concentration of very high inferred dipole fields in systems showing evidence of delayed energy injection (e.g., double-peaked SLSNe-I) would lend support to this interpretation.

\item {\bf Hard emission in NS-powered SLSNe-I:}  The filtering of the HS emission by the QN ejecta into an LFBOT minimizes the effects of hard emission from spin-down power in the HS scenario. 
We predict that SLSNe-I powered by classical magnetars (i.e., when $t_{\rm NS, SpD} \ll t_{\rm QN}$, showing only a single bright peak) should exhibit spectral signatures of hard emission as well as evidence of continuous energy injection. 
In these events, the hydrodynamic age inferred from observations would be consistent with the spectroscopic age, unlike the case with delayed energy injection (see \S \ref{sec:age-difference}).

\item {\bf Undulations in LFBOT light curves:} As discussed in \S\ref{sec:undulations}, the late-time undulations observed in SLSNe-I light curves may arise from the precession or external torquing 
of the newly formed HS. In the delayed energy-injection framework, such modulations should first appear in the luminosity output of the central engine and thus be imprinted on the LFBOT emission
before being reprocessed by the surrounding SN ejecta. Consequently, LFBOTs occurring in isolation, namely after the SN ejecta have largely
dissipated (i.e., when $t_{\rm ej,d} < t_{\rm QN} < t_{\rm NS,SpD}$), should retain this undulation signature more directly in their light curves.

\item {\bf Multi-messenger probes:} Beyond electromagnetic signatures, gravitational waves and neutrinos may be emitted. For instance, a rapid reorganization of the stellar core during conversion could generate a faint, yet potentially observable, burst of gravitational waves, especially from nearby sources (see Appendix in \citealt{staff_2012}). Similarly, if deconfinement occurs in the presence of neutrino trapping, the transition might be accompanied by a short-lived neutrino signal (see \citealt{ouyed_2022a,ouyed_2022b} and references therein). Although detecting such signals is challenging, they would offer independent evidence for our model.

\end{itemize}

Finally, in Appendix~\ref{sec:SLSNe-QCD} we discuss how SLSNe-I data can be used to infer properties of quark matter, and the implications for QCD in general and for the hadron-to-quark matter phase transition in particular.

\section{Conclusion}
\label{sec:conclusion}

In the QN framework, SLSNe-I arise from delayed internal energy deposition into already expanding, relatively old, Type Ic SN ejecta, rather than from a prompt or continuously powered central engine. 

The defining ingredient of the model is the delayed formation of a HS (a QCD-magnetar in our model), which
resets the powering engine (a millisecond nascent NS) and  injects energy into dilute ejecta weeks to months after the Type Ic SN core collapse. This naturally produces reheating at large radii, high radiation temperatures near maximum light, and the spectroscopic conditions required for strong O\,\textsc{ii} absorption.  While the QN realization provides a concrete mechanism, the broader phenomenology applies to any scenario in which magnetar-strength energy injection is activated after a substantial delay and at the energy scales explored here. 

Importantly, in this framework the delay time is controlled primarily by quark nucleation physics (e.g., surface tension and critical density effects), rather than by 
 NS progenitor's or environmental conditions.  The delayed formation of the HS is rooted in the microphysics of dense matter and is therefore fundamentally connected to QCD which speculates on deconfined quark phases at the extreme densities expected in the  cores of massive NS capable of sustaining internal magnetic fields up to $\sim 10^{18}\,$G (see Appendix \ref{sec:SLSNe-QCD}). This corresponds to a surface magnetic field of $\sim 10^{15}$~G, providing an additional pathway for magnetic amplification beyond classical mechanisms such as the convective dynamo and MRI during the proto-NS stage.

 In this picture, the transition from a NS  to a quark-containing HS triggers rapid magnetic-field amplification and a dramatic shortening of the NS spin-down timescale. The NS’s rotational energy is thus partitioned into two physically distinct eras. This is not a phenomenological decomposition of a magnetar light curve, but a physically triggered second engine phase governed by microphysics.

The model accommodates events with clear double-peaked light curves as well as those without an obvious first peak, depending on the relative ordering of $t_{\rm QN}$, $t_{\rm NS,SpD}$
and $t_{\rm ej, d}$.When $t_{\rm QN} < t_{\rm NS,SpD}$, a substantial fraction of the NS rotational energy is deposited into already expanded ejecta, producing a luminous second heating episode under conditions favorable for strong O\,\textsc{ii} absorption.   In particular, when $t_{\rm QN} \sim t_{\rm NS,SpD}$, the two peaks are equally important
but not hot enough to accommodate the W-shaped O\,\textsc{ii}  absorption. 

Our model predicts systematic offsets between spectroscopic expansion ages and photometric ages, particularly when the initial explosion epoch (with $L_{\rm Tot}< 10^{42}$ erg s$^{-1}$) is unobserved because pre-QN emission falls below survey detection limits. Late-time spectra should therefore provide the most reliable estimates of the true explosion time. Coordinated early-time photometric and spectroscopic monitoring offers a direct observational test of this prediction and provides a means of distinguishing delayed-engine scenarios from prompt or continuously powered magnetar models.

The QN framework, in which a NS undergoes a transition to a HS hosting deconfined quark matter, provides a unifying physical picture for a wide range of high-energy astrophysical phenomena. In this scenario, the newly formed HS acts as a QCD-magnetar whose rotational and magnetic energy can power diverse transients depending primarily on the surrounding environment and evolutionary stage of the system. When the conversion occurs within a dense SN envelope, the interaction between the QN ejecta and the SN ejecta can produce SLSNe. If the transition occurs after the envelope has dissipated, or in a more tenuous environment, the same central engine can manifest through other high-energy phenomena such as GRBs, FRBs, or the activity associated with AXPs and SGRs. The diversity of observed outcomes therefore reflects differences in the external conditions rather than the presence of fundamentally different central engines.

Within this framework, the QCD-magnetar represents an evolutionary pathway distinct from that of the classical magnetar formed directly in a core-collapse SN without a phase transition. Both objects may exhibit strong magnetic fields and rapid rotation, but the QCD-magnetar arises from the conversion of hadronic matter to quark matter and is therefore associated with the additional energy release and dynamical effects of the QN event. Considering both channels together provides a more natural and comprehensive framework for interpreting the wide range of still enigmatic high-energy transients observed in the universe. An important challenge for future work will be to determine whether observational signatures exist that can distinguish QCD-magnetars from their classical counterparts in astrophysical data. As we have emphasized earlier, one key differentiating factor is the presence of the QN ejecta, which is absent in standard magnetar models. In addition to moderating the hard emission from the HS spin-down, the QN ejecta can alter the composition and energetics of the SN ejecta, potentially producing observable features (such as late-time bumps or spallation signatures in the spectra)  that provide a direct probe of the QCD-magnetar engine.

If the QN-driven delayed-energy scenario is correct, SLSNe-I provide a unique astrophysical probe of dense QCD matter. The requirement that the HS  can sustain magnetic fields approaching $\sim 10^{18}$~G at core densities of a few times the nuclear saturation density imposes stringent constraints on viable quark-matter phases at play. Moreover, the observed delay times $t_{\rm QN}$ may directly reflect the stochastic nucleation timescale for deconfinement in NS cores. Measuring $t_{\rm QN}$ across a statistical sample of SLSNe-I could therefore provide empirical constraints on the physics of the hadron-to-quark phase transition, including the critical quark deconfinement density, and  the surface tension of quark matter.

%

\section*{DATA AVAILABILITY}

No new data were generated or analysed in support of this research.

\section{SLSNe as Probes of the Hadron-Quark Matter Phase Transition}
\label{sec:SLSNe-QCD}

Here we discuss the connection between our model and QCD, and how observations of SLSNe-I can be used to place constraints on fundamental properties of quark matter. We emphasize that the workflows presented here are intended as an exploratory inference scheme rather than a precise inversion, as they rely on several simplifying assumptions in the modeling of light curves and spectra. Nevertheless, they illustrate how multi-wavelength observations of double-peaked SLSNe-I can provide a potential avenue for probing the internal structure of compact stars and the physics of the hadron-to-quark phase transition. For convenience, in this appendix we adopt natural units with $\hbar = c = 1$.

\subsection{QCD scale Magnetic fields and quark deconfinement density}
\label{sec:estimating-nB}

Some QCD-based models involving up and down quarks suggest that extremely strong magnetic fields can be spontaneously generated in supranuclear quark matter (e.g., \citealt{iwazaki_2003,iwazaki_2005,ebert_2005,dvornikov_2016,efrain_2021}). In these phases, the magnetization is primarily determined by the quark chemical potential, $\mu_{\rm q}$.
For a degenerate two-flavor quark matter core, the quark chemical potential is related to the baryon number density through
$\mu_{\rm q} \simeq (\pi^2 n_B)^{1/3}$. 
 Given the QCD energy scale of $\sim 260$ MeV, compared to $\sim 8$ MeV for hadronic matter, one would generally expect such a quark matter to sustain magnetic fields of order $\sim 10^{18}$ G in the core of the HS in our model.  Fields of $\sim 10^{15}$ G at the stellar surface are therefore easily achievable if the field adopts a dipolar configuration.
~\\~\\
Here, we focus on color ferromagnetism (e.g., \citealt{iwazaki_2005}), though similar considerations apply to other models. 
If the quark core enters this phase, a color magnetic (i.e., a chromomagnetic) field, $B_{\rm g}$, can be spontaneously generated by gluon dynamics 
 followed by the observable magnetic field, $B_{\rm q}$, produced by the quarks responding to the color field. 
This is because  the number of negative  charged quarks is different from the number of positive
charged quarks. The difference induces a current yielding (see Eq. (8) in \citealt{iwazaki_2005})
\begin{equation}
B_{\rm q} \sim 4\times 10^{17}~{\rm G}~  \frac{100~{\rm MeV}} {\sqrt{g B_{\rm g}}}\, \frac{n_B}{1~{\rm fm}^{-3}}\ ,
\label{eq:iwazaki-M}
\end{equation}
where  $\sqrt{gB_{\rm g}}$ is  the energy scale (here in units of 100 MeV) associated with the spontaneously generated chromomagnetic field $B_{\rm g}$;  $g$ is
 the fundamental QCD gauge coupling related to the  strong coupling constant $\alpha_{\rm s} = g^2/4\pi$.
~\\~\\
Appendix~\ref{appendix:workflow} outlines how fits to the light curves and spectra of SLSNe can be used to estimate the internal properties of the HS, including the core magnetic field $B_{\rm q}$ and the corresponding baryon density $n_{\rm B}$, i.e., the quark deconfinement density. The procedure is as follows: first, determine the HS surface magnetic field $B_{\rm HS}$ required to reproduce the post-$t_{\rm QN}$ phase of the light curve. This can subsequently be used to infer the internal core field $B_{\rm HS, c}$. Finally, setting $B_{\rm q} = B_{\rm HS, c}$ in Eq.~(\ref{eq:iwazaki-M}) allows one to estimate the quark deconfinement density in the HS core.

\subsection{Surface tension from measured $t_{\rm QN}$}
\label{sec:estimating-sigma}

The delay time $t_{\rm QN}$ is associated with the quantum nucleation time of a critical quark-matter droplet in the cold, deleptonized NS  core (see \S4 of \citealt{ouyed_2025a}; also \citealt{ouyed_2022a} and references therein). The formation of a quark bubble requires overcoming the surface-energy barrier at the hadron–quark interface. A detailed discussion of the nucleation physics can be found in the literature (e.g., \citealt{horvath_1992,olesen_1993,heiselberg_1993,iida_1997,glendenning_1997,iida_1998,alford_2001,harko_2004,bombaci_2004,voskresensky_2003}). Here, we focus on the essential aspects relevant for connecting $t_{\rm QN}$ to the nucleation parameters.~\\~\\

The free-energy change for a quark droplet of radius $R_{\rm D}$ is given by 
\begin{equation}
\Delta F(R_{\rm D}) = -\frac{4\pi}{3}R_{\rm D}^3\Delta P + 4\pi R_{\rm D}^2\sigma \ ,
\end{equation}
where $\sigma$ is the hadron–quark surface tension and 
$\Delta P$ is the pressure difference between quark and hadronic matter. 

The critical radius above which the droplet becomes energetically favored is obtained from $d\Delta F/dR_{\rm D}=0$, yielding
\begin{equation}
R_{\rm D, c} = \frac{2\sigma}{\Delta P}\ .
\end{equation}
The corresponding critical free-energy barrier is then 
\begin{equation}
\Delta F_c = \frac{16 \pi \sigma^3}{3 (\Delta P)^2}\ .
\end{equation}

The nucleation rate (number of critical bubbles formed per unit volume per unit time) can be expressed in terms of 
Euclidean action for tunnelling   $S_E \sim \Delta F_c/\mu_q$, assuming a spherical droplet. It is,
\begin{equation}
\Gamma \sim \mu_q^4 \, e^{-S_E} \sim \mu_q^4 \exp\left(-\frac{16 \pi \sigma^3}{3 (\Delta P)^2 \mu_q}\right) \ .
\end{equation}

The nucleation timescale for forming a single critical bubble in the NS core of volume $V_c=\frac{4\pi}{3}R_c^3$  is
\begin{equation}
t_{\rm QN} \sim \frac{1}{\Gamma V_c}\ .
\end{equation}
Once a critical droplet forms, the subsequent growth of a macroscopic quark core proceeds rapidly on a dynamical timescale $\sim R_c/c\sim10^{-6}$-$10^{-5}$ s.
 This gives us

\begin{equation}
\label{eq:sigma}
\frac{\sigma^3}{(\Delta P)^2}\sim \frac{3\mu_{\rm q}}{16\pi} \ln{(t_{\rm QN}\mu_{\rm q}^4 V_{\rm c})}\ .
\end{equation}
 
 This relation shows that an observational estimate of the nucleation time $t_{\rm QN}$, together with a typical NS core volume $V_c$ and quark chemical potential $\mu_q$, constrains the combination $\sigma^3/(\Delta P)^2$ characterizing the hadron–quark interface. Because the dependence is logarithmic, the inferred ratio is only weakly sensitive to uncertainties in $t_{\rm QN}$ or $\mu_q$, even though the nucleation time itself depends exponentially on this ratio.  
Remarkably, the weeks-to-months delays suggested by our SLSN light-curve and spectral fits imply that this ratio must lie within a relatively narrow range, hinting at a quasi-universal value for the hadron–quark interface under NS core conditions.~\\~\\

Although $\mu_{\rm q}$ can be inferred from fits to double-peaked SLSNe-I light curves (see Appendix~\ref{sec:estimating-nB}), the size of the HS core, $R_{\rm c}$, must still be determined; we discuss this estimate below.

\subsection{Conversion energy release and late-time bumps in SLSNe}

In our model, the late-time bumps observed in SLSNe light curves is related to the collision of QN ejecta with the previously ejected SN material. The total energy of the bump, $E_{\rm bump}$, provides an estimate of the kinetic energy of the QN ejecta. This kinetic energy is only a fraction ($\sim 0.01$) of the total conversion energy released during the neutron-to-quark transition; i.e.   $E_{\rm conv} \sim 100 E_{\rm bump}$ (\citealt{keranen_2005,ouyed_2022a}). ~\\~\\  

The total number of neutrons converted in the core is then
\begin{equation}
N_{\rm n} \sim \frac{E_{\rm conv}}{\Delta E_{\rm n}} \,,
\end{equation}
where $\Delta E_{\rm n} \sim 100~{\rm MeV}$ is the energy released per neutron. With an independent estimate of the baryon density $n_{\rm B}$ (see Appendix~\ref{sec:estimating-nB}), the effective volume of the HS core can be inferred as
\begin{equation}
V_{\rm c} = \frac{N_{\rm n}}{n_{\rm B}} \,.
\end{equation}

The total number of quarks is also related to the quark chemical potential via
$N_{\rm n} \simeq \frac{\mu_{\rm q}^3}{\pi^2} V_{\rm c}$ which 
 allows one to connect the observed SLSN bump energy to the nucleation parameters of the hadron–quark interface. The combination $\sigma^3/(\Delta P)^2$ appearing in the nucleation rate can then be expressed  as
\begin{equation}
\label{eq:sigma2}
\frac{\sigma^3}{(\Delta P)^2} \sim \frac{3 \mu_{\rm q}}{16 \pi} 
\ln \left( \pi^2 \, \mu_{\rm q} \, t_{\rm QN} \, E_{\rm bump, MeV} \right) \,,
\end{equation}
where $E_{\rm bump, MeV}$ is the bump energy in MeV.  ~\\~\\

In summary, SLSNe-I observations provide a potential route to constraining the physical properties of the HS core: the surface magnetic field, core size, baryon density, and, through Eq.~(\ref{eq:sigma2}), the hadron–quark interface parameters such as surface tension and overpressure. These quantities in turn can inform fundamental QCD parameters, including the quark deconfinement density and the characteristics of supranuclear quark matter.

\section{Inferring Hybrid-Star magnetic filed from double-peaked SLSNe}
\label{appendix:workflow}

Magnetar models of SLSNe are known to be prone to parameter degeneracies when fitting a single light curve peak. In particular, the peak luminosity and diffusion timescale can often be reproduced by multiple combinations of the NS initial spin period $P_{\rm NS}$ and surface magnetic field $B_{\rm NS}$. As a result, fits to single-peaked SLSNe typically yield broad allowed ranges for the compact-object parameters.~\\~\\

In our model, double-peaked SLSNe provide an additional constraint that significantly reduces this degeneracy. The key observable is the delay $t_{\rm QN}$ between the first and second bumps of the light curve. The spin period of the HS at the time of its formation is uniquely related to the initial NS spin period through

\begin{equation}
P_{\rm HS}=P_{\rm NS}\left(1+\frac{t_{\rm QN}}{t_{\rm NS,SpD}}\right)^{1/2} \ .
\label{appendix:PHS-vs-PNS}
\end{equation}

I.e., $t_{\rm QN}$ acts as a critical anchor that links the two phases of the explosion and alleviates the degeneracy present in standard single-peak magnetar models.

\subsection*{Step 1: Constraining the ejecta properties}

The ejecta properties must first be constrained from the data. The diffusion timescale governing each bump depends on the ejecta mass and velocity through

\begin{equation}
t_{\rm ej,d} \propto 
\left(\frac{\kappa M_{\rm ej}}{v_{\rm ej} c}\right)^{1/2}.
\end{equation}

Spectroscopic observations provide measurements of the ejecta velocity from line widths, while the light curve rise time constrains the corresponding diffusion timescale. In our model the ejecta mass remains constant between the two phases, but the ejecta velocity may change after the re-heating in the post-$t_{\rm QN}$ epoch. As a result, the diffusion timescale of the second bump may differ from that of the first phase.

\subsection*{Step 2: Fit the first bump}

The first bump is modeled as energy injection from NS spin-down deposited into the SN ejecta, with a possible contribution from $^{56}$Ni decay. Using the ejecta properties derived above, this phase constrains combinations of the NS spin period $P_{\rm NS}$ and magnetic field $B_{\rm NS}$ that reproduce the observed luminosity and temporal evolution of the first bump. As in standard magnetar models fitting methods, this step generally yields a family of degenerate solutions in the $(P_{\rm NS},B_{\rm NS})$ parameter space.

\subsection*{Step 3: Use the observed delay}

The delay between the two bumps, $t_{\rm QN}$, is measured directly from the light curve. For each allowed $(P_{\rm NS},B_{\rm NS})$ solution obtained in Step 2, the NS spin-down evolution determines the spin period of HS  at $t_{\rm QN}$ through Eq.~(\ref{appendix:PHS-vs-PNS}). This fixes the corresponding HS spin period $P_{\rm HS}$.

\subsection*{Step 4: Model the second bump}

In our model the second bump is not powered directly by magnetar emission. Instead, the newly formed HS injects energy into the QN ejecta that produces the LFBOT (see \S \ref{sec:model}). The LFBOT radiation is subsequently reprocessed by the expanding SN ejecta, producing the observed optical emission of the second bump.~\\~\\

Given $P_{\rm HS}$,  the temporal evolution of the second bump primarily constrains the HS surface magnetic field $B_{\rm HS}$ together with the efficiency with which the LFBOT luminosity is reprocessed by the SN ejecta.

\subsection*{Step 5: Selecting the consistent solution}

Only a subset of the $(P_{\rm NS},B_{\rm NS})$ solutions obtained from the first bump will produce a second bump consistent with the observed luminosity and temporal evolution once the constraint on $P_{\rm HS}$ is applied. The requirement that both bumps be reproduced simultaneously therefore selects a narrow region of parameter space including $B_{\rm HS}$.

\subsection*{Step 6: Inferring the HS core magnetic field}

Finally, the internal magnetic field of the HS can be estimated assuming magnetic flux conservation and a dipolar field geometry. This estimate can then be used in Eq.~(\ref{eq:iwazaki-M}) to infer the HS core baryon density.~\\~\\

Additional self-consistency checks (e.g., using spectroscopic or multi-band photometric constraints) could be applied, but are left for future work.

\end{appendix}


\begin{thebibliography}{99}

\bibitem[Aamer et al.(2025)]{aamer_2025}  Aamer, A., Nicholl, M. Gomez, S. et al.\ 2025, \mnras, 541, 2674

\bibitem[Akiyama et al.(2003)]{akiyama_2003} Akiyama, S., Wheeler, J. C., Meier, D. L., \& Lichtenstadt, I.\ 2003, \apj, 584, 954

\bibitem[Alford et al.(2001)]{alford_2001} Alford, M., Rajagopal, K., Reddy, S. \& Wilczek, F.\ 2001, \prd 64, 074017

\bibitem[Anderson et al.(2018)]{anderson_2018} Anderson, J. P., Pessi, P. J., Dessart, L., et al. 2018, A\&A, 620, A67

\bibitem[Angus et al.(2019)]{angus_2019} Angus, C. R., Smith, M., Sullivan, M. et al.\ 2019, \mnras, 487, 2215

\bibitem[Arnett(1982)]{arnett_1982} Arnett W. D. 1982, ApJ, 253, 785

\bibitem[Barbary et al.(2009)]{barbary_2009} Barbary, K, Dawson, K. S., Tokita, K. et al.\ 2009, \apj, 690, 1358

\bibitem[Blanchard  et al.(2020)]{blanchard_2020} Blanchard, P. K., Berger, E., Nicholl, M. \& Villar V. A.\ 2020, \apj, 897, 114

\bibitem[Bombaci et al.(2004)]{bombaci_2004}  Bombaci, I., Parenti, I. \& Vida{\~n}a, I.\ 2004, \apj,  614, 314

\bibitem[Branch \& Wheeler(2017)]{branch_2017} Branch, D. \&. Wheeler, J. C.\ 2017, ``Supernova Explosions". Series: Astronomy and Astrophysics Library  (Springer-Verlag GmbH Germany) 

\bibitem[Chatzopoulos et al.(2012)]{chatzopoulos_2012} Chatzopoulos, E., Wheeler, J. C. \& Vink\'o, J.\ 2012, \apj,  746 121

\bibitem[Chen et al.(2013)]{chen_2013} Chen, T.-W., Smartt, S. J., Bresolin, F., et al.\ 2013, \apjl, 763, L28

\bibitem[Chen et al.(2016)]{chen_2016} Chen, K.-J., Woosley, S. E. \& Sukhbold, T.\ 2016, \apj, 832, 73

\bibitem[Chen et al.(2023)]{chen_2023} Chen Z. H., et al.,2023, \apj, 943, 41

\bibitem[Chevalier(1977)]{chevalier_1977} Chevalier, R.~A.\ 1977, \araa, 15, 175 

\bibitem[Chevalier \& Irwin(2011)]{chevalier_2011} Chevalier R. A. \& Irwin C. M.\ 2011, \apj, 729, L6

\bibitem[Contopoulos et al.(1999)]{contopoulos_1999} Contopoulos I., Kazanas D. \& Fendt C., 1999, \apj, 511, 351

\bibitem[Chomiuk, et al.(2011)]{chomiuk_2011} Chomiuk, L., Chornock, R., Soderberg, A. M., et al. 2011, \apj, 743, 114

\bibitem[Dessart et al.(2012)]{dessart_2012} Dessart L., Hillier D. J., Waldman R., Livne E. \& Blondin S., 2012, \mnras, 426, L76

\bibitem[Dexter \& Kasen(2013)]{dexter_2013} Dexter J. \& Kasen D.\ 2013, \apj, 772, 30

\bibitem[Drout et al.(2014)]{drout_2014} Drout, M. R., Chornock, R., Soderberg, A. M et al., 2014, \apj, 794, 23

\bibitem[Duncan \& Thompson(1992)]{duncan_1992} Duncan, R. C. \& Thompson, C.\ 1992, \apj, 392, L9

\bibitem[Dvornikov(2016)]{dvornikov_2016} Dvornikov., M.\ 2016, Phys. Lett. B, 760, 406

\bibitem[Ebert et al.(2005)]{ebert_2005} Ebert, D.,  Zhukovsky, V. Ch. \&  Tarasov, O. V.\ 2005, \prd 72, 096007

\bibitem[Efrain \& de la Incerra(2021)]{efrain_2021} Efrain, E. J. \& de la Incerra, V.\ 2021, Universe, 7, 458

\bibitem[Frohmaier et al.(2021)]{frohmaier_2021} Frohmaier C., et al.\ 2021, \mnras, 500, 5142

\bibitem[Gal-Yam et al.(2009)]{galyam_2009} Gal-Yam, A. et al.\ 2009, Nature, 462, 624

\bibitem[Gal-Yam(2012)]{galyam_2012} Gal-Yam, A. 2012, Sci, 337, 927

\bibitem[Gal-Yam(2018)]{galyam_2018} Gal-Yam, A.\ 2018, \apj, 866, L7

\bibitem[Gal-Yam(2019)]{galyam_2019} Gal-Yam, A.\ 2019, ARA\&A, 57, 305

\bibitem[Glendenning(1997)]{glendenning_1997} Glendenning, N. K.\ 1997, Compact Stars. Springer-Verlag, New York

\bibitem[Gomez et al.(2020)]{gomez_2020} Gomez S., Berger E., Blanchard P. K. et al. 2020, \apj, 904, 74

\bibitem[Gomez et al.(2021)]{gomez_2021} Gomez S., Berger E., Hosseinzadeh, G.\ et al. 2021, \apj, 913, 143

\bibitem[Gomez et al.(2024)]{gomez_2024} Gomez S., et al.\ 2024, \mnras, 535, 471

\bibitem[Guti\'errez et al.(2022)]{gutierez_2022} Guti\'errez, C. P., et al.\ 2022, \mnras, 517, 2056

\bibitem[Haensel \& Zdunik(1989)]{haensel_1989} Haensel P., Zdunik J. L. \& Schaeffer R.\ 1989, A\&A, 217, 137

 \bibitem[Harko et al.(2004)]{harko_2004}  Harko, T.,  Cheng, K. S. \&  Tang, P. S.\ 2004, \apj, 608, 945
 
 \bibitem[Haskell, et al.(2008)]{haskell_2008} Haskell, B., Samuelsson, L., Glampedakis, K. \& Andersson, N.\ 2008, \mnras, 385, 531
 
 \bibitem[Hatsukade et al.(2018)]{hatsukade_2018} Hatsukade, B., Tominaga, N., Hayashi, M., et al. 2018, ApJ, 857, 72
 
 \bibitem[Heger \& Woosley(2002)]{heger_2002} Heger, A., \& Woosley, S. E.\ 2002, \apj, 567, 532

 \bibitem[Heiselberg, et al.(1993)]{heiselberg_1993} Heiselberg, H.,  Pethick, C. J. \& Staubo, E. F.\ 1993, \prl,  70, 1355 

\bibitem[Horvath et al.(1992)]{horvath_1992} Horvath, J. E., Benvenuto, O. G. \&, Vucetich, H.\ 1992, \prd,  45, 3865

\bibitem[Howell et al.(2013)]{howell_2013} Howell D. A., et al.\ 2013, \apj, 779, 98

 \bibitem[Iida \& Sato(1997)]{iida_1997} Iida, K., \& Sato, K. 1997, Prog. Theor. Phys., 98, 277
 
 \bibitem[Iida \& Sato(1998)]{iida_1998} Iida, K. \& Sato, K.\ 1998, \prc, 58, 2538 

\bibitem[Inserra et al.(2013)]{inserra_2013}  Inserra, C., Smartt, S. J., Jerkstrand, A., et al.\ 2013, \apj, 770, 128

\bibitem[Inserra et al.(2017)]{inserra_2017} Inserra, C.\ et al.\ 2017, \mnras, 468, 4642

\bibitem[Inserra (2019)]{inserra_2019}  Inserra C., 2019, Nature Astronomy, 3, 697

\bibitem[Iwazaki(2003)]{iwazaki_2003}  Iwazaki, A. \& Morimatsu, O.\ 2003, Phys. Lett. B, 571, 61 

\bibitem[Iwazaki(2005)]{iwazaki_2005} Iwazaki, A.\ 2005, Phys. Rev. D, 72, 114003

\bibitem[Jaikumar et al.(2007)]{jaikumar_2007} Jaikumar, P., Meyer, B. S., Otsuki, K. \& Ouyed, R.\ 2007, A\&A, 471, 227

\bibitem[Kasen \& Bildsten(2010)]{kasen_2010} Kasen D., Bildsten L. 2010, ApJ, 717, 245

\bibitem[Kerzendorf \& Sim(2014)]{kerzendorf_2014} Kerzendorf, W. E. \& Sim, S. A. 2014, \mnras, 440, 387

\bibitem[Ker\"anen et al.(2005)]{keranen_2005} Ker\"anen, P., Ouyed, R. \& Jaikumar, P.\ 2005, \apj, 618, 485

\bibitem[K\"onyves-T\'oth \& Vink\'oo(2021)]{konyves-toth_2021}  K\"onyves-T\'oth, R. \& Vink\'oo, J. 2021, \apj, 909, 24

\bibitem[K\"onyves-T\'oth(2022)]{konyves-toth_2022}  K\"onyves-T\'oth, R.\ 2022, \apj, 940, 69

\bibitem[Kumar et al.(2020)]{kumar_2020} Kumar, A., Pandey, S. B.,  K\"onyves-T\'oth, R., et al.\ 2020, \apj, 892, 28

\bibitem[Kurucz \& Bell(1995)]{kurucz_1995} Kurucz, R. L. \& Bell, B., 1995, Atomic line list, KurCD, 23

\bibitem[Lander(2021)]{lander_2021} Lander, S.~K. 2021, MNRAS, 507, L36

\bibitem[Langer(2012)]{langer_2012} Langer, N.\ 2012, ARA\&A, 50, 107

\bibitem[Lattimer \& Schutz(2005)]{lattimer_2005} Lattimer, J. M. \& Schutz, B. F., 2005, \apj, 629, 979

\bibitem[Law et al.(2009)]{law_2009} Law, N. M., Kulkarni, S. R., Dekany, R. G., et al.\ 2009, PASP, 121, 1395

\bibitem[Leahy \& Ouyed(2008)]{leahy_ouyed_2008} Leahy, D., \& Ouyed, R.\ 2008, \mnras, 387, 1193

\bibitem[LeBaron et al.(2026)]{lebaron_2026} LeBaron, N.,  Margutti, R., Chornock, R. et al.\ 2026, \apjl, 997, L10

\bibitem[Leloudas et al.(2012)]{leloudas_2012} Leloudas, G. et al.\ 2012, A\&A, 541, A129

\bibitem[Liu et al.(2017)]{liu_2017} Liu, Y.-Q., Modjaz, M. \& Bianco, F. B.\ 2017, \apj, 845, 85

\bibitem[Lucy(1991)]{lucy_1991} Lucy, L. B.\ 1991, \apj, 383, 308

\bibitem[Lucy(1999)]{lucy_1999} Lucy, L. B., 1999, A\&A, 345, 211

\bibitem[Lunnan et al.(2014)]{lunnan_2014} Lunnan, R., Chornock, R., Berger, E., et al.\ 2014, \apj, 787, 138

\bibitem[Lyne, Pritchard  \& Smith(1993)]{lyne_1993} Lyne, A. G., Pritchard, R. S. \& Smith, F. G.\ 1993, \mnras, 265, 1003

\bibitem[Manchester \& Taylor,(1977)]{manchester_1977} Manchester, R. N. \& Taylor, J. H.\ 1977, in Pulsars, ed. R. N. Manchester \& J. H. Taylor (San Francisco, CA: W. H. Freeman), 281

\bibitem[Masada, et al.(2022)]{masada_2022} Masada, Y., Takiwaki, T., \& Kotake, K.\ 2022, \apj, 924, 75

\bibitem[Mazzali \& Lucy(1993)]{mazzali_1993} Mazzali, P.~A. \& Lucy, L. B.\ 1993, A\&A, 279, 447

\bibitem[Mazzali et al.(2016)]{mazzali_2016} Mazzali, P.~A., et al.\ 2016, \mnras, 458, 3455

\bibitem[Metzger(2022)]{metzger_2022} Metzger, B. D., 2022, \apj, 932, 84

\bibitem[Michel \& Goldwire(1970)]{michel_1970} Michel, F. C. \& Goldwire, H. C. J., 1970, \apj, L5, 21

\bibitem[Mihalas(1978)]{mihalas_1978} Mihalas, D. 1978, Stellar Atmospheres (Freeman and Company)

\bibitem[Moriya(2013)]{moriya_2013} Moriya, T. J., Blinnikov, S. I., Baklanov, P. V., Sorokina, E. I., \&
Dolgov, A. D.\ 2013, \mnras, 430, 1402

\bibitem[Moriya, Sorokina \& Chevalier(2018)]{moriya_2018} Moriya, T. J., Sorokina, E. I. \& Chevalier, R. A.\ 2018, Space Sci. Rev., 214, 59

\bibitem[Moriya(2024)]{moriya_2024} Moriya, T. 2024, ``Superluminous supernovae", a chapter for the Encyclopedia of Astrophysics (edited by I. Mandel, section
editor F.R.N. Schneider) [arXiv:2407.12302]

\bibitem[Nahar(1999)]{nahar_1999} Nahar, S. N. 1999, ApJS, 120, 131

\bibitem[Neill et al.(2011)]{neill_2011} Neill, J. D. et al.\ 2011, \apj, 727, 15

\bibitem[Nicholl et al.(2013)]{nicholl_2013} Nicholl M. et al.\ 2013, Nature, 502, 346

\bibitem[Nicholl et al.(2016)]{nicholl_2016}  Nicholl, M. Smartt, S. J., Jerkstrand, A.\ et al.\ 2016, \mnras, 452, 3869

\bibitem[Nicholl, et al.(2019)]{nicholl_2019} Nicholl, M., Berger, E., Blanchard, P. K. et al.\ 2019, \apj, 871, 102

\bibitem[Nicholl(2021)]{nicholl_2021} Nicholl, M.\ 2021, ``Superluminous supernovae: an explosive decade", Astronomy and Geophysics, vol. 62, no. 5,
pp. 5.34–5.42. https://doi.org/10.1093/astrogeo/atab092

\bibitem[Nomoto et al.(2007)]{nomoto_2007} Nomoto, K., Tominaga, N., Tanaka, M., Maeda, K., \& Umeda, H.
2007, in American Institute of Physics Conference Series, Vol. 937, Supernova 1987A: 20 Years After: Supernovae and
Gamma-Ray Bursters, ed. S. Immler, K. Weiler, \& R. McCray, 412-426

\bibitem[Obergaulinger et al.(2009)]{obergaulinger_2009} Obergaulinger, M., Cerdá-Dur\'an, P., M\"uller, E., \& Aloy, M. A.\ 2009, A\&A, 498, 241

\bibitem[Ofek et al.(2007)]{ofek_2007} Ofek, E. O., Cameron, P. B., Kasliwal, M. M., et al.\ 2007, \apj, 659, L13

 \bibitem[Olesen \& Madsen(1993)]{olesen_1993} Olesen, M. L. \& Madsen, J.\ 1993, \prd  47, 2313 

\bibitem[Ostriker \& Gunn(1971)]{ostriker_1971} Ostriker, J. P. \& Gunn, J. E.\ 1971, \apj, 164, L95

\bibitem[Ouyed et al.(2007a)]{ouyed_2007a} Ouyed, R., Leahy, D. \& Niebergal, B.\ 2007a, A\&A 473, 357

\bibitem[Ouyed et al.(2007b)]{ouyed_2007b} Ouyed, R., Leahy, D. \& Niebergal, B.\ 2007b, A\&A, 475, 63

\bibitem[Ouyed \& Leahy(2009)]{ouyed_leahy_2009} Ouyed, R. \& Leahy, D.\ 2009, \apj, 696, 562

\bibitem[Ouyed et al.(2009)]{ouyed_2009} Ouyed, R., Pudritz, R. E \& Jaikumar, P.\ 2009, \apj, 702, 1575

\bibitem[Ouyed et al.(2011)]{ouyed_2011} Ouyed, R.,   Leahy, L.,  Ouyed, A. \& Jaikumar, P., J.\ 2011, \prl, 107, 151103

\bibitem[Ouyed \& Leahy(2013)]{ouyed_leahy_2013} Ouyed, R. \& Leahy, D. 2013, RAA, 13, 1202

\bibitem[Ouyed et al.(2013)]{ouyed_2013} Ouyed, R. Koning, N. \& Leahy, D\ 2013, RAA 13, 1463

\bibitem[Ouyed et al.(2014)]{ouyed_2014} Ouyed, R., Leahy, D. \& Koning, N.\ 2014, Astrophysics and Space Science, 352, 715

\bibitem[Ouyed et al.(2015)]{ouyed_2015} Ouyed, R..  Leahy, D. \& Koning, N.\ 2015, RAA, 15, 483

\bibitem[Ouyed et al.(2016)]{ouyed_2016} Ouyed, R., Leahy, D. \& Koning, N. \ 2016, \apj, 818, 77

\bibitem[Ouyed et al.(2020)]{ouyed_2020} Ouyed, R., Leahy, D. \& Koning, N.\ 2020, RAA, 20, 027

\bibitem[Ouyed(2022a)]{ouyed_2022a} Ouyed R. 2022a, ``The Micro-physics of the Quark-nova: Recent Developments”. Astrophysics in the XXI Century with Compact
Stars", 53-83 [eISBN 978-981-12-2094-4]

\bibitem[Ouyed(2022b)]{ouyed_2022b} Ouyed R. 2022b, Universe, 8, 322

\bibitem[Ouyed et al.(2025)]{ouyed_et_al_2025} Ouyed, R., Leahy, D. \& Koning, N.\ 2025, \mnras, 537, I2876

\bibitem[Ouyed(2025a)]{ouyed_2025a} Ouyed, R.\ 2025a, \mnras, 543, 3885

\bibitem[Ouyed(2025b)]{ouyed_2025b} Ouyed, R.\ 2025b   [arXiv:2506.20540]

\bibitem[Pastorello et al.(2010)]{pastorello_2010} Pastorello, A., Smartt, S. J., Botticella, M. T., et al. 2010, \apjl, 724, L16

\bibitem[Perley et al.(2016)]{perley_2016} Perley, D. A. et al.\ 2016, \apj, 830, 13

\bibitem[Pursiainen et al.(2018)]{pursiainen_2018} Pursiainen, M., Childress, M., Smith, M. et al., 2018, \mnras, 481, 894

\bibitem[Quimby et al.(2007)]{quimby_2007} Quimby, R. M., Aldering, G, Wheeler, J. C. et al.\ 2007,  \apjl, 668, L99

\bibitem[Quimby et al.(2011)]{quimby_2011} Quimby R. M., Kulkarni S. R., Kasliwal M. M. et al. 2011, Nature, 474, 487

\bibitem[Quimby et al.(2013)]{quimby_2013} Quimby, R. M., Yuan, F., Akerlof, C. \& Wheeler, J. C.\ 2013, \mnras, 431, 912

\bibitem[Quimby et al.(2018)]{quimby_2018} Quimby R. M., et al., 2018, \apj, 855, 2

\bibitem[Rau et al.(2009)]{rau_2009} Rau, A., Kulkarni, S. R., Law, N. M., et al.\ 2009, PASP, 121, 1334

\bibitem[Raynaud et al.(2020)]{raynaud_2020} Raynaud, R., Guilet, J., Janka, H.-T., \& Gastine, T.\ 2020, Science Advances, 6

\bibitem[Reboul-Salze et al.(2021)]{reboul-salze_2021} Reboul-Salze, A., Guilet, J., Raynaud, R., \& Bugli, M.\ 2021, A\&A, 645, A109

\bibitem[Reynolds \& Chevalier(1984)]{reynolds_1984} Reynolds, S. P., \& Chevalier, R. A. 1984, ApJ, 278, 630 

\bibitem[Saito et al.(2024)]{saito_2024} Saito, Y., Tanaka, M., \& Mazzali, P.~A.\ 2024, \apj, 967,13

\bibitem[Schulze et al.(2018)]{schulze_2018} Schulze, S., Kr\"uhler, T., Leloudas, G., et al.\ 2018, \mnras, 473, 1258

\bibitem[Shapiro \& Teukolsky(1983)]{shapiro_1983} Shapiro, S. L. \& Teukolsky\ S. A.\ 1983, Black holes, white dwarfs, and neutron stars, John Wiley \& Sons, New York

\bibitem[Shivvers et al.(2019)]{shivvers_2019} Shivvers, I., Filippenko, A. V., Silverman, J. M. et al.\ 2019, \mnras, 482, 1545

\bibitem[Smith et al.(2008)]{smith_2008} Smith, N, Foley, R. J, Bloom, J. S. et al.\ 2008, \apj, 686, 485

\bibitem[Sobolev(1957)]{sobolev_1957} Sobolev, V. V.\ 1957, Soviet Ast., 1, 678

\bibitem[Soker \& Gilkis(2017)]{soker_2017} Soker, N. \& Gilkis, A., 2017, \apj, 851, 95

\bibitem[Staff et al.(2006)]{staff_2006} Staff, J. E., Ouyed, R. \& Jaikumar, P.\ 2006, \apjl,  645, L145

\bibitem[Staff et al.(2012)]{staff_2012} Staff, J. E.,  Jaikumar, P., Chan, V.  \& Ouyed, R.\ 2012, \apj, 751, 24

\bibitem[Stoll et al.(2011)]{stoll_2011} Stoll, R., Prieto, J. L., Stanek, K. Z. et al.\ 2011, \apj, 730, 34

\bibitem[Taddia et al.(2015)]{taddia_2015}  Taddia, F. Sollerman, J., Leloudas, G.\ et al.\ 2015, A\&A 574, A60 

\bibitem[Thompson \& Duncan(1993)]{thompson_1993} Thompson, C. \& Duncan, R. C.\ 1993, \apj, 408, 194

\bibitem[Vogl et al.(2019)]{vogl_2019} Vogl, C., Sim, S. A., Noebauer, U. M. et al.\ 2019, A\&A 621, A29 

\bibitem[Voskresensky et al.(2003)]{voskresensky_2003} Voskresensky, D. N., Yasuhira, M. \& Tatsumi, T.\ 2003, Nucl. Phys. A ,723, 291

\bibitem[Vreeswijk et al.(2014)]{vreeswijk_2014} Vreeswijk, P. et al.\ 2014, \apj, 797, 24

\bibitem[Weber(2005)]{weber_2005} Weber, F.\ 2005, Progress in Particle and Nuclear Physics, 54, 193

\bibitem[Wheeler et al.(2015)]{wheeler_2015} Wheeler, J. C., Johnson, V. \& Clocchiatti, A. 2015, \mnras, 450, 1295

\bibitem[Wiese(1996)]{wiese_1996} Wiese, W. L. 1996, in AIP Conf. Proc. 381, Atomic Processes in Plasmas
(Tenth), ed. A. L. Osterheld \& W. H. Goldstein (Melville, NY: AIP), 177

\bibitem[Woosley(2010)]{woosley_2010} Woosley, S. E., 2010, \apj, 719, L204

\bibitem[Yaron \& Gal-Yam(2012)]{yaron_2012} Yaron, O. \& Gal-Yam, A. 2012, PASP, 124, 668

\bibitem[Young et al.(2010)]{young_2010} Young, D. R., Smartt, S. J., Valenti, S., et al. 2010, A\&A, 512, A70



\begin{appendix}
\end{thebibliography}
\end{document}